\documentclass[12pt]{article}
\usepackage{amssymb}

\newcommand{\bmat}{\left(\begin{array}}
\newcommand{\emat}{\end{array}\right)}
\def\NPB#1#2#3{Nucl. Phys. B{#1} (19#2) #3}

\def\yzero{\smash{\hbox{$y\kern-4pt\raise1pt\hbox{${}^\circ$}$}}}

\def\g{\gamma}

\def\-{\hphantom{-}}

\def\s2{\frac{1}{2}}

\def\oh{\frac{1}{2}}
\def\beq{\begin{equation}}
\def\eeq{\end{equation}}
\def\beqa{\begin{eqnarray}}
\def\eeqa{\end{eqnarray}}

\def\IF{\relax{\rm I\kern-.18em F}}
\def\II{\relax{\rm I\kern-.18em I}}
\def\IP{\relax{\rm I\kern-.18em P}}
\def\IC{\relax\hbox{\kern.25em$\inbar\kern-.3em{\rm C}$}}
\def\IR{\relax{\rm I\kern-.18em R}}

\def\Dsl{\,\raise.15ex\hbox{/}\mkern-13.5mu D} %this one can be subscripted

\def\IC{\bf C}

\newcommand{\drawsquare}[2]{\hbox{%
\rule{#2pt}{#1pt}\hskip-#2pt%  left vertical
\rule{#1pt}{#2pt}\hskip-#1pt%  lower horizontal
\rule[#1pt]{#1pt}{#2pt}}\rule[#1pt]{#2pt}{#2pt}\hskip-#2pt%  upper horizontal
\rule{#2pt}{#1pt}}% right vertical

\newcommand{\fund}{\raisebox{-.5pt}{\drawsquare{6.5}{0.4}}}%  fund
\newcommand{\Ysymm}{\raisebox{-.5pt}{\drawsquare{6.5}{0.4}}\hskip-0.4pt%
        \raisebox{-.5pt}{\drawsquare{6.5}{0.4}}}%  symmetric second rank
\newcommand{\Yasymm}{\raisebox{-3.5pt}{\drawsquare{6.5}{0.4}}\hskip-6.9pt%
        \raisebox{3pt}{\drawsquare{6.5}{0.4}}}%  antisymmetric second rank
\newcommand{\antifund}{\overline{\fund}}

\catcode`\@=11
\newdimen\@rotdimen
\newbox\@rotbox

\def\@vspec#1{\special{ps:#1}}%  passes #1 verbatim to the output
\def\@rotstart#1{\@vspec{gsave currentpoint currentpoint translate
   #1 neg exch neg exch translate}}% #1 can be any origin-fixing transformation
\def\@rotfinish{\@vspec{currentpoint grestore moveto}}% gets back in synch
\def\@rotr#1{\@rotdimen=\ht#1\advance\@rotdimen by\dp#1%
   \hbox to\@rotdimen{\hskip\ht#1\vbox to\wd#1{\@rotstart{90 rotate}%
   \box#1\vss}\hss}\@rotfinish}
\def\@rotl#1{\@rotdimen=\ht#1\advance\@rotdimen by\dp#1%
   \hbox to\@rotdimen{\vbox to\wd#1{\vskip\wd#1\@rotstart{270 rotate}%
   \box#1\vss}\hss}\@rotfinish}%
\def\@rotu#1{\@rotdimen=\ht#1\advance\@rotdimen by\dp#1%
   \hbox to\wd#1{\hskip\wd#1\vbox to\@rotdimen{\vskip\@rotdimen
   \@rotstart{-1 dup scale}\box#1\vss}\hss}\@rotfinish}%
\def\@rotf#1{\hbox to\wd#1{\hskip\wd#1\@rotstart{-1 1 scale}%
   \box#1\hss}\@rotfinish}%
\def\rotate{\@ifnextchar[{\@rotate}{\@rotate[l]}}
\def\@rotate[#1]#2{\setbox\@rotbox=\hbox{#2}\@nameuse{@rot#1}\@rotbox}

\catcode`\@=12
\topmargin
-1.5cm
\textwidth
15.5cm
\textheight
23.5cm
\oddsidemargin
0.7cm
\evensidemargin
1.2cm

\begin{document}
%----------------------------------------------------------------------%
%  numbering equations with section number
%----------------------------------------------------------------------%
\makeatletter
\@addtoreset{equation}{section}
\makeatother
\renewcommand{\theequation}{\thesection.\arabic{equation}}
%----------------------------------------------------------------------%
%  title page
%----------------------------------------------------------------------%
\pagestyle{empty}
\rightline{CAB-IB/2902603} \rightline{\tt hep-th/0307183}
\vspace{0.5cm}

\begin{center}
{\LARGE {Type IIB orientifolds on Gepner points \\[10mm]
} }{\large {Gerardo Aldazabal $^1$, Eduardo C. Andr{\'e}s $^1$, Mauricio
Leston $^2 $ and \\[2mm]
 Carmen N{\'u}{\~n}ez $^2$
\\[2mm]
} }{\small {\ $^1$ Instituto Balseiro, CNEA, Centro At{\'o}mico Bariloche,\\[%
-0.3em]
8400 S.C. de Bariloche, and CONICET, Argentina.\\[1mm]
$^2$ Instituto de Astronom\'{\i}a y F\'{\i}sica del Espacio, CONICET \\[%
-0.3em]
C.C. 67 - Suc. 28, 1428 Buenos Aires, Argentina, and \\[-0.3em]
Departamento de F\'{\i}sica, FCEN-UBA, Argentina \\[9mm]
} {\bf Abstract} \\[7mm]
}

\begin{minipage}[h]{14.0cm}
We study various aspects of  orientifold projections of Type IIB closed
string
theory on Gepner points in different dimensions.
The open string sector is introduced, in the usual constructive  way, in
order to
cancel RR charges carried by orientifold planes.
Moddings by cyclic permutations of the internal $N=2$ superconformal
blocks as well as
 by discrete phase symmetries are implemented. Reduction
in the number of generations, breaking or enhancements of gauge symmetries
and topology changes are shown to be induced by such moddings.
Antibranes sector is also considered;
in particular we show how non supersymmetric models with antibranes and
free of closed and open tachyons do appear in this context.
A systematic study of consistent models
in $D=8$  dimensions and some illustrative examples in $D=6$ and $D=4$
dimensions are presented.
\end{minipage}
\end{center}

\newpage \setcounter{page}{1} \pagestyle{plain}
%
%EndExpansion
\setcounter{footnote}{0}
%----------------------------------------------------------------------%
%  Paper begins
%----------------------------------------------------------------------%

\section{Introduction}
Compactifications of ten dimensional $E_8 \times E_8$ heterotic string down to four
dimensional theories with close resemblance to the Standard Model
(or extensions of it)  defined the
scenario for the so called string phenomenology, since the middle
eighties.
The guide lines were established in \cite{chsw} where  $E_8 \times E_8$
heterotic
string compactified on a Calabi-Yau manifold was shown to lead to an $N=1$
three generations $E_6$ model. Further breaking of $E_6$ may  be achieved,
for instance,
by turning on Wilson lines.
Quite soon, also other compactifications were considered according to
these  ideas, such as those
 involving compact orbifolds or free fermionic string
models in $E_8 \times E_8$ and, in a less degree,  $SO(32)$ heterotic string \cite{pheno}. For
 pioneer work on Type I model construction see \cite{review} and references therein.

Particularly  relevant  for our following discussion  are the Gepner
models  proposed in \cite{gepner}. That work provides an algebraic construction of
supersymmetric string theory in even $D$ smaller than 10
 dimensions,
in terms of solvable rational  conformal internal theories,
without any reference to the original theory in ten dimensions.
Some evidence for identification of such constructions with Calabi-Yau
compactifications was also advanced.

Since the middle nineties, the irruption of {\it dualities} has  marked a
drastic
 change in our view of string theories, both from  theoretical and
 phenomenological perspectives. D-branes play a prominent role in this new approach.

The fact that branes localize gauge interactions on their world volumes
sets   a new scenario for string phenomenology, where  particle
theories are confined into  world branes. Since several  features of
such theories appear to depend  upon just the local behavior of strings in
the vicinity of D-branes ( without even considering compactifications),
the appealing  possibility of a {\it bottom-up} \cite{botup}
  approach opens up. In this approach a local world
 Type II brane model (resembling the Standard Model) is built up in a
first step and is successively
embedded in a global consistent string model. It
 proves to be very powerful for toroidal like compactifying manifolds
with either 
branes at orbifold like singularities or at angles \cite{intersec},
needed in order to achieve chirality (see also  \cite{bbkl} for intersecting branes  and
 Calabi-Yau).

The relevant  Type I string theory on generic Calabi-Yau manifolds is much
more
cumbersome. In particular, the geometry of D-branes becomes fuzzy and a
{\it bottom-up} like construction  becomes somewhat out of control.
However, we would like to stress that important steps towards the
understanding of the algebraic and geometrical interpretation of D-branes
in such generic cases, have been achieved \cite{fuchs2, huis, ABbranes}.
This might, hopefully,  lead to a model building {\it bottom-up} like
 procedure in a near future.

Beyond their  phenomenological interest, Type I Calabi-Yau
compactifications
provide a fruitful arena to study Type I - heterotic dualities. As we know, this is
an essential ingredient in the program to realize the nature of M-theory.
In particular, it appears  worth studying compactifications in different
dimensions
as a  relevant step pointing to establish connections within the  intricate web of
string dualities.

The present article deals with the building of  such kind of
models in the special  points of Calabi-Yau  moduli space
described by Gepner models. Previous work on this subject has been
developed in recent years. Open descendants of Gepner models have
been discussed in \cite{gepnertv1,gepnert2}. D-branes
\cite{recknagels,recknagelpb, satoh} and orientifold planes in these
models have been considered in \cite{horibr,goma}.

Type I open plus closed unoriented  superstring theory in ten dimensions can
be constructed from Type IIB superstrings. The IIB theory is known to be
invariant under the exchange of left and right moving sectors. When such a
symmetry is modded out the resultant theory seems to be inconsistent. Such
inconsistency manifests, for instance, through the appearance of unphysical
tadpoles in string amplitudes and can be interpreted as an unbalanced charge
under RR closed string fields carried by orientifold 9-planes  \cite{polchi}.
Full consistency is recovered by adding an open string sector with open
strings ending on D9-branes carrying opposite RR charge. It is in this sense
that the open string sector appears as a twisted sector for the left-right
exchange projection. The solution of tadpole cancellation conditions fix the
Chan Paton gauge groups. The same scheme is valid in lower dimensions
whenever left and right movers are coupled symmetrically.

In  the present paper we  follow these steps by starting with   Type IIB theories
where the internal sector is built up from Gepner models.
Our main aim here is to develop a systematic procedure to handle such models. 
In particular,
 we show  how moddings by phase symmetries or by cyclic permutations may be
implemented  in order to  achieve  partial control on the number of generations, the
breaking or enhancement of gauge symmetries and supersymmetry, etc.
The introduction of antibranes sectors is also discussed.
 Several examples in $D=8,6$ and $4$ dimensions are constructed and
presented for illustrative purposes. A more biased study towards
 phenomenologically interesting models or the finding of heterotic duals, for which
the methods developed here should be helpful, is postponed for future work.

The paper is organized as follows. In Sections 2 and 3, which  contain
brief reviews of
the partition function in Type I superstring theory and of Gepner models
respectively,  the main ideas of the
construction are developed and  notation is set up. The vacuum
amplitude in Type I theory at Gepner points is discussed in Section 4
and it is illustrated through explicit examples in $D=8, 6$ and $4$
spacetime
dimensions
 in Sections 5, 6 and 7,  where the matter content
and gauge groups of Chan Paton factors leading to consistent theories are
specified. Moddings by
cyclic permutations and by discrete phase symmetries are considered in
Section 8. Introduction of antibranes is briefly considered in section 9.
Section 10 offers a brief summary and outlook.
In order to keep track of the essential aspects of the construction many
details are relegated to appendices. Appendix A summarizes
explicit expressions and properties of the characters of the N=2
superconformal minimal models and the modular transformation properties
of supersymmetric characters of N=2 strings. In Appendix B we list the
spectrum of
states contained in the relevant characters of the Gepner models
constructed in the main body of the article.

\section{Vacuum amplitude in Type I superstring}

Consider the Type IIB torus partition function in
$D$ dimensions (We leave details for  reference \cite{polchi} and
explicit
examples for next section).
It is schematically defined as

\begin{equation}\label{2bpf}
{\cal Z}_{T} (\tau, {\bar \tau})=   \sum_{a,b} \chi_a(\tau )
{\cal N}^{ab} {\bar \chi}_b({\bar \tau })
\end{equation}
where the characters $\chi_a(\tau )= {\rm Tr}_{{\cal H}a}
q^{L_0-\frac{c}{24} }$, with $q= e^{2i\pi \tau}$, span a representation
of the modular group of the torus
generated by {\sf S}: $\tau \to -\frac1{\tau}$ and {\sf T}: $\tau
\to\tau+1$ transformations.
${\cal H}_a$ is the Hilbert space of a conformal field theory with central
charge
$c=15$
generated from
a conformal primary state $\phi _a$
(similarly for the right moving algebra).

In particular $\chi_a( -\frac1{\tau} )= S_{aa'}\chi_{a'}(\tau )$
and modular invariance requires
$ S {\cal N} S^{-1}= {\cal N}$.
Generically the characters can be split into a spacetime piece,
contributing with
$c_{st}= {\bar c }_{st}= {\frac32} D$ and an internal sector with
$c_{int} = {\bar c}_{int}= {\frac32}(10-D)$.
We are looking for left-right symmetric theories
and therefore we must also
require ${\cal N}^{ab}={\cal N}^{ba}$.

Let $\Omega $ be the reversing order (orientifolding)  operator permuting
right and left movers.
Modding by  order reversal symmetry is then
implemented by introducing the projection operator
$\frac12 (1+\Omega)$ into the torus partition function.
The resulting vacuum amplitude reads
\begin{equation}\label{clomega}
{\cal Z}_{\Omega} (\tau, {\bar \tau})= {\cal Z}_{T} (\tau, {\bar \tau})
+ {\cal Z}_{K} (\tau- {\bar \tau}) .
\end{equation}

 The first contribution is just the symmetrization
(or anti-symmetrization in
case states anticommute) of left and right sector contributions
indicating that two states differing by a left-right ordering must be
counted once.
The second term is the
Klein bottle contribution and takes into account states that are
exactly the same in both sectors.
In such case, the operator
$e^{2i \pi \tau {L_0}}
e^{-2i \pi {\bar \tau}{{\bar L}_0}}$,
when acting on the same states, becomes
$e^{2i \pi 2it_K {L_0}}$ with $\tau -{\bar {\tau}}= 2it_K$ and thus
\begin{equation}\label{kbd}
{\cal Z}_{K} (2it_K) = \frac 12 \sum_{a} {\cal K}^{a}  \chi_a(2it_K) ,
\end{equation}
where $|{\cal K}^{a}|={\cal N}^{aa}$ (there is a sign freedom in this definition which we fix by
imposing consistency
conditions \cite{fps,pss}).
The Klein bottle
amplitude in the {\it transverse channel} is obtained by performing
an {\sf S} modular transformation
such that

\begin{equation}\label{kbo}
\tilde {\cal Z}_{K} (i l)= \frac 12 \sum_{a} O^2_a  \chi_a(il )
\end{equation}
with $l=\frac1{2t_K}$ and
\beq
O^2_a= 2^D {\cal K}^{b}S_{ba}
\eeq
This notation for the closed channel coefficients highlights the fact
that the Klein bottle
 transverse channel represents a closed string propagating
between two
crosscaps (orientifold planes) which act  like boundaries. This amplitude
must still
be integrated over
the tube length.
Since closed string states of mass $m$ contribute as   $e^{-lm^2}$ in
 the
character, one concludes that
generically massless states will lead to tadpole
like divergences in the limit  $l \to \infty $.
While NSNS  massless states could presumably be interpreted as background
redefinitions \cite{fs, dudasmour},
 RR tadpoles lead, as mentioned,  to unavoidable inconsistencies.
Note that for such fields propagating in $ \chi_a$, $O_a$ represents the
charge of the orientifold
 plane (crosscap)  under them.

Inclusion of an open string sector with D-branes carrying $-O_a$ RR
 charge provides
a way for having a consistent theory \cite{pol,gp,pchj} with net 
vanishing charge.
\footnote{ In section 8 we consider the possibility of including 
antibranes with the consequent breaking of supersymmetry.}

An open string cylinder amplitude,
representing strings propagating between two D-branes, and
a M\"obius strip amplitude with strings propagating between orientifold
planes and D-branes must be
included.
In the long tube limit the sum of the contributions from the
Klein bottle, cylinder and M\"obius strip
in the transverse channel must then  factorize as
\begin{equation}
\tilde {\cal Z}_{K} (il)
+\tilde {\cal Z}_M (il) +
\tilde {\cal Z}_C (i l)\to
\sum_{a} (O_a +D_a)^2 \frac1{{m_a}^2}= \sum_{a} (O_a^2
+2 O_a D_a+ D_a^2) \frac1{{m_a}^2}
\label{factori}
\end{equation}
where ${m_a}$ is the mass of the state in  $\chi_a$.
For massless RR fields  $ D_a$ is the D-brane RR charge
and  absence of divergences requires
\begin{equation}
O_a +D_a=0 .
\label{tadpolecg}
\end{equation}
The generic form of the cylinder amplitude in the direct channel should
read
\begin{equation}\label{cild}
{\cal Z}_{C} (it_{C}) =\oh  \sum_{a} {\cal C}_{a}  \chi_a(it_{C} ) ,
\end{equation}
where
\beq
{\cal C}_{a} = C_{jka} n_j n_k
\eeq
represents the multiplicity of states contained in $ \chi_a(it )$ and
$n_j$, $n_k$ are Chan-Paton multiplicities.  $n_j$ can be interpreted as
 the number of branes of type $j$ where the string endpoints
must be attached \footnote{$j$ should, presumably, have a
topological interpretation, as it is the case for branes at
orbifold singularities, where it labels the monodromy at the
singularity.}. $ \sum_j n_j=N_B$ is the total number of D-branes.
Let us mention that, in general, while the index $a$ runs over the
different conformal highest weight representations defining   the
characters $\chi_a $,  $j$ indices are not necessarily in a one to
one correspondence with them. However, there is such
correspondence for charge conjugation modular invariants
\cite{bs1,bs2}.

 $C_{ija}$ must thus be positive integers.
Actually, as we discuss below, ${C_{ija}}= 0,1,2$.
The  transverse channel representation of this amplitude reads
\begin{equation}\label{ciltr}
{\tilde {{\cal Z}}}_{C} (i l)= \frac 12 \sum_{a} D^2_a  \chi_a(il )
\end{equation}
with  $ D_a=D_{ja} n_j $ and
\beq
(D_{ja} n_j)^2= {\cal C}_{b} S_{ba}= C_{jkb} n_j  n_k S_{ba}
\label{da}
\eeq

The M\"obius strip amplitude presents some additional subtleties since the
modulus
$it_M +\frac 12$ is not purely imaginary. Indeed the characters
are given by
\beq
\chi_a^{\Omega }(it_{M}) \equiv {\rm Tr}_{{\cal H}_a}(e^{\pi
it(L_{0}-\frac{%
c}{24})}\Omega )=\chi_a (it_{M}+\frac{1}{2}) ,
\eeq
and this introduces relative signs for
the oscillator excitations at the various mass levels leading to
complex characters.
The amplitude in the direct channel takes the form
\begin{equation}\label{msd}
{\cal Z}_{M} (it_M) =\oh  \sum_{a} {\cal M}_{a}  {\hat \chi}_a(it_{M}+\oh
)
\end{equation}
where now
\beq
{\cal M}_{a} = M_{ja}n_j
\eeq
are integer numbers and
the ``hat'' in the characters indicates that the phase $e^{i\pi
(h-c/24)}$ has been extracted  to make them real.

The characters in the direct and transverse channels of the M\"obius strip
are related by the transformation \cite{bs1}
{\sf P}: $it_M+\frac{1}{2}
\to \frac{i}{4t_M}+\frac{1}{2}$.
This can be generated
from the modular transformations {\sf S} and {\sf T} as
${\rm \sf P}={\rm {\sf TST}}^{2}{\sf S}$.
The transverse channel representing a
closed string propagating between a D-brane and an orientifold plane must read
\begin{equation}\label{mstr}
{\tilde {{\cal Z}}}_{M} (i l)= \frac 12\sum_{a} O_a (D_{ja} n_j) {\hat
\chi}_a(il+\oh )
\end{equation}
with
\beq
O_a (D_{ja} n_j)= 2^{\frac D2}{\cal M}_{b} P_{ba}= 2^{\frac D2}M_{jb}n_j
P_{ba}
\eeq
Notice that the tube length $l=\frac{1}{2t_K}
=\frac1{t_C}=\frac1{4t_M}=-{\frac1{2\pi}}\ln q$
for the different string amplitudes
 must be the same in order for them to be comparable.
We thus see that in the long length limit
( $ { \hat  \chi }_a (il+\oh ) \to  \chi_a(il )$)
the correct closed string channel  factorization (\ref{factori}) is
obtained.

In forthcoming sections we will apply this open descendant construction to
IIB theories where
the internal sector is built up from Gepner models.

Before closing this section let us make a few comments about the open
string spectrum.  Notice that the coefficients of $q^{m_a^2}$ in a $q$
expansion of cylinder $ + $ M\"obius strip direct channel
amplitudes, which are proportional to 
\beq 
\oh[{(C_{jka} n_j n_k )} \pm {(M_{ja}n_j ) }] , 
\label{multip} 
\eeq
are nothing but the
multiplicities of open string fields of mass $m_a$.  In principle,
such multiplicities and the spacetime transformation properties of the
corresponding characters should allow us to reconstruct the spectrum.
Notice that even and odd levels in the  M\"obius strip differ in sign,
due to the
$1/2$ term in the argument of the character. 
 
Since  open string gauge group representations are generated by
the Chan Paton indices in  the two string endpoints,
 we can infer that  only symplectic, orthogonal and/or unitary groups
are allowed  \cite{polchi, sagcarg}. Moreover, only adjoint ($
{\bf Adj } $),
symmetric ($\Ysymm $) or
antisymmetric ( $\Yasymm  $ ) (and their conjugate) representations can
be built from
Chan-Paton
factors ending on the same
type of
brane. The quadratic part
of such representations comes from cylinder contributions and
 therefore we must expect that $C_{iia}= 0,1,2$.
Recall also  that, had we obtained a  
symmetric (antisymmetric) representation at some mass level then the next
level would contain an antisymmetric (symmetric) one and so on, due to the
alternate
signs in the M\"obius strip contribution.

If two  gauge groups $G_i \times G_j$ were present,
bi-fundamental $ \big[\, (\fund_i,\antifund_j) \big]\, $
representations  corresponding to endpoints ending on two different sets
of $n_i$ and $n_j$
D-branes respectively could also arise and thus $C_{jia}= 0,1$.
  Linear terms in $n_j$ in (\ref{multip}) coming  from the M\"obius strip
amplitude
must complete
the two index representations and thus ${M_{ja}}= 0,1,-1$.

Moreover,  $C_{iia}= 2$ for the character $\chi_a$,  
with the corresponding null M\"obius strip coefficient
 $M_{ia}= 0$, would indicate a $U(n_i)$
adjoint representation. \footnote{Actually, once a unitary group is identified, it proves
useful to rewrite the term $n^2$ as $n \bar n$ (see for instance
\cite{review}). Even if
numerically   $n =\bar n$, this allows us
to distinguish complex representations.}
Similarly if $C_{jia}= 1$ with $i\ne j$, then  $M_{ja}=M_{ia} =0$.

Once the Klein bottle partition function is obtained from the left-right
symmetric
type IIB torus partition function, our construction of the open string
sector
will completely rely on

1. Factorization

2. Massless RR tadpole cancellation

3. Consistency restrictions on the integer coefficients $C_{jia}$ and
$M_{ja}$.

\section{Review of Gepner models}

\medskip

Gepner has shown how to construct supersymmetric closed string theories in
four spacetime dimensions replacing the geometrical notion of curling up the
extra dimensions into a compact internal manifold by an algebraic procedure
where the internal sector consists of tensor products of N=2 superconformal
minimal models with total central charge $c_{int}=9$ \cite{gepner}.
Spacetime supersymmetry and modular invariance are implemented by keeping in
the spectrum only states for which the total $U(1)$ charge is an odd
integer.
Let us briefly review Gepner's construction to set up
notation.

A consistent string theory in $D$ spacetime dimensions requires an
internal
conformal field theory with
$c_{int}=12 -\frac 32 (D-2)$
in the light cone gauge. N=1 spacetime supersymmetry is achieved if the
internal CFT has N=2 supersymmetry. The $(D-2)$ spacetime bosons and
fermions $X^\mu$, $\psi^\mu$ define a CFT with $c_{st}=\frac 32 (D-2)$ and
they
realize an N=2 superconformal algebra for even $D$.

Gepner models represent an explicit algebraic construction of supersymmetric
string vacua where the internal sector is given by a tensor product of $r$
copies of N=2 superconformal minimal models with levels $k_j$, $j=1,...,r$
and central charge
\begin{equation}
c = \frac{3k}{k+2} \quad , \quad k=1,2,...  \label{mm}
\end{equation}

\noindent
{\bf N=2 Superconformal Minimal models}

Let us recall the N=2 superconformal algebra here for completeness, namely
\begin{eqnarray}
[L_m, L_n] & = & (m-n)L_{m+n} + \frac c{12} (m^3 - m) \delta_{m+n,0}
\nonumber \\
\left [ L_m, J_n \right ] & = & -n J_{m+n}  \nonumber \\
\left [ L_m, G^\pm_r \right ] & = & (\frac m2 - r) G^\pm_{m+r}  \nonumber \\
\left [J_m, J_n \right ] & = & \frac {nc}3 \delta_{n+m, 0}  \nonumber \\
\left [J_m, G_r^\pm \right ] & = & \pm G^\pm_{m+r}  \nonumber \\
\{G_r^+, G_s^-\} & = & 2L_{r+s} + (r-s) J_{r+s} + \frac c3 (r^2 - \frac
14)\delta_{r+s,0}  \nonumber \\
\{G_r^+, G_s^+\} & = & \{G_r^-, G_s^-\} = 0 .
\end{eqnarray}
Integer or semi-integer modding $r, s$ correspond to the R or NS sector
respectively. The spectral flow symmetry of the algebra allows to consider
twisted sectors interpolating between R and NS. In fact the following
operators
\begin{eqnarray}
&&\tilde L_m = L_m +\frac n2 J_m + \frac c6 n^2\delta_{m,0}  \nonumber \\
&&\tilde G_r^{\pm} = G^\pm_{r\pm \frac n2}  \nonumber \\
&&\tilde J_m = J_m + \frac c6 n \delta_{m,0} , \label{twist}
\end{eqnarray}
generate an isomorphic N=2 algebra with the same central charge and modified
$G_r^\pm \rightarrow G_{r\pm \frac n2}^\pm$ modding.

As is well known unitary representations of the N=2 superconformal algebra
are found for discrete values of the central charge. For $c<3$ the discrete
minimal series is given by (\ref{mm}). The primary fields of the minimal
models are labelled by three integers $(l,q,s)$ such that $l=0,1,...,k$; $%
l+q+s=0$ mod 2 and they belong to the NS or R sector when $l+q$ is
even or
odd respectively. The conformal dimensions and charges of the highest
weight
states are given by
\begin{eqnarray}
\Delta_{l,q,s} = \frac{l(l+2)-q^2}{4(k+2)} + \frac {s^2}{8} \quad {\rm mod}
~1  \label{peso} \\
Q_{l,q,s} = -\frac{q}{k+2}+\frac s2 \quad {\rm mod}~ 2 .  \label{carga}
\end{eqnarray}

Two representations labelled by $(l^{\prime},q^{\prime},s^{\prime})$ and $%
(l,q,s)$ are equivalent, $i.e.$ they correspond to the same state, if
\begin{equation}
l^{\prime}=l  \quad , \quad
q^{\prime}= q ~{\rm mod} ~ 2(k+2)\quad , \quad s^{\prime}=s ~ {\rm mod}~4
\label{id1}
\end{equation}
or
\begin{equation}
l^{\prime}=k-l \quad , \quad q^{\prime}=q+k+2 \quad , \quad s^{\prime}=s+2
\label{id2}
\end{equation}
The exact conformal dimension and
charge of the highest weight state in the representation $(l,q,s)$ are
obtained from equations (\ref{peso}) and (\ref{carga}) using the
identifications above to bring $(l,q,s)$ to the $standard ~ range$ given by
\begin{equation}
l=0,1,...,k \quad ; \quad |q-s| \le l \quad ; \quad l+q+s=0 ~ {\rm mod} ~ 2
\end{equation}
and $|s|$ is the minimum value among those in (\ref{id1}) and (\ref{id2}).

The primary fields obey the inequalities
\begin{equation}
\Delta_{l,q,s}\ge \frac{|Q_{l,q,s}|}{2}
\end{equation}
for representations with even $s$ which belong to the NS sector, while for
odd $s$, which belong to the R sector, they satisfy
\begin{equation}
\Delta_{l,q,s}\ge \frac c{24} \quad .
\end{equation}

The twisting operation (\ref{twist}) corresponds to applying $n$ times the
transformations $q \rightarrow q+1, ~ s
\rightarrow s+1$.
The conformal dimensions and charges of the fields in the $n$-th twisted
sector, when both $l, q, s$ and $l, q+n, s+n$ are in the standard range,
are
obtained as
\begin{eqnarray}
\Delta _{l,q+n,s+n} ~ = ~ \Delta _{l,q,s}^{n} &=&\Delta _{l,q,s}^{0} +\frac{n%
}{2}Q_{l,q,s}^{0}+{n^{2}} \frac c{24}  \nonumber \\
Q_{l,q+n,s+n} ~ = ~ Q_{{l},{q},{s}}^{n} &=&Q_{{l},{q},{s}}^{0}+n \frac c6\;%
\label{cdc}
\end{eqnarray}
Notice that even $n$ interpolates between the same sector
whereas odd $n$ exchanges R and NS sectors.
When the transformations $q \rightarrow q+n, s\rightarrow s+n$ take them
outside the standard range the expressions for the conformal dimension
and charge differ from (\ref{cdc}) by an integer
and an even number respectively.
Moreover the identifications (%
\ref{id1}) imply that one comes back to the original representation after
twisting by $n=2(k+2)$ for even $k$ and by $n=4(k+2)$ for odd $k$. (A
particular case
is given by $l=k/2$ for even $k$
where the identification (%
\ref{id2}) implies that  the original representation
 is re-obtained after twisting by $n=k+2$).

The partition function of the minimal models on the torus can be written in
terms of the characters of the irreducible representations as
\begin{equation}
{\cal Z}^{(m.m.)}_T(\tau) = \sum_{(l, q, s),(\bar l, \bar q, \bar s)} {\cal N%
}_ {(l, q, s),(\bar l, \bar q, \bar s)} \chi_{(l, q, s)}(\tau,0) \chi_{(\bar
l, \bar q, \bar s)}^*(\bar\tau,0)
\end{equation}
where the coefficients ${\cal N}_{(l, q, s),(\bar l, \bar q, \bar s)}$ are
non negative integer numbers which count the number of times the irreducible
representation $(l, q, s)\otimes(\bar l, \bar q, \bar s)$ is contained in $%
{\cal H}$. The existence of a unique ground state requires ${\cal N}%
_{(0,0,0),(0,0,0)} = 1$.
The characters in the sector ${\cal H}_{(l,q,s)}$ are given by
\begin{equation}
\chi_{(l, q, s)}(\tau,z) = {\rm Tr}_{{\cal H}_{(l, q, s)}} \left ( e^{2\pi i
\tau (L_0 - \frac c{24})} e^{2\pi i z J_0}\right )  \label{cardef}
\end{equation}
in the holomorphic untwisted sector, and by
\begin{equation}
\chi_{(l, q+n, s+n)}(\tau,z) \equiv {\rm {\bf Q}}^n \chi_{(l,q,s)}(\tau, z)
\end{equation}
in the holomorphic sector twisted by $n$, where {\bf Q} is the operator
interpolating between NS and R sectors, and similarly for the
antiholomorphic part.

The Hilbert space can be decomposed into two subspaces ${\cal H}_{(l,q)}=%
{\cal H}_{(l,q,s)}+{\cal H}_{(l,q,s+2)}$, where each subspace is generated
by an even number of $G_{r}^{\pm }$ on the primary fields $|\Psi
_{(l,q,s)}>$ and $|\Psi _{(l,q,s+2)}>$. Both subspaces are related by the
action of one $G^{\pm }$. Therefore it is convenient to define
\beq
\chi _{l,q}(\tau ,z)\equiv {\rm Tr}_{{\cal H}_{l,q}}(e^{2\pi i(L_{0}-%
\frac{c}{24})\tau }e^{2\pi iJ_{0}z})=\chi _{(l,q,s)}(\tau ,z)+\chi
_{(l,q,s+2)}(\tau ,z)
\label{carss}
\eeq
 Explicit expressions and properties of the characters of
the N=2 superconformal minimal models are included in Appendix A.

The character $\chi _{l,q}$ with even $l+q$ contains two families of
states with primaries $(l,q,s)$ and $(l,q,s+2)$ whose conformal dimensions
differ by $1/2$.
It is always possible to choose $s=0$ for the primary with the smaller
conformal dimension and $(l,q,s)$ in the standard range. This state
 has conformal dimension and charge
given by (\ref{peso}) and (\ref{carga}), namely
\begin{eqnarray}
\Delta _{l,q} &=&\frac{l(l+2)-q^{2}}{4m}-\frac{c}{24}\qquad   \nonumber \\
Q_{l,q} &=&-\frac{q}{m}\qquad (\left| q\right| \leq l,l+q=even)
\end{eqnarray}
where $\Delta _{l,q}\equiv
\Delta _{l,q,0}$ and $Q_{l,q}$ $\equiv Q_{l,q,0}$.
For the other primary,  $\left| s\right| \geq 2$.

In the R sector the two highest weight states have the same conformal
dimension and their charges differ by one, except when one of
the conformal dimensions is $c/24$.
It is always possible to choose $s=-1$ for the representation in which the
highest weight state has smaller charge.
Thus the conformal dimension and charge of this state are
given by
\begin{eqnarray}
\Delta _{l,q} &=&\frac{l(l+2)-q^{2}}{4(k+2)}+\frac{1}{8}\quad ;\quad
Q_{l,q}=-\frac{q}{k+2}-\frac{1}{2}  \nonumber \\
l+q &=&odd,\; \quad -l-1\leq q\leq l-1
\end{eqnarray}

For  $q\neq -l-1$  ($\Delta _{l,q}>\frac{c}{24}$)
the highest weight state $(l,q,s+2)$ can be obtained choosing $s=1$; it
has the same conformal dimension as $(l,q,s)$ but its charge is
$Q_{l,q}$ $+1\,$ whereas for
$q=-l-1$ ($\Delta _{l,q}=\frac{c}{24}$),
the conformal dimension is $\Delta _{l,q}+1$.

{}From the equivalence relations (\ref{id1}) and (\ref{id2}) one can
show the following identities
\begin{equation}
\chi _{l,q}(\tau ,z)=\chi _{k-l,q+k+2}(\tau ,z) =
\chi _{l,q+2(k+2)}(\tau ,z)
\end{equation}

Charge conjugate characters contain highest weight states with charges
$Q_{l,q}$ and $Q_{l,-q}$ and they satisfy the following equality
\begin{equation}
\chi _{l,q}(\tau ,z)=\chi _{l,-q}(\tau ,-z) . \label{chcon}
\end{equation}

\medskip

Modular transformations map one sector into another
and thus modular
invariance requires considering the four sectors NS$^{\pm }$, R$^{\pm }$,
defined as
\begin{eqnarray}
\chi _{l,q}^{NS^{\pm }} &=&\frac{1}{2}(\chi _{(l,q,0)}\pm \chi _{(l,q,2)})
\label{carssns} \\
\chi _{l,q}^{R^{\pm }} &=&\frac{1}{2}(\chi _{(l,q-1,-1)}\pm \chi
_{(l,q-1,1)})  \label{carssr}
\end{eqnarray}
Note that $\chi_{l,q}^{NS^+}=\chi_{l,q}$.
They can all be written in terms of $\chi _{l,q}^{NS^{+}}$ by shifting $z$
as
\begin{eqnarray}
\chi _{l,q}^{NS^{-}}(\tau ,z) &=&e^{-\pi iQ}\chi _{l,q}^{NS^{+}}(\tau ,z+%
\frac{1}{2})  \nonumber \\
\chi _{l,q}^{R^{+}}(\tau ,z) &=&e^{2\pi i\tau c/24}e^{2\pi izc/6}\chi
_{l,q-1}^{NS^{+}}(\tau ,z+\frac{1}{2}\tau )  \nonumber \\
\chi _{l,q}^{NS^{-}}(\tau ,z) &=&e^{-\pi iQ}e^{2\pi i\tau c/24}e^{2\pi
izc/6}\chi _{l,q-1}^{NS^{+}}(\tau ,z+\frac{1}{2}\tau +\frac{1}{2}).
\label{shift}
\end{eqnarray}
Under {\sf S:} $\tau \rightarrow -\frac{1}{\tau }$, the minimal characters $%
\chi _{l,q}$ with even $l+q$ transform as

\begin{equation}
\chi _{l,q}(-\frac{1}{\tau },\frac{z}{\tau })=e^{2\pi i\frac{z^{2}c}{6\tau }%
}\sum\limits_{l^{\prime },q^{\prime }}S_{l,q;l^{\prime },q^{\prime }}\chi
_{l^{\prime },q^{\prime }}(\tau ,z) \quad , \label{tm}
\end{equation}
or equivalently as
\begin{equation}
\chi _{l,q}(-\frac{1}{\tau },\frac{z}{\tau })=e^{2\pi i\frac{z^{2}c}{6\tau }%
}\sum\limits_{l^{\prime },q^{\prime }}S_{l,q;l^{\prime },q^{\prime
}}^{-1}\chi _{l^{\prime },q^{\prime }}(\tau ,-z)\quad
\end{equation}
where the $S$ matrix is given by
\begin{equation}
S_{l,q;l^{\prime },q^{\prime }}\equiv \frac{2}{(k+2)}e^{\pi i\frac{%
qq^{\prime }}{k+2}}\sin (\pi \frac{(l+1)(l^{\prime }+1)}{k+2})
\end{equation}
(for even ($l+q$)). It verifies the following equality
\beq
S_{l,q;l^{\prime },q^{\prime }}=S_{k-l,q+k+2;l^{\prime },q^{\prime }}
\quad .
\label{eqs}
\eeq
Applying  the
{\sf S} transformation twice one gets
\begin{equation}
\chi _{l,q}(\tau ,z)=\sum\limits_{l^{\prime },q^{\prime }}S_{l,q;l^{\prime
},q^{\prime }}^{2}\chi _{l^{\prime },q^{\prime }}(\tau ,-z)
\end{equation}
which leads to the {\em charge conjugation} matrix
\[
S^{2}=C\quad ,\quad C_{l,q;l^{\prime },q^{\prime }}=\delta _{l,q;l,-q}
\quad .
\]
Using the explicit expressions for the characters it is easy to verify the
following action of the {\sf S} transformation:
NS$^+ \rightarrow$ NS$^+$, NS$^- \leftrightarrow$ R$^+$,
R$^- \rightarrow$ R$^-$.

Under ${\sf T}:\tau \rightarrow \tau +1$ the characters transform as
\begin{eqnarray}
\chi _{l,q}^{NS^{+}}(\tau +1,z) &= & e^{2\pi i(\Delta _{l,q}
-\frac{c}{24})}\chi _{l,q,0}(\tau ,z)+e^{2\pi
i(\Delta _{l,q}+\frac{odd}{2}-\frac{c}{24})}
\chi _{l,q,2}(\tau ,z+\frac{1}{%
2})\nonumber
\\
 &=&e^{2\pi i(\Delta _{l,q}-\frac{c}{24})}\chi _{l,q}^{NS^{-}}(\tau ,z)
\end{eqnarray}

\begin{equation}
\chi _{l,q}^{NS^{-}}(\tau +1,z)=e^{2\pi i(\Delta _{l,q}-\frac{c}{24})}\chi
_{l,q}^{NS^{+}}(\tau ,z)
\end{equation}

\beq
\chi _{l,q}^{R^{\pm }}(\tau +1,z) =e^{2\pi i(\Delta _{l,q-1,-1}-\frac{c}{24%
})}\chi _{l,q}^{R^{\pm }}(\tau ,z)
=e^{2\pi i(\Delta _{l,q}-\frac{Q_{l,q}}{2})}\chi _{l,q}^{R^{\pm }}(\tau ,z)
\eeq
Therefore {\sf T}: NS$^{\pm }\rightarrow $ NS$^{\mp }$ ; R$^{\pm
}\rightarrow $ R$^{\pm }$.

\medskip
\noindent {\bf N=2 strings}

A $D$ dimensional string theory is obtained by taking the tensor product of $%
r$ internal N=2 SCFTs such that $\sum_{i=1}^r c_i = 12 -\frac 32 (D-2)$ and
appending the spacetime contribution. Let us start by reviewing the
spacetime part.

The $(D-2)$ spacetime bosons and fermions realize a (2,2) superconformal
algebra. The fermions $\psi^\mu(z)$ $(\mu = 1,..., D-1)$ exhibit a $SO(D-2)$
symmetry which require the states to be in unitary representations of the
affine transverse Lorentz algebra at level $k=1$. These are the scalar,
vector, spinor and conjugate spinor representations labelled respectively by $%
\lambda =0, 2, 1, -1$. The contribution of each pair of transverse
dimensions to the spacetime characters is given by
\begin{eqnarray}
\Upsilon_{0(2)}(\tau,z)&=&\frac 1{2\eta(\tau)^3}\left
(\vartheta{{0 \atopwithdelims[] {0}}} (\tau,z)\pm
\vartheta{{0 \atopwithdelims[] \oh  }} (\tau,z)\right )
\nonumber \\
\Upsilon_{1(-1)}(\tau,z)& =&\frac 1{2\eta(\tau)^3}\left
(\vartheta{\oh \atopwithdelims[] {0}} (\tau,z) \mp \vartheta{\oh \atopwithdelims[] {\oh}}(\tau,z)\right )
\quad .
\label{stcartheta}
\end{eqnarray}
where the upper (lower) sign corresponds to first (second)
subindex in the character.
The conformal spacetime dimensions and charges of the states are
\begin{equation}
\Delta_{st} = \frac{\lambda^2}{8} \quad , \quad Q_{st} = \frac \lambda 2
\quad ,  \label{stdc}
\end{equation}
and the field identification is $\lambda^{\prime} = \lambda$ mod 4.

Similarly as in (\ref{carssns}) and (\ref{carssr}) we define for each pair
of transverse dimensions
\begin{equation}
\chi^{NS^\pm}(\tau, z)=\Upsilon_0(\tau, z)\pm \Upsilon_2(\tau, z) \quad ;
\quad \chi^{R^\pm}(\tau, z)=\Upsilon_{-1}(\tau, z)\pm \Upsilon_1(\tau, z)
\quad ,
\end{equation}
and it can be seen from (\ref{stcartheta}) that $\chi^{R^-}(\tau, 0)\equiv 0$. We
denote the spacetime characters in the $\nu$-th twisted sector as
\begin{equation}
\chi_{\nu}(\tau, z)=\Upsilon_{0+\nu}(\tau, z)+\Upsilon_{2+\nu}(\tau, z)
\quad .
\end{equation}
Note that $\chi_{\nu}(\tau, z) = \chi^{NS^+}(\tau, z)$ if $\nu$ is even and $%
\chi_{\nu}(\tau, z) = \chi^{R^+}(\tau, z)$ if $\nu$ is odd.

The {\sf S} modular transformation on these spacetime characters reduces to
\begin{eqnarray*}
\chi^{NS^+}(-\frac{1}{\tau },\frac{z}{\tau }) &=&\frac{e^{2\pi i\frac{%
z^{2}c_{st}}{6\tau }}}{-i\tau }S_{st}\chi^{NS^+}(\tau ,z) \quad ,
\end{eqnarray*}
where $c_{st}=\frac 32 (D-2)$ and $S_{st}=1$.

Putting everything together the character associated to a primary state of
the full theory is given by the product of the contributions of 
spacetime times the $r$ internal theories. In order to achieve N=1
supersymmetry on the world sheet all the states in the product must belong
to a definite sector, $i.e.$ NS (R) states must be tensored only with NS (R)
states. Modular invariance requires odd total U(1) charge $Q$.
 These conditions lead to the following character
\begin{equation}
\chi_{\vec \alpha}(\tau, z) 
\equiv \hat{{\cal P}}_{GSO}\{ \chi^{\prime}_{\vec\alpha}(\tau, z)\}
\equiv \hat{{\cal P}}_{GSO}\{ 
[\chi_{\nu}(\tau,
z)]^d \prod_{i=d+1}^{d+r} \chi_{\alpha_i}(\tau, z) \}
\label{gso}
\end{equation}
where $[\chi_{\nu}(\tau, z)]^d$ is the $D$ dimensional spacetime character
with $d=\frac {(D-2)}2$. Here $\vec \alpha$ is a $(d+r)$-component vector
with
entries $\alpha_i = \nu$ for $i=1,...,d$ and $\alpha_i = (l_i, q_i)$ for
$i=d+1,...,d+r$ denoting the full primary state of the
product of internal and spacetime theories such that both $l_i+q_i$ and $\nu$
are even or odd. $\hat{{\cal P}}_{GSO}$ denotes the generalized GSO
projection over states with odd U(1) charge.

The action of the supersymmetry operator on the product theory can be
expressed more conveniently introducing a vector $\vec\alpha^{(n)}$ with
components
$\alpha_i^{(n)}=\nu + n$ for $i=1,...,d$ and
$\alpha_i^{(n)}=(l_i,q_i+n)$ for $i=d+1,...,d+r$ as
\begin{equation}
{\rm {\bf Q}}^n \chi_{\vec \alpha}(\tau, z) = \chi_{\vec\alpha^{(n)}}(\tau,
z).
\end{equation}
Notice that if $\alpha_i$ denotes a state in the NS sector then $%
\alpha_i^{(n)}$ and therefore $\vec\alpha$ correspond to the NS sector.

N=1 supersymmetry in spacetime requires summing over all twisted sectors.
The identifications among characters allow to sum over $n$ mod $2m$ where $%
m $ is the $l.c.m.$ of all the $k_i+2$ in the product. The \textit{supersymmetric}
character is finally given by (see Appendix A)
\beqa
& & \chi^{susy}_{\vec\alpha}(\tau, z) = \sum_{n=0}^{2m-1} (-1)^{n}
\chi_{\vec \alpha^{(n)}} (\tau, z)=  \label{susych} \\
& &= \frac{1}{2m}\sum\limits_{n,p\;{\rm mod\;2m}}(-1)^{n+p}
e^{2\pi i(n^{2}\frac{c}{24}\tau +n\frac{c}{6}z)}
\left [\chi_0(\tau
,z+\frac n2 \tau+\frac p2)
\right ]^{d}
\prod\limits_{i=1}^{r}\chi _{l_{i},q_{i}}(\tau ,z+\frac{n%
}{2}\tau +\frac{p}{2}) \nonumber
\eeqa
with $c=12$. NS or R sectors are obtained when
summing over even or odd $n$, respectively and periodic ($+$) or
antiperiodic ($-$)
characters arise when summing over even or odd $p$, respectively.

This is a useful result. The open sector will be easily written down, in
an
explicit supersymmetric expression,  as linear combinations of these characters.

Let us discuss some properties of the states contained in $\chi _{\vec{\alpha%
}}^{susy}(\tau )$. It is convenient to introduce the following notation:
 $\vec{\beta}$ is
a $d+r$ component vector with entries $\beta _{i}=\lambda _{i}$ for $%
i=1,...,d$ and $\beta _{i}=(l_{i},q_{i},s_{i})$ for $i=d+1,...,d+r$ and such
that $l_{i}+q_{i}+s_{i}$ and $\lambda _{i}$ are both even or odd.
Analogously we can define the vector $\vec{\beta}^{(n)}$ obtained as $%
\lambda _{i},q_{i},s_{i}\rightarrow \lambda _{i}+n,q_{i}+n,s_{i}+n$.
Due to the generalized
GSO projection the states contained in the characters
$\chi _{\vec{\alpha}}^{NS}$ and $\chi _{\vec{\alpha}}^{R}$
carry an index $\vec\beta$ such that
$Q_{\vec{\beta}^{(n)}}$ is odd
($Q_{\vec{\beta}}=\sum_{i=1}^{d+r}Q_{\beta _{i}}$).
Before GSO projecting, the charges of the states contained in the
character
are all related by
\begin{equation}
Q_{\vec{\beta}^{(n)}}=Q_{\vec{\beta}}+2n ~~{\rm mod} ~2
\end{equation}
and therefore all the GSO projected products of states in
$\chi _{\vec{\alpha}^{(n)}}$ for a given $n$ can be obtained by twisting
the GSO projected product of states for $n=0$.

The full conformal dimensions of the fields in the $n$-th twisted sector
are
given by
\begin{equation}
\Delta _{\vec{\beta}^{(n)}}=\Delta _{\vec{\beta}}+\frac{n}{2}Q_{\vec{\beta}}+%
\frac{n^{2}}{2} ~~{\rm mod} ~1
\end{equation}
where $\Delta _{\vec{\beta}}=\sum_{i=1}^{d+r}\Delta _{\beta _{i}}$.
Notice that the sum of the
conformal dimensions of the untwisted states after GSO projecting
differs by an integer from the sum of the conformal dimensions of the states
twisted by $n$ (since $\frac{n}{2}(Q_{\vec{\beta}}+n)$ is integer for odd
$Q_{%
\vec{\beta}}$). Therefore all the states obtained by twisting an odd $U(1)$
charge state have conformal dimension differing by an integer from that of
the untwisted state.

Let us now compare GSO projected states in the NS sector labelled with
vectors $\vec{\beta}$ and $\vec{\beta}^{\prime }$ having $%
l_{i}=l_{i}^{\prime },q_{i}=q_{i}^{\prime }$, $\lambda _{i}\neq \lambda
_{i}^{\prime }$ and $s_{i}\neq s_{i}^{\prime }$. \ The difference in their
conformal dimensions and charges is given by the number of states with $%
s_{i}\neq 0$. Indeed considering that
\begin{eqnarray}
\Delta _{\lambda +2}-\Delta _{\lambda } =\frac{1+\nu }{2}~~{\rm mod}%
~1\quad &;&\quad Q_{\lambda +2}-Q_{\lambda }=1~~{\rm mod}~2  \nonumber \\
\Delta _{l,q,s+2}-\Delta _{l,q,s}=\frac{1+l+q+s}{2} ~~{\rm mod} ~1\quad
&;&\quad
Q_{l,q,s+2}-Q_{l,q,s} =1 ~~{\rm mod} ~2 \nonumber
\end{eqnarray}
it is easy to see that the states with odd $U(1)$ charge verify
\begin{equation}
\Delta _{\vec{\beta}^{\prime }}-\Delta _{\vec{\beta}}\in Z\quad ;\quad
Q_{\vec{\beta}^{\prime }}-Q_{\vec{\beta}}=0 ~~ {\rm mod} ~2
\end{equation}
We conclude that all the states contained in a given $\chi _{\vec{\alpha}%
}^{susy}$ have conformal dimensions given by $\Delta _{\vec{\beta}^{(n)}}-%
\frac{1}{2}=\Delta _{\vec{\beta}_{0}}-\frac{Q_{_{\vec{\beta}_{0}}}}{2}~{\rm %
mod}~$ 1
 ($\vec{\beta}_{0}$ is the vector $\vec{\beta}$
with $s_{i} =0$  for all $i$). Since $\Delta _{\vec{\beta}%
_{0}}=\Delta _{\vec{\alpha}}$ and $Q_{\vec{\beta}_{0}}=Q_{\vec{\alpha}}$,
finally
\begin{equation}
\Delta _{\vec{\beta}^{(n)}}-\frac{1}{2}=\Delta _{\vec{\alpha}}-\frac{Q_{_{%
\vec{\alpha}}}}{2}~{\rm mod}~1  \label{pesossusy}
\end{equation}
where $\vec{\alpha}\in $ NS. This is an important relation because it
gives the conformal dimensions of the product of GSO projected states
(modulo an integer number) from the conformal dimensions and charges of
the highest weight states in the non-projected character
 $\prod_i\chi _{\alpha_{i}}$.

Taking into account that $\chi _{\vec{\alpha}}^{susy}$ contains
the sum over all twisted sectors and that all $\chi _{\vec\alpha
^{(n)}}^{susy}$ with even $n$ contain the same representations, the
following identities hold
\begin{equation}
\chi _{{\vec{\alpha}}^{(n)}}^{R/NS}(\tau ,z)=\chi _{\vec{\alpha}%
}^{R/NS}(\tau ,z)\quad ;\quad \chi _{{\vec{\alpha}}^{(n)}}^{\pm }(\tau
,z)=\chi _{\vec{\alpha}}^{\pm }(\tau ,z).  \label{identifi}
\end{equation}
One may thus choose a representative $\vec{\alpha}$ for all the equivalent
vectors under twisting and the number of independent characters is then
reduced by $m$ for each R or NS sector. There is an important exception
when some of the $k_i$ are even: if $\vec\alpha$ contains
 $l_{i}=\frac{k_{i}}{2}$ for all even $k_{i}$ then the number of
supersymmetric characters related by twisting is $m/2$ and
the states with $l_i=\frac {k_i}2$ are obtained twice in the sum over $n$
from 0 to $2m-1$. We shall refer to these as {\it short vectors}.

The following relation between supersymmetric characters follows
from the identity (\ref{chcon}) between charge conjugate characters
for each minimal model
\begin{equation}
\chi _{\vec{\alpha}}^{NS/R}(\tau ,z)=\chi _{\overline{\left( \vec{\alpha}%
\right) }}^{NS/R}(\tau ,-z) \label{chconmm}
\end{equation}
where  $\overline{\left( \vec{\alpha}\right) }$ is the vector obtained
replacing $q_{i}$ by $-q_{i}$ in $\vec{\alpha}$.

Modular transformations of the supersymmetric characters are discussed in
Appendix A.
\medskip

\section{Type I superstring at Gepner points}

The spectrum of perturbative Type II
closed string states in Gepner models is contained in the full
supersymmetric and modular invariant partition function for N=2 strings on
the torus which is obtained combining the right and left sectors as
\begin{equation}
{\cal Z}_T(\tau, \bar \tau) = \sum_ {\vec \alpha ;
\vec {\bar \alpha}} {\cal N}_{\vec \alpha ; \vec {\bar \alpha}}
\chi^{susy}_{\vec\alpha}(\tau,0) \chi^{susy ~ *}_{\vec {\bar
\alpha}}(\bar\tau,0) \quad ,
\end{equation}
and integrating over $\tau$ with the appropriate measure
\begin{equation}
{\sf Z}_T = \int \frac{d\tau d\bar\tau}{({\rm Im} \tau)^2} {\cal
Z}_T(\tau,
\bar\tau) \quad .
\end{equation}
Here ${\cal N}_{\vec\alpha , \vec{\bar\alpha}}$ are positive integer
coefficients obtained from the product $\prod_{i=1}^r{\cal N}_{\alpha_i;\bar
\alpha_i}$ of the individual minimal models
such that the partition function is modular invariant.

In the following section we construct the
partition functions for the unoriented and open descendants of type IIB
N=2
superstrings, $i.e.$ the vacuum amplitudes from the Klein bottle, the
M\"obius
strip and the cylinder, with special attention to the possible contributions
to tadpoles and their cancellation.

\subsection{Klein Bottle Amplitude}

The partition function from the Klein bottle can be
obtained from that of the torus as discussed in Section 2. Integrating
over $t$ with the appropriate measure, the vacuum amplitude for the Klein
bottle in the direct (open) channel is given by
\begin{eqnarray}
{\sf Z}_K &=& \frac 12\int_0^\infty \frac{dt}{4t} {\rm Tr}_{{\cal H}_{cl}}
\left
\{\Omega ~ {\rm exp}\left [2\pi i (\tau (L_0-\frac c{24})-\bar\tau (\bar
L_0-\frac c{24}))\right ]
\right \}  \nonumber \\
&=& \frac 12 \int_0^\infty \frac{dt}{4t} {\ (}\frac
1{4\pi^2\alpha^\prime t}
{\
)}^{\frac D2}
{\rm Tr}^{\prime}_{{\cal H}_{cl}} \left \{ {\rm exp}%
\left [-4\pi t (L_0-\frac c{24})\right ] \right \}
\end{eqnarray}
where Tr$^\prime$ denotes the trace over the discrete oscillator modes and
the factor \break
$(4\pi^2\alpha^\prime t)^{-D/2}$ comes from the integral over the bosonic
zero modes. The trace
 can be written in terms of the supersymmetric
characters $\chi^{susy}_{\vec \alpha}$ as
\begin{equation}  \label{kbdir}
{\cal Z}_K (it) = \frac 12 {\rm Tr}^{\prime}_{{\cal H}_{cl}} \left \{ {\rm
exp}%
\left [-4\pi t (L_0-\frac c{24})\right ] \right \} =
\frac12 \sum_{\vec\alpha} {{\cal K}_{{\vec \alpha}}
\chi^{susy}_{{\vec \alpha }}(2it)}
\end{equation}
where $|{\cal K}_{\vec\alpha}|={\cal N}_{\vec\alpha\vec\alpha}$.

 The Klein bottle amplitude in the {\it
transverse channel}  is obtained by performing an {\sf S} modular
transformation
\begin{eqnarray}
{\sf Z}_{K} &=&\frac 12 \int\limits_{0}^{\infty
}\frac{dt}{4t}(\frac{1}{4\pi
^{2}\alpha
^{\prime }t})^{\frac{D}{2}}\sum_{\vec\alpha} {{\cal K}_{{\vec \alpha}}
\chi^{susy}_{{\vec \alpha }}(2it)}\nonumber \\
&=&\frac 12 \frac{1}{(8\pi ^{2}\alpha
^{\prime })^{\frac{D}{2}}}\int\limits_{0}^{\infty }\frac{dl}{4l}2^{D}l^{%
\frac{D}{2}}\sum_{\vec\alpha} {{\cal K}_{{\vec \alpha}}
\chi^{susy}_{{\vec \alpha }}(\frac il)} \nonumber  \\
 &=&\frac 12 \frac{1}{(8\pi ^{2}\alpha ^{\prime })^{\frac{D}{2}}}
\int\limits_{0}^{\infty }
\frac{dl}{4}
{\tilde { {\cal Z } }_{K} (il)}
\label{kbtrans}
\end{eqnarray}
where $l=1/2t$.
\begin{equation}  \label{kbtr}
{\tilde {{\cal Z}}}_{K} (i l)= \frac 12 \sum_{\vec\alpha} O^2_{\vec\alpha}
\chi^{susy}_{\vec\alpha}(il )
\end{equation}
represents a closed string propagating between
two crosscaps (orientifold planes) and
 \begin{equation}
O^2_{\vec\alpha}= 2^D{\cal K}^{\vec\beta}{S}_{\vec\beta\vec\alpha}
\end{equation}
As mentioned, generically, massless $RR$ fields will be present such that
${\tilde {{\cal Z}}}_{K} (i l)\propto
O^2_{\vec\alpha} q^0+...$ will lead to undesired tadpole divergencies when
integrated over $l$.
Therefore one has to include
D-brane amplitudes and the corresponding
open string sectors to cancel such divergences.

\subsection{The open sector: Cylinder and M\"obius strip amplitudes}

Let us discuss the open sector of the theory. Here the partition function
takes
the form
\begin{equation}
{\sf Z}_{open} = \int_0^\infty \frac {dt}{4t}{\rm Tr}_{{\cal H}_o} \left
[ \left (\frac{1+\Omega
}{2}\right )
e^{2\pi i(L_{0}-\frac{c}{24})it} \right ]=
{\sf Z}_C+{\sf Z}_M
\end{equation}
where
\begin{eqnarray}
{\sf Z}_C &=& \frac 12\int_0^\infty \frac{dt}{4t} {\rm Tr}_{{\cal H}_{o}}
\left
[ e^{2\pi
i(L_{0}-\frac{c}{24})it}\right ]   \nonumber \\
&=& \frac 12\int_0^\infty \frac{dt}{4t} {\ (}\frac
1{8\pi^2\alpha^\prime t}
{\
)}^{\frac D2}
{\rm Tr}^{\prime}_{{\cal H}_{o}}
\left [ e^{2\pi
i(L_{0}-\frac{c}{24})it}\right ]   \nonumber
\eeqa
and
\beqa
{\sf Z}_M &=& \frac 12 \int_0^\infty \frac{dt}{4t} {\rm Tr}_{{\cal H}_{o}}
\left
[ \Omega e^{2\pi
i(L_{0}-\frac{c}{24})it}\right ]   \nonumber \\
&=& \frac 12 \int_0^\infty \frac{dt}{4t} {\ (}\frac
1{8\pi^2\alpha^\prime t}
{\
)}^{\frac D2}
{\rm Tr}^{\prime}_{{\cal H}_{o}} \left \{ {\rm exp}%
\left [-2\pi t (L_0 - \frac c{24})\right ] \Omega \right \}
\eeqa
Similarly as above Tr$^\prime$ denotes the trace over the discrete
oscillator modes and
the factor
$(8\pi^2\alpha^\prime t)^{-D/2}$ comes from the integral over the bosonic
zero modes. The traces
 can be written in terms of the supersymmetric
characters $\chi^{susy}_{\vec \alpha}$ as
\begin{equation}  \label{cildir}
{\cal Z}_C (it) = \frac 12 {\rm Tr}^{\prime}_{{\cal H}_{o}}
{\ [}e^{2\pi
i(L_{0}-\frac{c}{24})it}
{\ ]}= \frac 12
\sum_{\vec\alpha} {{\cal C}_{{\vec \alpha}}
\chi^{susy}_{{\vec \alpha }}(it)}
\end{equation}
\beq
{\cal Z}_M (it)= \frac 12 {\rm Tr}^\prime_{{\cal H}_o}
{\ [}e^{2\pi
i(L_{0}-\frac{c}{24})it}\Omega
{\ ]}= \frac 12
\sum_{\vec\alpha}{\cal M}_{\vec\alpha}
\hat\chi^{susy}_{\vec\alpha}(it+\frac 12)
\eeq
where
\begin{equation}
{\cal C}^{\vec \alpha} = C^{{\vec \alpha }}_{{\vec \alpha ^{\prime}},{\vec
\alpha ^{\prime \prime}}} n^{{\vec \alpha ^{\prime}}} n^{{\vec \alpha
^{\prime \prime}}} \quad ; \quad
{\cal M}_{\vec\alpha}=\sum_{\vec\alpha^\prime}M_{\vec\alpha
{\vec\alpha}^\prime} n^{\vec\alpha^\prime}
\eeq
represent the multiplicity of states contained in
the characters and $n^{\vec\alpha}$  are Chan-Paton multiplicities
(see discussion below (\ref{cild})).
${C^{\vec\alpha}_{\vec\alpha^{\prime}\vec\alpha^{\prime\prime}}}$
must thus be positive integer numbers whereas
$M_{\vec\alpha\vec\alpha^{\prime}}$ are integer numbers.
\textit{Hatted} characters from the  M\"obius strip are defined as
\beq
 \hat\chi^{susy}_{\vec\alpha}
(it+\frac
12) =
e^{-i\pi(\Delta_{\vec\alpha}-\frac{Q_{\vec\alpha}}2)}
\chi^{susy}_{\vec\alpha}(it+\frac 12)
\label{tphase}
\eeq
where a phase has been extracted to make them real
 (see Appendix A).

We can therefore proceed to write down such amplitudes in the transverse
channel in order to study factorization and tadpole cancellation.
A rescaling of the parameter $t$
in each amplitude is needed
in order to express such amplitudes in terms of the common tube length
$l = -\frac 1{2\pi}~ {\rm log}~ q$.
While supersymmetric characters in
the amplitude from the cylinder
involve only ${\sf S}$ transformations relating the open and closed
channels
(see (\ref{ciltr})), the
characters in the M\"obius strip are evaluated at $i t +\frac12 $, and
thus
expressing them  in terms of the parameter $l$ requires a combined action
of both ${\sf T}$ and ${\sf S}$ \cite{bs1}. In fact, for our characters,
it is shown in Appendix A that  such transformation is achieved by
a matrix
\begin{equation}
\hat{P}=T^{(-1/2)}{S}T^{2}{S}^{-1}T^{(1/2)}
\end{equation}
where
\begin{equation}
T_{\vec{\alpha}\vec{\alpha}}^{(1/2)}=e^{\pi i(\Delta _{\vec{\alpha}}-\frac{%
Q_{\vec{\alpha}}}{2})}
\end{equation}
is the phase introduced  in (\ref{tphase}) such that
characters in the direct and
transverse channels are related as
\beq
\hat\chi^{NS/R} _{\vec{\alpha}} (it+\frac{1}{2})
=(2it)^{d}\hat P_{\vec{\alpha}%
\vec{\alpha}^{\prime }}\hat\chi _{\vec{\alpha}^{\prime
}}^{NS/R}(\frac{i}{4t}+%
\frac{1}{2}) .
\eeq

Thus

\begin{eqnarray}
{\sf Z}_{C}&=&\frac 12 \int\limits_{0}^{\infty
}\frac{dt}{4t}(\frac{1}{8\pi
^{2}\alpha
^{\prime }t})^{\frac{D}{2}}\sum_{\vec\alpha} {{\cal C}_{{\vec \alpha}}
\chi^{susy}_{{\vec \alpha }}(it)}\nonumber \\
&=& \frac 12\frac{1}{(8\pi ^{2}\alpha ^{\prime
})^{\frac{D}{2}}}\int\limits_{0}^{%
\infty }\frac{dl}{4}\sum_{\vec\alpha} {{\cal C}_{{\vec \alpha}}
{S}_{\vec\alpha\vec\alpha^\prime}\chi^{susy}_{{\vec \alpha^\prime }}(
il)}
\eeqa

\begin{eqnarray}
{\sf Z}_{M}&=& \frac 12\int\limits_{0}^{\infty
}\frac{dt}{4t}(\frac{1}{8\pi
^{2}\alpha
^{\prime }t})^{\frac{D}{2}}\sum_{\vec\alpha} {{\cal M}^{{\vec \alpha}}
\hat\chi^{susy}_{{\vec \alpha }}(it+\frac 12)}\nonumber \\
&=& \frac 12\frac{1}{(8\pi ^{2}\alpha ^{\prime
})^{\frac{D}{2}}}\int\limits_{0}^{%
\infty }\frac{dl}{4}2\times 2^{\frac D2}\sum_{\vec\alpha\vec\alpha^\prime}
{{\cal M}^{{\vec \alpha}} i^d\hat P_{\vec\alpha\vec\alpha^\prime}
\hat\chi^{susy}_{{\vec \alpha^\prime }}(il+\frac 12)} .
\eeqa
where $d=(D-2)/2$.

The sum of all three amplitudes in the transverse channel thus reads

\beqa
 {\sf Z}_{K}+{\sf Z}_{C}+{\sf Z}_{M}=-\frac 12\frac{1}{(8\pi ^{2}\alpha
^{\prime })^{\frac{D}{2}}}\int\limits_{0}^{\infty }\frac{dl}{4}%
\sum\limits_{\vec\alpha }&\{&{O}^2_{\vec\alpha}\chi
_{\vec\alpha }(il)~ +~
D^2_{\vec\alpha}\chi _{\vec\alpha}(il) \\
&+& 2 \times 2^{\frac D2}\tilde{{\cal M}}_{\vec\alpha}\hat\chi
_{\vec\alpha}(il+\frac{1}{2})\}
\eeqa
Therefore the requirement to reconstruct a perfect square for
$l\rightarrow \infty$, the
factorization property sketched in equation (\ref{factori}), amounts to
\begin{equation}
D^2_{\vec\alpha}+ 2 \times 2^{\frac D2}\tilde{{\cal M}}%
_{\vec\alpha}\;+O^2_{\vec\alpha} ={\rm perfect}\;{\rm %
square}
\end{equation}
Namely, $2^{\frac D2}\tilde{{\cal M}}%
_{\vec\alpha} = \pm D_{\vec\alpha}O_{\vec\alpha}$,
recalling that $D^2_{\vec\alpha}$ is a quadratic polynomial
in $n_{\vec\alpha}$ whereas $\tilde{{\cal M}}_{\vec\alpha}$ is a linear
polynomial in $n_{\vec\alpha}$.

Moreover, for transverse characters containing $RR$ fields, $i.e.$
those originated
from the periodic blocks in the direct
channel of the Klein bottle and cylinder and the $R$ sector in the direct
channel of the M\"obius strip, zero $RR$ charge condition (\ref{tadpolecg})
must be satisfied.

\section{Examples in 8 dimensions}

We illustrate our construction
through explicit examples in $D=8$ dimensions.
In this case $c_{int} = 3$ and  there are  only three Gepner models:
 $1^3$, $1~4$ and $2^2$. Such models are known to correspond to toroidal
compactifications on, specific, rational tori \cite{gepner}.
Due to their simplicity, since they involve few blocks and low $k$
(thus a manageable number of states), it proves useful to study them
in detail and look for an exhaustive set of solutions.
In this section we concentrate  in the open sector.
Open descendants from toroidal compactification have been
discussed in \cite{bps} (see also \cite{witten,park} for other perspectives).
Other examples in 6 and 4 spacetime dimensions are then presented in
forthcoming sections.

\medskip
$\bullet ~~{\bf 1^3}$

\medskip
For the $k=1$ minimal
model, the labels $(l,q,s)$ in the standard range of the NS sector
are
(0,0,0); (0,0,2); (1,1,0); (1,1,2);
$(1,-1,0)$ and $(1,-1,2)$.
The corresponding conformal dimensions and charges
in all the inequivalent twisted
sectors are contained in the following table
\begin{equation}
.
\begin{tabular}{|l|l|l|l|}
\hline
$n$ & $
\mathop{}
{\rm Representations}$ & $\Delta $ & $Q$ \\ \hline
$0$ & $(0,0,0)$ & $0$ & $0$ \\ \hline
$1$ & $(0,1,1)$ & $\frac{1}{24}$ & $\frac{1}{6}$ \\ \hline
$2$ & $(0,2,2)\sim (1,-1,0)$ & $\frac{1}{6}$ & $\frac{1}{3}$ \\ \hline
$3$ & $(0,3,3)\sim (1,0,1)$ & $\frac{3}{8}$ & $\frac{1}{2}$ \\ \hline
$4$ & $(0,4,4)\sim (1,1,2)$ & $\frac{2}{3}$ & $\frac{2}{3}$ \\ \hline
$5$ & $(0,5,5)\sim (1,2,3)\;$ & $\frac{25}{24}$ & $\frac{5}{6}$ \\ \hline
\end{tabular}
\begin{tabular}{|l|l|l|l|}
\hline
$n$ & $
\mathop{}
{\rm Representations}$ & $\Delta $ & $Q$ \\ \hline
$0$ & $(0,0,2)\;\sim (1,\pm 3,\pm 4)$ & $\frac{3}{2}$ & $\pm 1$ \\ \hline
$1$ & $(0,1,3)\sim (1,-2,-3)\;$ & $\frac{25}{24}$ & $-\frac{5}{6}$ \\ \hline
$2$ & $(0,2,4)\;\sim (1,-1,-2)$ & $\frac{2}{3}$ & $-\frac{2}{3}$ \\ \hline
$3$ & $(0,3,5)\;\sim (1,0,-1)$ & $\frac{3}{8}$ & $-\frac{1}{2}$ \\ \hline
$4$ & $(0,4,6)\sim (1,1,0)$ & $\frac{1}{6}$ & $-\frac{1}{3}$ \\ \hline
$5$ & $(0,5,7)\sim (0,-1,-1)$ & $\frac{1}{24}$ & $-\frac{1}{6}$ \\ \hline
\end{tabular}
\end{equation}
There are three Gepner models containing only products of $k=1$ minimal
models, namely $1^r$ with $r=3, 6$ and $9$, which define string
theories
in 8, 6 and 4 dimensions respectively.
$\chi^{susy}_{\vec \alpha}$
contains $N_1, N_2$ and $N_3$ factors of
$\chi_{(0,0)}$, $\chi_{(1,-1)}$ and
$\chi_{(1,1)}$ respectively, such that $N_1+N_2+N_3 = R$ and 
$Q_{\vec\alpha} = \frac{N_2 - N_3}{3} \in {\mathbb Z}$
. Any cyclic permutation of $(N_1, N_2,
N_3)$ leads to the same $\chi^{susy}_{\vec\alpha}$.

According to (\ref{pesossusy}) the conformal dimensions of the states
contained in
a given $\chi^{susy}_{\vec \alpha}$ are
\beq
 \Delta_{\vec \beta^{(n)}} - \frac 12
= \frac{N_3}{3} \quad {\rm mod} ~ 1 .
\eeq
The general form of the states is

\[
\lbrack \prod\limits_{i=1}^{d
}(0)^{1-d_{i}}(2)^{d_{i}}](0,0,0)^{N_{1}-n_1}
(0,0,2)^{n_1}(1,-1,0)^{N_{2}-n_2}(1,-1,2)^{n_2}(1,1,0)^{N_{3}-n_3}(1,1,2)^{n_3}
\]
where the first two entries refer to the label $\lambda$ of the spacetime
contribution, $d_i=0,1$ and $n_i=0,...,N_i$.
The odd $U(1)$ charge condition leads to
\beq
\sum_{i=1}^{d} d_i + n_1 - n_2 + n_3 + \frac{N_2-N_3}{3} =
{\rm
odd}
\eeq

The relation (\ref{identifi}) in the case $1^3$ implies that two
characters are
identical if the following replacements are performed in each internal
theory
$(0,0)\rightarrow (1,-1)  ; (1,-1) \rightarrow (1,1) ;
 (1,1) \rightarrow (0,0)$.
Therefore
there are   characters for this model, namely
$\chi_A=\chi^{susy}_{(0,0)^3}$,
 $\chi_B=\chi^{susy}_{(0,0)(1,-1)(1,1)}$
and  $\chi_C=\chi^{susy}_{(0,0)(1,1)(1,-1)}$.
Where, for instance, (see (\ref{susych}))
\begin{equation}
\chi_A=\chi^{susy}_{(0,0)^3}
=\sum\limits_{n,p\;{\rm mod\;6}}\;\frac{%
(-1)^{n+p}}{6}
e^{2\pi i n^{2}\frac{c}{24}\tau }
\left [\chi_0(\tau, \frac n2 \tau + \frac p2)
\right ] ^{3} \chi _{(0,0)}^3(\tau ,\frac{n%
}{2}\tau +\frac{p}{2})
\label{susych13}
\end{equation}
Notice that $\chi _{(0,0)}^3$ is a short hand notation indicating
that the same
character $\chi _{(0,0)}$ is being considered in each internal block.
 The conformal dimensions
of the highest weight states
are $\Delta_A=\frac 12$ and $\Delta_B=\Delta_C=\frac 56$, respectively
and the
GSO projected
combinations of states contained in these characters are listed in
Appendix B. Notice, for instance, that $\chi^{susy}_{(0,0)^3}$ massless
states span the $D=8$, $N=1$ vector 
representation~\footnote{Namely, it contains $1+1+6+4+4^\prime$ $SO(6)$ little group
representations}~\cite{strathdee}.

The matrices ${S}^{(1^3)}$ and $\hat P^{(1^3)}$ are
\begin{equation}
{S}^{(1^3)}=\frac{1}{\sqrt{3}}\left(
\begin{array}{ccc}
1 & 1 & 1 \\
1 & e^{\frac{2\pi i}{3}} & e^{-\frac{2\pi i}{3}} \\
1 & e^{-\frac{2\pi i}{3}} & e^{\frac{2\pi i}{3}}
\end{array}
\right) \quad ; \quad
\hat P^{(1^3)}=\frac {i^{-d}}{\sqrt{3}}\left(
\begin{array}{lll}
- 1 & 1& 1 \\
1 & e^{-\frac{\pi i}{3}} & e^{\frac {\pi i}{3}}  \\
1 & e^{\frac{\pi i}{3}} & e^{-\frac{\pi i}{3}}
\end{array}
\right) \label{p1}
\end{equation}
with $d=(D-2)/2=3$

There are two modular invariant combinations of characters in the torus to
be considered, diagonal and charge conjugation. We discuss them
separately.

\noindent
{{\it i) Diagonal} {${\bf( 1_ A)^3}$}

There are several possibilities for the Klein bottle
partition function in the direct channel, namely
\begin{equation}
{\cal Z}_{K}(it)= \frac 12 \left [\pm \chi _{A}(2it) \pm
\chi _{B}(2it) \pm
\chi _{C}(2it) \right ] .
\label{knd}
\end{equation}

Let us  start with all positive signs. The partition function
in the transverse channel reads
\begin{equation}
\tilde{\cal Z}_{K}(il)=\frac 12 2^8 {\sqrt[2]{3}}\chi _{A}(il) ,
\end{equation}
so only the term
\beq
\tilde{\cal Z}_{M}(il) = \frac 12 \times 2 \times 2^4 \sqrt[2]{3}
A\hat{\chi}_{A}(il+\frac{1}{ 2}) \label{zt111}
\eeq
 must be present in
the M\"obius strip sector. The transverse cylinder amplitude must read
\beq
\tilde{\cal Z}_{C}(il) = \frac 12 {\sqrt[2]{3}}[A^{2}\chi
_{A}(il)+B^{2}\chi _{B}(il)+C^{2}\chi _{C}(il)] \nonumber
\label{cyld}
\eeq
to ensure factorization. 
Since massless RR tadpoles (and also NSNS here) are all contained in
$ \chi_{A}$, $A=-16$ is needed in order to achieve tadpole cancellation.

When rewriting the amplitudes in the direct channel (using the
transformation matrices (\ref{p1})) the linear combinations of
coefficients multiplying the three characters must satisfy the
following consistency conditions:

$a)$ they must be integer polynomials in $n_i$;

$b)$ the coefficients of $\chi_A$, the character containing the massless
vector in the open sector, must be $\frac 12 n_i(n_i +1)$
for $Sp(n_i)$,  $\frac 12 n_i(n_i -1)$ for $SO(n_i)$ and $n_i^2$ for
$U(n_i)$ (or rather $n_i{\bar n}_i$, see footnote 3 in page 7).
 Consistency conditions $a)$ and $b)$ imply in
this case $A=-n$ and $B=C=0$ leading to the direct channel amplitudes
\beqa
{\cal Z}_C(it)&= &\frac 12 n^2 [  \chi_A (it)+\chi_B (it) + \chi_C (it)\ ]\\
 {\cal Z}_M(it) & =&
\frac 12 n[ \hat\chi_A(it+\frac 12)-\hat\chi_B(it+\frac 12)
-\hat\chi_C(it+\frac 12)]
\eeqa

Then the tadpole cancellation condition is $n=2^4 $, and we are lead to an
$Sp(16)$ gauge group with massive matter transforming in the
antisymmetric~\footnote{Actually in $\textbf{119}+\textbf{1} $ since
the antisymmetric representation is reducible} and symmetric
representations (recall that the change in sign in MS for different
levels changes the symmetry of the corresponding representation).

It is easy to check that other possible combinations of signs in the partition function from
the Klein bottle (\ref{knd}) do not admit solutions satisfying conditions
$a)$ and $b)$.

\medskip

\noindent
{\it ii) Charge conjugation} ${ \bf {(1_C)^3 }}$ }

The Klein bottle partition function in the direct channel is in this case
\begin{equation}
{\cal Z}_{K}(it)=\frac12 \chi _{A}(2it)
\end{equation}
which, written in the transverse channel, reads
\begin{equation}
\tilde{\cal Z}_{K}(il)=\frac 12 2^8\frac 1{\sqrt[2]{3}}(\chi _{A}(il)+\chi
_{B}(il)+\chi _{C}(il))
\end{equation}
Generic expressions for cylinder and M\"obius strip amplitudes 
in the transverse channel are given by
\beqa
\tilde{\cal Z}_{M}(il)& =
&\frac 12 \times 2 \times 2^4\frac{1}{\sqrt[2]{3}}[A\hat{\chi}_{A}(il+\frac{1}{%
2})+B\hat{\chi}_{B}(il+\frac{1}{2})+C\hat{\chi}_{C}(il+\frac{1}{2})]
\nonumber \\
\tilde{\cal Z}_{C}(il)& = &\frac 12\frac{1}{\sqrt[2]{3}}[A^{2}\chi
_{A}(il)+B^{2}\chi _{B}(il)+C^{2}\chi _{C}(il)] \nonumber
\label{mscylcc}
\eeqa
where, $A, B, C$ are linear (complex) combinations of $n_i$
($i= A,B,C$). Tadpole cancellation thus requires $A=-16$.
When rewriting them in the
 open string direct  channel we obtain
\beqa
3 {\cal Z}_M(it) = \frac 12 \ [(-A+B+C) \hat\chi_A(it+\frac12)
+ (A+B e^{\frac {\pi i}3}+Ce^{\frac {-\pi i}3}) \chi_B (it+\frac12) \nonumber\\
+ (A+B e^{\frac {-\pi i}3}+Ce^{\frac{\pi i}3}) \chi_C (it+\frac12) \ ]
 \nonumber\\
3{\cal Z}_C(it) = \frac 12 \ [(A^2 + B^2 + C^2) \chi_A (it)
+ (A^2+B^2e^{-\frac {2\pi i}3}+C^2e^{\frac {2\pi i}3}) \chi_B (it) \nonumber\\
+ (A^2+B^2e^{\frac {2\pi i}3}+C^2e^{-\frac{2\pi i}3}) \chi_C (it)\ ]
.
\label{mscyldir}
\eeqa

A solution satisfying the consistency conditions $a)$ and $b)$ is
  $B=-n e^{\frac{i}3 \pi}-{\bar n} e^{-\frac{i}3 \pi}+m=C^* $,
$A =-n -{\bar n}-m $, ( numerically $n=\bar n$ ) leading to
\beqa
{\cal Z}_{C}(it) &=&(\frac12 m^2+n {\bar n})
\chi_{A}(it) +(\frac12 n^2 +n m) \chi _{B}(it)+(\frac12 {\bar n}^2 +{\bar n} m)
\chi_{C}(it)  \nonumber \\
{\cal Z}_{M}(it)  &=&
\frac12 m\hat{\chi}_{A}(it+\frac{1}{2})
-\frac12 {\bar n} \hat{\chi}_{B}(it+\frac{1}{2})-\frac12 {n}
\hat{\chi}_{C}(it+\frac{1}{2})
 \label{zmzc}
\eeqa
with the tadpole cancellation
condition
\beq
-A =n +{\bar n}+m =2n + m=16 .
\eeq

The interpretation here is less clear.  $n$ and
$\bar{n}$ have been interchanged in ${\cal Z}_{M}$ with respect to
what appears in ${\cal Z}_{C}$. On the other hand, the characters
$\chi_B$  and $\chi_C$  are the same (considered as functions of $\tau$
and $z$) so, expansion in powers of $q$ seems to indicate an 
$Sp(2(8-n))\times{}U(n)$ $N=1$, $D=8$ vector
multiplet with descendant massive fields in the corresponding adjoint
representations and extra massive matter transforming in $ \bf
{(2(8-n), n)} +\bf {(2(8-n),{\bar n})} +(1,\Yasymm) +(1,{\overline{\Yasymm } }) $
 (or $ (1,\Ysymm) +(1,{\overline{\Ysymm }})$ depending on the mass level).

Such exchange of $n$ and $\bar n$ might indicate that linear
combinations (symmetric and antisymmetric) of fields $|$B$>$ and $|$C$>$,
which have similar conformal weight and charge, must be considered.

Notice that $n=0$ leads to  $Sp(16)$ and there  is no contribution from
massive characters  $\chi_{B} $ and $\chi_{C}$.

A particular prescription leading to a consistent theory,
$i.e.$ a theory verifying the requirements of factorization
and tadpole cancellation, was found in \cite{bs1} for the
charge conjugation torus partition function. In this case, Cardy
\cite{cardy} has shown that,  when there are the same number of characters
than Chan Paton factors,
 a natural solution for the partition
function from the cylinder
is given by $C_{ab}^c = N_{ab}^c$, where
$N_{ab}^c$, the number of conformal blocks of a rational CFT,  can
be written
in terms of the $S$ matrix as
\beq
N_{ab}^c = \sum_l \frac{S_{al}S_{bl}(S^\dag)^{lc}}{S_{0l}} . \label{cardy}
\eeq
This is the celebrated Verlinde theorem \cite{verlinde} arising as a
consequence of the
well established fact that the modular matrix $S$ diagonalizes the fusion
rules. The proof of the theorem relies on the technical assumption that
both left and right extended chiral algebras consist only of generators
with integral conformal dimension and thus evidently excludes the
superconformal case. A generalized Verlinde formula which describes the
fusion rules in all sectors of $N=1$ superconformal theories was obtained
in \cite{nosss} whereas the $N=2$ case was dealt with in reference
\cite{waki}. Cardy's extension of the theorem to surfaces with boundaries
was generalized to $N=1$ in reference \cite{nepo}.

It was shown in \cite{pss} that one can define another integer valued
object \cite{fuchs}
\beq
Y_{ab}^c = \sum_l\frac {S_{al}P_{bl}(P^\dag)^{lc}}{S_{0l}}
\eeq
leading to
\beq
{\cal K}_a = Y^0_{a0} \quad , \quad M_a^b = Y_{a0}^b \label{ansatz}
\eeq
This ansatz requires symmetric $S$ and $P$ matrices, and therefore
 non trivial solutions must reproduce some of the models found above.
Actually, the partition functions (\ref{zmzc})
can be obtained with this prescription.

There are other solutions satisfying the factorization and
tadpole cancellation conditions but they do not verify the requirement
$b)$ above, $i.e.$ they do not lead to massless vectors transforming in
either $Sp$, $SO$ or $U$ groups.

 \medskip
 {$\bullet ~~{\bf 2^2}$}

\medskip
All the representations of the $N=2$ superconformal $k=2$ minimal model
can be obtained by twisting the pairs $(0,0,0) ; (0,0,2)$ and $(1,-1,0) ;
(1,-1,2)$. They are listed in Appendix B.
The characters of the Virasoro algebra are
contained in the following table

\[
\begin{tabular}{||l||l||l||}
\hline\hline
Character  & Weight& Charge \\ \hline\hline
$(0,0)$ & $0$ & $0$ \\ \hline\hline
$(2,-2)$ & $\frac{1}{4}$ & $\frac{1}{2}$ \\ \hline\hline
$(2,0)$ & $\frac{1}{2}$ & $0$ \\ \hline\hline
$(2,2)$ & $\frac{1}{4}$ & $-\frac{1}{2}$ \\ \hline\hline
$(1,-1)$ & $\frac{1}{8}$ & $\frac{1}{4}$ \\ \hline\hline
$(1,1)$ & $\frac{1}{8}$ & $-\frac{1}{4}$ \\ \hline\hline
\end{tabular}
\]
and the independent supersymmetric characters in the $2^2$ product
theory are
$\chi_A = \chi^{susy}_{(0,0)^2}$,
$\chi_B = \chi^{susy}_{(1,-1);(1,1)}$,
$\chi_C = \chi^{susy}_{(0,0);(2,0)}$ with
$\Delta_A = \frac 12$,
$\Delta_B = \frac 34$,
$\Delta_C = 1$, respectively. The GSO projected combinations of states
contained in
these
characters are listed in Appendix B.

This model admits only the diagonal modular invariant. The
Klein bottle
 partition
function is either
\beqa
&1)& \quad {\cal Z}_K(it) = \frac 12 \left [\chi_A(2it) + \chi_B(2it) +
 \chi_C(2it)\right ],
\nonumber\\
&2)& \quad {\cal Z}_K(it) = \frac 12 \left [-\chi_A(2it) + \chi_B(2it) +
 \chi_C(2it) \right ] \quad {\rm or} \nonumber \\
&3)& \quad {\cal Z}_K(it) = \frac 12 \left [\chi_A(2it) + \chi_B(2it)
- \chi_C(2it)\right ].
\label{22pf}
\eeqa
The matrices ${S}$ and $\hat P$ are in this case
\begin{equation}
{S}^{(2^2)}=\frac{1}{2}\left(
\begin{array}{ccc}
1 & 1 & 1 \\
1 & 1 & -1 \\
2 & -{2} & 0
\end{array}
\right) \quad ; \quad
\hat{P}%
^{(2^2)}
=i^{-d}\left(
\begin{array}{ccc}
0 & 1 & 0 \\
1 & 0 & 0 \\
0 & 0 & 1
\end{array}
\right)
\end{equation}
with $d=(D-2)/2=3$

They may be used to rewrite the Klein bottle amplitude in the transverse
channel as
\beqa
&1)& \quad
\tilde{\cal Z}_K(il) =  2^8 \chi_A(il) \\
 \quad &2)& \quad
\tilde{\cal Z}_K(il) =  \frac 12 2^8 \left [
 \chi_A(il) - \chi_B(il) -  \chi_C(il) \right ]\\
&3)&  \quad \tilde{\cal Z}_K(il) =  2^8  \chi_B(il)
\eeqa
Let us discuss case $1)$ first.
The contribution from the open sector must then be
\beqa
\tilde{\cal Z}_{M}(il+\frac{1}{2})
&=& - 2 \times 2^4 A\;\hat{\chi}_{A}(il+\frac{1}{2}%
) \\
\tilde{\cal Z}_{C}(il) &=& A^{2}\;\chi _{A}(il)+B^{2}\chi
_{B}(il)
+C^{2}\chi _{C}(il)
\eeqa
Consistency conditions lead to $A=\frac12 (n+\bar n)= B$,
$C= \frac i2(n-\bar n)$
(recall $n=\bar n$) and
$n=16$ from tadpole cancellation.
Therefore
\beq
{\cal Z}_C(it)= n \bar n \chi_A(it) +
(\frac12 n^2 + \frac12 {\bar n}^2) \chi_B(it)
\eeq
and
\beq
{\cal Z}_M(it+\frac 12) =
-\frac12 (n+{\bar n}) \hat\chi_B(it+\frac 12)\nonumber
\eeq
 Here we obtain a massless  $N=1, D=8$ $U(16)$ vector multiplet (with
 massive descendants in the adjoint)
and massive states in the antisymmetric
representation (with descendants in the symmetric or antisymmetric
according to the level).

Let us now consider case $2)$. Since all characters contribute to 
the Klein bottle amplitude, it appears that complex multiplicities are not
permitted and therefore no unitary groups are allowed.
The resultant  partition functions in the transverse
channel are 
\beqa
\tilde{\cal Z}_{M}(il+\frac{1}{2})
&=&-2\times 2^4n\;\hat{\chi}_{A}(il+\frac{1}{2}%
) \\
\tilde{\cal Z}_{C}(il) &=&2n^{2}\;\chi _{A}(il)
\eeqa
whereas in the direct channel they are
\beq
{\cal Z}_M(it+\frac 12) =-n\hat\chi_B(it+\frac 12)\nonumber
\eeq
\beq
{\cal Z}_C(it)=n^2 [\chi_A(it) +
\chi_B(it)+\chi_C(it)]
\eeq
 The tadpole cancellation condition is now $n=8$.  Even though the
massless character ({\em i.e.} the one corresponding to the massless
state) does not appear in the M\"obius strip amplitude, there is no
consistent
solution in terms of $n$ and $\bar n$ multiplicities and, therefore,
no unitary group seems to be allowed. The amplitude can be interpreted
as corresponding to $SO(8)\times Sp(8)$ vector multiplet plus massive
descendant states in symmetric and antisymmetric representations and
in bi-fundamentals.  

Starting with ${\cal Z}_K = \chi_A - \chi_B + \chi_C$ the 
same result is obtained.

The third case is interesting since the Klein bottle amplitude has no
massless
tadpoles. Therefore, closed unoriented theory is  consistent
with no need of an open string sector. (See \cite{gepnertv1,gj} for other
examples).

\medskip
 {$\bullet ~~ \textbf{4 ~ 1}$}
\medskip

The states of the $k=4$ minimal model may be classified in two sets,
one with $l=0$ and $l=4$ and the other one with $l=2=\frac k2$ (short).
They are listed in Appendix B.

Taking  into account the spectrum of the $k=1$ minimal model, the only
possible combinations of states in the $4 ~ 1 $ Gepner model are

\begin{equation}
\begin{tabular}{||l||l||}
\hline\hline
$\chi _{\vec{\alpha}}^{Susy}$ & $\Delta _{\vec{\beta}^(n)}%
\mathop{\rm mod}
1$  \\ \hline\hline
$\chi _{(0,0)(0,0)}^{Susy}$ & $\frac{1}{2}$  \\ \hline\hline
$\chi _{(2,0)(0,0)}^{Susy}$ & $\frac{5}{6}$  \\ \hline\hline
\end{tabular}
\end{equation}

Let us introduce the following notation:
$\chi _{A}\equiv \chi
_{(0,0)(0,0)}^{susy},\chi _{B}\equiv \chi _{(2,0)(0,0)}^{susy}$ where the
first pair of indices refer to $k=4$ and the second one to $k=1$.

In fact, it may be seen by comparing the tables in Appendix B  that the
spectrum of states in $\chi_A$ is identical to
that of $\chi_{(0,0)^3}$ in the $1^3$ model and similarly for $\chi_B$
and $\chi_{(0,0)(1,-1)(1,1)}$ (or equivalently $\chi_{(0,0)(1,1)(1,-1)}$).
Actually, this is in agreement with the conjectured equivalence
between conformal models ${\bf 4} \equiv {\bf{1_A^2} } $
\cite{equiv}.

Moreover, this identification remains valid for the open string.
This can be easily checked by noticing that there is only one
modular invariant combination of characters in this case that
leads to the following Klein bottle amplitude  in the direct channel
\begin{equation}
{\cal Z}_{K}(it) =\frac 12 \left [\chi
_{A}(2it)+{\chi}_{B}(2it)\right ]
\end{equation}
Matrices ${S}$ and  $\hat{P}$ in this case are
\begin{equation}
{S}^{(4~1)}=\frac{1}{\sqrt[2]{3}}\left(
\begin{array}{cc}
1 & 1 \\
{2} & -1
\end{array}
\right) \quad ; \quad  \hat{P}^{(4~1)}=\frac
{i^{-d}}{\sqrt[2]{3}}\left(
\begin{array}{cc}
-1 & 1 \\
~{2} & 1
\end{array}
\right) \label{p14}
\end{equation}
where $d=(D-2)/2=3$, the first row and column refer to $A$ and the second ones
to $B$. They allow us to find  the transverse channel contribution from
the Klein bottle
\begin{equation}
\tilde{\cal Z}_{K}(il)=\frac 12 2^8 \sqrt[2]{3}\chi _{A}(il)
\end{equation}
(compare with (\ref{zt111})) and to find that consistency requirements
lead to the
same  $Sp(16)$ theory found in the ${\bf (1_A)^3} $ model.

Before concluding this section let us stress some aspects of our results.

Notice that, while rank 16 groups
can be
obtained $- e.g.$ in the $2^2$ model, case $1)$ $-$,
gauge groups of rank 8 are also  found.
Naively, a rank $N_D=16$  group could have been
expected from a toroidal compactification of $SO(32)$
string with 16 pairs of 9 branes.

This subject has been discussed from several perspectives in the literature.
In fact, rank reduction can be explained from the presence of a discrete
$NSNS$
antisymmetric field (see \cite{bianchi} and references therein).
Indeed, while generically orientifold projection
kills antisymmetric field fluctuations, in some cases
a discrete vacuum expectation value for a rank $b$ antisymmetric
tensor field can still be preserved, leading to
$N_D= 32 \times 2^{-\frac{b}2} $.
The  issue of rank reduction was addressed in \cite{witten}
(see also \cite{park}) from a more topological point of
view. There, the presence of four fixed  orientifold seven planes
of different kind,   $O^+$ ($ O^-$) carrying $ -8 ~(+8)$ units of
D-brane charge,  is identified depending on the way
the orientifold acts on the $T^2$ torus (with or without
vector structure). While
D-branes on smooth  points generate unitary groups, when they sit on
the top of a $O^-$ $(O^+)$ orientifold plane an $Sp$ ($SO$) gauge
group is produced.

Thus, the  $Sp(2(8-n)) \times U(n)$ group obtained in the $ 1^3$  example
above would correspond
to the case of $3O^+ + O^-$ points,
 where $8-n $ pairs sit at $O^-$ points
and the rest coincide at smooth points.

Also, the $U(16)$ group in $2^2$ would correspond to branes at smooth
points
in  a  $4 O^+$  configuration. Recall that sign change in the Klein bottle
amplitude
for this case leads to rank reduction, here with $Sp(8)\times SO(8)$,
 which
would correspond to two groups of
four pairs of branes distributed between the   $O^+$ and $O^-$ points.
The third case in (\ref{22pf}), with no tadpoles, would correspond in this
description to a  $2O^++ 2O^-$ configuration.
Notice that correspondence between our results and
this description is indicative and would require further investigation.

\section{Examples in 6 dimensions}
$N=1$, $D=6$ supersymmetric models have potential
chiral and gravitational anomalies.
Anomaly cancellation is thus a strong check on the consistency
of the whole construction.

The massless representations in $D=6$ are labelled by the
representations
of the little group   $SO(6)\sim SU(2)_1 \times SU(2)_2$
and they gather into  gravity, vector, tensor and hyper multiplets.
Clearly gravity and tensor multiplets are present only in the closed sector.
They are obtained by plugging left and right moving massless states,
which are invariant under
$\Omega $ orientation reversal exchange.
Consider, as an example, the coupling of $RR$ states. Massless Ramond
states in $D=6$ are conformal weight $1/2$ states
of the form (see table $B.10$) $|(+,+)_{1},R _0>, |(-,-)_{-1},R_0>$
which organize into  $(\textbf{2},\textbf{1})R$ Lorentz representation
or  $|(+,-)_0,R_1>, |(-,+)_0,R_{-1}>$ leading to
$(\textbf{1},\textbf{ 2})$ representations.
We indicate with a subindex the corresponding spacetime or internal
charge and  $R$ summarizes the internal sector fermionic  content.
In particular, if left and right $R,R'$ Ramond states
couplings $(\textbf{2},\textbf{1}) R \times (\textbf{2},\textbf{1}) R'$
are allowed by modular invariant coefficients, orientifolding will lead to
$ (\textbf{3}, \textbf{1}) \frac12 (R \times R'+ R' \times R)$ triplets and
$ (\textbf{1}, \textbf{1}) \frac12 (R \times R'- R' \times R)$ singlets.
Together with $NSR$ fermions such states group in a tensor supermultiplet
 $T =(\textbf{3}, \textbf{1}) + (\textbf{1},
\textbf{1}) + 2(\textbf{2}, \textbf{1})$.
Recall that, due to anticommutation of fermion
fields, only the scalar will be present if $R=R'$.
The other closed states are constructed following  the same steps.

As an illustration of a specific computation, let us consider the simple
$1^6$ models.
Left and right moving sectors can be coupled in several manners.
In particular, each left block can be coupled either diagonally or to its
charge conjugate left state (we choose positive ${\cal K}_{\vec\alpha} $
signs
here).
Thus, ${\bf (1_A)^6}$, ${\bf (1_A)^51_C}$, ${\bf 1_A (1_C)^5}$, ${\bf
(1_A)^3 (1_C)^3 }$, ${\bf (1_A)^2(1_C)^4}$, ${\bf (1_ A)^4(1_C)^2}$
and ${\bf (1_C)^6}$ modular invariant couplings are possible. The
first four cases have $0 T+ 21 H$, the last \textit{conjugate }
invariant has $10 T+ 11 H$, while the other two contain $6 T + 15H $
multiplets in the closed sector.  Simple solutions for some of these
cases have been found in \cite{gepnertv1}.
For instance, by noticing that in $\bf{(1_A)^6} $ the
Klein bottle amplitude in the transverse channel contains just the $q=0$
states
\footnote{Actually, it is a general result for any $k$ model that
diagonal invariant coupling leads to zero charge characters in the
transverse channel.}, namely
\beq
\tilde{\cal Z}_K(il)=
9 \frac12 2^6 \chi_{(0,0)^6}^{susy}(il) ,
\label{klein16}
\eeq
it is immediate to see that $ \tilde{\cal Z}_C(il)=
n^2 \tilde{\cal Z}_K(il)$ and $\tilde{\cal Z}_M(il)= - 2 \times 2^3
 n {\tilde{\cal Z}}_K(il+\frac12)$ give a consistent theory with $n=8$
corresponding to
$SO(8)$ gauge group and 10 hypermultiplets in the $\textbf{28}$
representation.

As another illustration, with a more general solution,
let us consider the
  ${\bf (1_A)^4 (1_C)^2}$ model
with  Klein bottle amplitude
\beq
{\cal Z}_K(it) = \frac 12
\{
\left ( \chi_{(0,0)}  +
\chi_{(1,-1) } + \chi_{(1,1)}\right )^4
\chi_{(0,0)} ^2  \}^{susy}
\eeq
where only internal blocks are displayed and ${susy}$ means that sums over
$n,p$,
 as indicated in (\ref{susych}), must be performed.
The following
contributions from the direct channel of the cylinder
\beqa
 {\cal Z}_C(it) =(\oh n_1^2+n_2^2  ) ~ \{ \chi_{(0,0)}^6 &+&
{\underline {\chi_{(1,-1)}\chi_{(1,1)}\chi_{(0,0)}^2}} \chi_{(0,0)}^2  +
{\underline {\chi_{(1,-1)}^3 \chi_{(0,0)}}}\chi_{(0,0)}^2 +  \nonumber\\
&+ &
{\underline {\chi_{(1,1)}^3  \chi_{(0,0)}}} \chi_{(0,0)}^2 +
{\underline {\chi_{(1,-1)}^2 \chi_{(1,1)}^2}} \chi_{(0,0)}^2 ~\}^{susy}
+   \nonumber\\
+ ~ (\oh {n_2}^2 + n_1 n_2 ) &\{&
\chi_{(0,0)}^4  \chi_{(1,-1)}\chi_{(1,1) } +
{\underline {\chi_{(0,0)}^2
\chi_{(1,-1)}\chi_{(1,1)}}} ~\ {\underline { \chi_{(1,-1)}\chi_{(1,1)}}} \nonumber\\
&+& {\underline {\chi_{(1,1)}^2 \chi_{(0,0)}^2}} ~\
 {\underline {\chi_{(1,1)} \chi_{(0,0)}}} ~\}^{susy}
\eeqa
and from the
 M\"obius strip
\beq
{\cal Z}_M (it) =
- \sum_{all~characters}[(-1)^{N_{(1,1)}} (\oh n_1 \hat X_1 +
\oh n_2 \hat X_2)]^{susy} ,
\eeq
where $X_1$ and $X_2$ refer to all the products of characters with the
last two factors being respectively $\chi_{(0,0)}^2$ and
$\chi_{(1,-1)}\chi_{(1,1)}$
or $\chi_{(1,1)}
\chi_{(1,-1)}$,
lead to a consistent solution provided
the tadpole cancellation condition
\begin{equation}
2n_2+n_1 = 8
\label{numbranes}
\end{equation}
is satisfied.

{}From $\chi_{(0,0)^6}^{susy}$ we read the $SO(n_1) \times U(n_2)$ gauge
group.
The other massless
characters in $X_1$ (namely  $\chi_{(1,-1)}^3
\chi_{(0,0)}^3$, $\chi_{(1,1)}^3\chi_{(0,0)}^3$
and any permutation of the first four
characters in the products)
transform in the adjoint representation of $SO(n_1)$ and
 $U(n_2)$.
 The massless characters in $X_2$ (namely ${\underline
{\chi_{(1,-1)}^2\chi_{(1,1)}^2}} \chi_{(1,1)}\chi_{(1,-1)}$ and
${\underline
{\chi_{(1,-1)}^2\chi_{(1,1)}^2}}\chi_{(1,-1)}\chi_{(1,1)}$) have
massless states transforming in the antisymmetric representation
(descendants will be in the symmetric or antisymmetric, according to
the level) of $U(n_2)$ and a bi-fundamental representation.  

Thus, the massless multiplets are
\begin{eqnarray}
&{\rm Vector}&  SO(n_1) \times U(n_2) \nonumber\\
&{\rm Hypers} &  4(\Yasymm,1 ) + 4(1,\rm Adj)+  6(1,\Yasymm )+6 (\textbf{n}_1,\textbf{n}_2)
\end{eqnarray}

It is easy to check  that all gauge and gravitational anomalies cancel
(recall that 6 tensor multiplets and 15 hypermultiplets are present
in the closed sector) whenever the tadpole cancellation
condition (\ref{numbranes})
is satisfied.

It is interesting to compare this computation
 with an orbifold like case. Consider, for instance,
a Type IIB orientifold on $T^4/ Z_3$. By looking at $Z_3$ twists
action on  D9-branes one finds (when no Wilson lines are turned
on) that the general, massless spectrum reads (see for instance
\cite{ukth})
\begin{eqnarray}
&{\rm Vector}&  SO(n_1) \times U(n_2) \nonumber\\
&{\rm Hyper} &  (\textbf{n}1,\textbf{n}_2)+ (1,\Yasymm )
\end{eqnarray}
If \textit{twisted} tadpole cancellation condition $n_2-n_1=8$ is
satisfied, the spectrum is free of both gauge and gravitational
anomalies (once the eleven hyper and the ten tensor multiplets
from the closed sector are included) independently of the total
number of branes. It is the global  untwisted RR tadpole
cancellation condition $n_1+2n_2=32$, which fixes the total number
of branes \footnote{Interestingly enough such conditions can be
understood, by considering D-brane probes, as consistency conditions  of the
effective theory in all topological sectors \cite{ukth}}. In our
case we see that factorization, in massless and massive transverse
channel sectors, leads to an effective field theory
\label{c4a2spec} which is inconsistent unless the number of branes
is restricted to $n_1+2n_2=8$. Thus global information appears to
be required at every step in our construction.

\section{Examples in 4 dimensions}
\medskip
\noindent
$\bullet ~~ {\bf 1^9}$

The analysis of this model follows very closely the $1^6$ case with
the obvious modifications. The diagonal modular invariant partition
function ${\bf (1_A)^9}$ leads to the usual tadpole cancellation condition,
$n=2^{D/2}=4$
and the massless vector belongs to a $Sp(4)$, $D=4$, N=1 vector multiplet.

The ${\bf (1_A)^7 (1_C) ^2}$ case is analogous to the ${\bf (1_A)^4(1_C)^2}$ example above.
There
is a
consistent model with Chan
Paton gauge group  $Sp(n_2)\times U(n_1)$ and tadpole cancellation
condition $2n_1+n_2=4$.

\medskip
\noindent $\bullet ~~{\bf 3^5}$

The representations of the $k=3$ minimal model are listed in Appendix B.
It is convenient to denote the supersymmetric characters in this theory
with the number
$N_i$ which refers to the multiplicity of the $i$-th character in the
product. The index $i=1,...,9$ refers  to the states
(0,0), (3,-3),  (3,-1), (3,1), (3,3), (2,0), (2,2),
(1,-1), (1,1), (2,-2) respectively.
The
combinations of characters with integer U(1) charge are the following:

$N_{1}+N_{6}=5$

$N_{1}+N_{6}=3;N_{2}+N_{7}=N_{5}+N_{10}=1$

$N_{1}+N_{6}=3;N_{3}+N_{8}=N_{4}+N_{9}=1$

$N_{1}+N_{6}=N_{2}+N_{7}=N_{3}+N_{8}=N_{4}+N_{9}=N_{5}+N_{10}=1$

$N_{1}+N_{6}=1;N_{2}+N_{7}=N_{5}+N_{10}=2$

$N_{1}+N_{6}=1;N_{3}+N_{8}=N_{4}+N_{9}=2.$

The diagonal modular invariant partition function in the
torus leads to the following expression for the
direct channel from the Klein bottle
\begin{equation}
{\cal Z}_{K}(it)=\frac 12 \frac{1}{5}\left [(\chi _{I}(it)+\chi
_{II}(it))^{5}\right ]^{susy}
\end{equation}
where
\begin{eqnarray}
\chi _{I} &=&\chi _{(0,0)}+\chi _{(3,-3)}+\chi _{(3,-1)}+\chi _{(3,1)}+\chi
_{(3,3)}  \nonumber \\
\chi _{II} &=&\chi _{(2,0)}+\chi _{(2,2)}+\chi _{(1,-1)}+\chi _{(1,1)}+\chi
_{(2,-2)}
\end{eqnarray}

Applying the $S$ matrix for the $k=3$ model (see Appendix B)
we find in the transverse channel
\begin{equation}
\tilde{\cal Z}_{K}(il)=
\frac 12 2^{4}\sqrt[4]{5}\left [(\kappa^{\frac{3}{2}}\tilde{\chi}%
_{(0,0)}(il)+\kappa^{-\frac{3}{2}}\tilde{\chi}_{(2,0)}(il))^{5}\right
]^{susy}
\end{equation}
where $\kappa\equiv \frac{1}{2}(1+\sqrt{5})$.

Following the same procedure as  in the 6
dimensional examples treated above,
we propose the following partition function for the transverse
channel from the cylinder
\begin{equation}
\tilde{\cal
Z}_{C}
(il)=\frac
12 \sqrt[4]{5}{\cal A}_{\vec{\gamma}}\left
[\prod\limits_{i=1}^{5}(\tilde{\chi}%
_{(0,0)}(il))^{1-\gamma _{i}}(\tilde{\chi}_{(2,0)}(il))^{\gamma
_{i}}\right ]^{susy}
\label{cil35}
\end{equation}
Here
$\vec\gamma=\vec{\gamma}( \vec{\alpha})$
 denotes a 5 component vector (one for each theory) taking values
$0$ or 1 if the state belongs to group I or II respectively. Note that
 $\vec{\gamma}( \vec{\alpha}_{(n)}) =\vec{\gamma}(
\vec{\alpha}) $ since the states remain in the same group under twisting.

Therefore the tadpole cancellation conditions are
\begin{equation}
{\cal A}_{\vec{0}} =2^{4}\kappa^{\frac{15}{2}}  \qquad ; \qquad
{\cal A}_{\vec{1}} =2^{4}\kappa^{-\frac{15}{2}}  \label{Tadpole1}
\end{equation}
where $\vec 0 \equiv (0,0,0,0,0)$ and $\vec 1 \equiv (1,1,1,1,1)$.

Applying the $S$ matrix we can transform (\ref{cil35}) to the direct channel
where the
partition function from the cylinder reads
\begin{equation}
{\cal Z}_{C}(it)
=\frac
12\sum\limits_{\vec{\gamma},\vec{\alpha}}{\cal 
C}_{\vec{\gamma}(\vec{\alpha}%
)}\chi _{\vec{\alpha}}
\eeq
with
\beq
{\cal C}_{\vec{\gamma}}
=\frac{1}{(\sqrt[2]{5}\kappa)^{\frac{5}{2}}}\sum\limits_{\vec{%
\gamma}^{\prime }}\frac{1}{\left( \sqrt[4]{5}\right)
^{5}}\kappa^{(\vec{\gamma}-%
\vec{\gamma}^{\prime })^{2}}(-1)^{\vec{\gamma}.\vec{\gamma}^{\prime
}}{\cal A}_{%
\vec{\gamma}^{\prime }}
\end{equation}

Let us denote
\beq
M_{\vec{\gamma}\;\vec{\gamma}%
^{\prime }}\equiv
\frac{1}{(\sqrt[2]{5}\kappa)^{\frac{5}{2}}}\kappa^{(\vec{\gamma}-%
\vec{\gamma}^{\prime })^{2}}(-1)^{\vec{\gamma}.\vec{\gamma}^{\prime }}
\eeq
the  $32\times 32$ real matrix relating
${\cal C}_{\vec\gamma}$ to ${\cal A}_{{\vec\gamma}^\prime}$.
It verifies $M^{-1}=M$ and $M=M^{\top }$ and thus it can be used to
compute the coefficients in the direct channel using Cardy's prescription
\cite{cardy} (${\cal C}_{\vec{\gamma}} \equiv
\sum\limits_{\vec{\alpha},\vec{\beta}}N_{\vec{%
\alpha}\;\vec{\beta}}^{_{\vec{\gamma}}}n^{\vec{\alpha}}n^{\vec{\beta}}
 $) with
\begin{equation}
N_{\vec{\alpha}\;\vec{\beta}}^{_{\vec{\gamma}}} =\sum\limits_{\vec{\delta}}%
\frac{M_{\vec{\alpha}\;\vec{\delta}}M_{\vec{\beta}\;\vec{\delta}}M_{\vec{%
\gamma}\;\vec{\delta}}}{M_{\vec{0}\;\vec{\delta}}}
\end{equation}
and then
\beq
\frac{1}{\left( \sqrt[4]{5}\right) ^{5}}{\cal A}_{\vec{\gamma}}
=\frac{\left(
\sum\limits_{\vec{\delta}}M_{\vec{\gamma}\vec{\delta}}n^{\vec{\delta}%
}\right) ^{2}}{M_{\vec{0}\vec{\gamma}}} .
\eeq
We find
\begin{eqnarray}
N_{\vec{\alpha}\;\vec{\beta}}^{_{\vec{\gamma}}}
&=&\prod\limits_{i=1}^{5}(1-\delta _{\alpha _{i}+\beta _{i}+\gamma _{i},1})
\nonumber \\
{\cal A}_{\vec{\gamma}} &=&\frac{1}{\kappa^{5/2}}\frac{\left(
\sum\limits_{\vec{\delta}%
}\kappa^{(\vec{\gamma}-\vec{\delta})^{2}}(-1)^{\vec{\gamma}.
\vec{\delta}}n^{\vec{%
\delta}}\right) ^{2}}{\kappa^{(\vec{\gamma})^{2}}}
\end{eqnarray}
and ${\cal A}_{\vec 1 - \vec\gamma}$ can be obtained from
${\cal A}_{\vec\gamma}$
inverting the sign of $\sqrt 5$ in all powers of $\kappa$.

The tadpole cancellation conditions become
\begin{eqnarray}
{\cal A}_{\vec{0}}
&=&2^{4}\kappa^{\frac{15}{2}}=\kappa^{-\frac{5}{2}}\left(
\sum\limits_{%
\vec{\delta}}\kappa^{(\vec{\delta})^{2}}n_{\vec{\delta}}\right) ^{2} \\
{\cal A}_{\vec{1}}
&=&2^{4}\kappa^{-\frac{15}{2}}=\kappa^{\frac{5}{2}}\left(
\sum\limits_{%
\vec{\delta}}(-\kappa)^{-(\vec{\delta})^{2}}n_{\vec{\delta}}\right) ^{2} ,
\end{eqnarray}
and they lead to the following equations for the Chan Paton coefficients
\begin{eqnarray}
n_{0}+n_{2}+n_{3}+2n_{4}+3n_{5} &=&12  \nonumber \\
n_{1}+n_{2}+2n_{3}+3n_{4}+5n_{5} &=&20  \label{Tadpoles35}
\end{eqnarray}
where $n_{i}=\sum\limits_{\vec{\gamma}\;/\;\left| \vec{\gamma}\right|
=i}n_{%
\vec{\gamma}}$ ($e.g.$,
$%
n_{4}=n_{(1,1,1,1,0)}
+n_{(1,1,1,0,1)}+n_{(1,1,0,1,1)}+n_{(1,0,1,1,1)}+n_{(0,1,1,1,1)}
$).

The coefficients (of the real characters $\hat \chi$)
$\tilde{\cal M}_{\vec{\gamma}}$ which complete a perfect square in the
transverse channel from the M\"obius
strip are given by
\begin{equation}
\tilde{\cal M}_{\vec{\gamma}}=\pm
\sqrt[4]{5}\sqrt{{\cal A}_{\vec{\gamma}}\kappa^{\frac{15}2 -3(%
\vec{\gamma})^{2}}}
\end{equation}
and one of the solutions to the tadpole cancellation conditions is
\begin{equation}
\tilde{\cal 
M}_{\vec{\gamma}}=-\sum\limits_{\vec{\delta}}\sqrt[4]{5}(-1)^{(\vec{%
\gamma})^{2}}\kappa^{(\vec{\gamma}-\vec{\delta})^{2}}(-1)^{\vec{\gamma}.\vec{%
\delta}}\kappa^{\frac 52 - 2(\vec{\gamma})^{2}}n_{\vec{\delta}} .
\end{equation}
Transforming to the direct channel with  $\hat P^{(k=3)}$  we find
\begin{equation}
M_{\vec{\beta}}^{\vec{\delta}}
=-\frac{(-1)^{N_{1}+N_{2}+1}}{5}\prod\limits_{i=1}^{5}(1-\delta _{2\delta
_{i}+\gamma _{i}(\beta _{i}),1})
\end{equation}
The factor $\frac{1}{5}$ cancels when summing over all twisted
sectors and therefore the partition function from the M\"obius strip
in the direct channel is
\begin{equation}
{\cal Z}_{M}=-\frac 12{\sum_{\vec{\alpha}}}^{\prime }\left(
\prod\limits_{i=1}^{5}(1-\delta _{2\delta _{i}+\gamma _{i}(\alpha
_{i}),1})\right)
(-1)^{N_{1}+N_{2}+1}\hat{\chi}^{susy}_{\vec{\alpha}}\;n_{\vec{%
\delta}}
\label{mob}
\end{equation}
Comparing with the contribution from the cylinder
\begin{equation}
{\cal Z}_{C}=\frac 12{\sum_{\vec{\beta}}}^{\prime }\left(
\prod\limits_{i=1}^{5}(1-\delta _{\gamma _{i}(\alpha _{i})+\beta _{i}+\delta
_{i},1})\right) \chi^{susy}
_{\vec{\alpha}}\;n_{\vec{\beta}}n_{\vec{\delta}}
\label{cili}
\end{equation}
it is easy to see that the term $n_{\vec{\delta}}$ appears in
(\ref{mob}) only  if the term $(n_{\vec{\delta}})^{2}$
appears in (\ref{cili}) and thus the
Chan-Paton contribution from  $n_{\vec{\delta}}$ is either vanishing or
$
\frac{1}{2}n_{\vec{\delta}}(n_{\vec{\delta}}\pm 1)$. The sign $\pm 1$
depends on the phase
$(-1)^{N_{1}+N_{2}+1}e^{-\pi i(\Delta _{\vec{\beta}}-\frac{Q_{%
\vec{\beta}}}{2})}e^{\pi i(\Delta _{{}}^{GSO}-\frac{1}{2})}$. It is
$-1$ for the massless vector character $\chi _{(0,0)^{5}}$ and therefore
the Chan Paton group is $SO(n_i)$ for all $n_i$. It is also $-1$ for the
characters
 $\chi
_{(0,0)^{3}(3,-3)(2,-2)}$, $%
\chi _{(2,0)^{5}}$, $\chi _{(0,0)(1,-1)^{3}(2,-2)}$, $\chi
_{(0,0)^{2}(3,3)(1,1)^{2}}$ and it is  $+1$ for $\chi
_{(0,0)^{3}(3,3)(2,2)}$,
$\chi_{(0,0)(1,1)^{3}(2,2)}$, $\chi _{(0,0)^{2}(3,-3)(1,-1)^2}$.
 In conclusion, the characters with phase $-1$ ($+1$) have lowest
level states transforming in the antisymmetric (symmetric)
representation of $SO(n_i)$.

Let us work out an example. Consider a particular family of solutions
to the
tadpole cancellation conditions, namely
\begin{equation}
n_{0} =12-n \quad ; \quad n_{1}=20-n  \quad ; \quad n_{2} =n
\label{35example}
\end{equation}
where
$n_{1}=n_{(1,0,0,0,0)};n_{2}=n_{(1,1,0,0,0)}$. The partition function from
the cylinder reads
\begin{eqnarray}
Z_{C} &=&\frac{1}{2}[(n_{A}^{2}+n_{B}^{2}+n_{C}^{2})\chi ^{A}+  \nonumber \\
&&(n_{B}^{2}+n_{C}^{2}+2n_{B}n_{A})\chi ^{B}+  \nonumber \\
&&(n_{C}^{2}+2n_{C}n_{A}+2n_{C}n_{B})\chi ^{C}+  \nonumber \\
&&(n_{C}^{2}+2n_{B}n_{C})\chi ^{D}]
\end{eqnarray}
where $
\chi ^{A} =\frac{1}{5}(\chi _{I})^{5} ; ~
\chi ^{B} =\frac{1}{5}(\chi _{II})\chi _{I}^{4} ; ~
\chi ^{C} =\frac{1}{5}(\chi _{II})^{2}\chi _{I}^{3} ; ~
\chi ^{D} =\frac{1}{5}\chi _{I}\chi _{II}\chi _{I}^{3}$,
and thus
$n_{0}=n_{(0,0,0,0,0)}\equiv n_{A};n_{1}=n_{(1,0,0,0,0)}\equiv
n_{B}$, $n_{2}=n_{(1,1,0,0,0)}\equiv n_{C}$.

$\chi^A, \chi^B, \chi^C$ and $\chi^D$ are the combinations of
characters contributing to the M\"obius strip. As may be seen from
Table B.12 in Appendix B, these contain the following massless
characters and their charge conjugate combinations: $(0,0)^{5}$,
$(1,-1)^{2}{\underline{(0,0)^{2}(3,-3)}}$ and
$(2,-2){\underline {(0,0)^{3}(3,-3)}}$ plus all possible permutations
with $(2,-2)$ in the second position.

\section{Modding by discrete symmetries}
Each of the  blocks defining the internal sector of the theory possesses
$Z_m$ phase symmetry \cite{gepner,gq}.
 Namely, conformal fields transform under such an action as
\beq
\Phi_{l,q,{\bar l},{\bar q}}
\to e^{2i \pi \gamma {\frac{(q+{\bar q})}{2m}}}\Phi_{l,q,{\bar l},{\bar q}}
\label{phsym}
\eeq with $\gamma \in Z$.
 Also, since in many cases the internal sector
contains  several identical conformal blocks, models should be
 invariant under the permutation of such blocks. Thus, generically
the models posses a $G=\otimes_{a=1}^rZ_{m_a} \times {\cal P}$
group of symmetries (${\cal P}$ denoting block permutations)
and
therefore it is worth considering  the possibility of dividing them out.
Such discrete symmetries  have been extensively studied in the context of
the $E_8 \times E_8$
heterotic string on Gepner and coset models (see for instance \cite{cyclic1,cyclic2,aaan,fimqr}).
Generically they lead to a reduction in the number of generations.

Here we show how such moddings can be implemented in the Type IIB orientifold
on Gepner points.
The general idea is to obtain expressions for {\it modded}
supersymmetric
characters with well defined modular transformation properties. Once this is
 achieved the closed sector is obtained by just plugging left and right
modded  characters and the Klein-bottle amplitude can be immediately
written  down in order to proceed to the construction of the D-brane sector.

\subsection{Modding out phase symmetries}

Consider, as a simple example,
the case of   just one block closed partition function. In order to mod out
the phase symmetry in (\ref{phsym})
the constraint $\gamma q= 0 \,  mod \,
m$ should be
implemented in the left moving sector (and similarly in the right moving one).
This can be achieved by introducing the projector
$ \frac1M \sum_{x=0}^{M-1} e^{2i \pi \gamma {\frac{q}{m}} x} $
in the character,
 where $M$ is the order of the cyclic  group $G$, $i.e.$ the least integer
such that
$M \gamma= 0 \, {\rm mod} \, m$. As usual,
such a truncation will generically produce a
non modular invariant partition function and G-twisted sectors must be added.
Twisted sectors can be included in order to ensure that the {\it modded}
character $ \chi^{G}_{l,q}(\tau)$ transforms as the original one.
Namely, by defining
\begin{equation}\label{cG}
  \chi^{G}_{l,q}(\tau)=\frac1M \sum_{x,y=0}^{M-1} e^{2i \pi \gamma^2y^2
{\frac{c \tau }{6}} }
e^{-2i \pi {\frac{\gamma^2 x y}{m}}} \chi_{l,q}(\tau,\gamma y \tau +\gamma x )
\end{equation}
 we can see, for instance,  by using the transformation
properties for the characters discussed
in Appendix A that
$ \chi^{G}_{l,q}(-1/\tau) =\sum_{l',q'}  S_{l,q;l',q'}
\chi^{G}_{l',q'}(\tau)$.

Notice the similarity with the ``supersymmetry projection'' ($n\equiv
y, \gamma \equiv 1/2 , p\equiv x$).
Therefore, when attached to the
antiholomorphic
part  $\chi^{G}_{{\bar l},{\bar q}}$ a modular invariant partition
function is recovered.

$\chi^{G}_{l,q}$ can be rewritten (see  (\ref{112}) and (\ref{115}) in
Appendix A) as
\beq
 \chi^{G}_{l,q }(\tau )= \frac1M \sum_{x,y}
e^{-2i \pi {x \frac{ \gamma}{m}(q_+ \gamma y}) }
\chi_{l,q +2 \gamma y}(\tau)
\label{modchiex}
\eeq

These considerations are  easily generalizable to  products of conformal
 theories in the internal sector of the string which are generically
invariant under a cyclic phase symmetry of the form $\otimes_a  Z_{M_a}$.

For the sake of simplicity let us mode out  by just one $Z_{M_a}$ symmetry.
The projection parameters will now be encoded into an $r$ dimensional
vector
\beq
{{\vec \Gamma  }}^{ a} = ( { \gamma_1}^{a }  ,{\gamma_2}^{a }, \dots,{\gamma_r}^{ a})
\eeq
where $M_a$ is the least integer such that
\beq
M_a { \gamma_i}^{a }= 0 \quad mod \, (k_i+2)
\eeq
and $a $ labels one of the different, inequivalent, moddings.

The product of  characters in the internal sector now reads
\beq
 \chi^{G}_{{\vec l},{\vec q }}(\tau )= \sum_{x,y}
e^{-2i \pi {x \frac{\gamma_i}{m}(q_i+ \gamma_i y}) }
\chi_{{\vec l},{\vec q }+2{\vec \gamma}y}(\tau)
\label{modchi}
\eeq
where $\vec l$, $\vec q$ are $r$-component vectors with entries $l_i$,
$q_i$ respectively.
{}From here we may obtain the  projection conditions on each twisted sector
 $y =0,\dots,M_a$
\begin{equation}\label{twistedpro}
   \sum_{i=1}^r   \frac1m_i { \gamma_i^{a }
(q_i+y \gamma_i^{a } )}
\in {\mathbb Z}
\end{equation}

Since  $\chi^{G}_{{\vec l},{\vec q }}$ transforms
as the original, non projected,
character, it is straightforward to write down the supersymmetrized
 projected character $\chi^{G, susy}_{\vec\alpha}$:  we must just
replace
 $\chi^{G}_{l_i,q_i}$ into expression
(\ref{jisusy}) in Appendix A.

Projection on integer  total charge leads to
\begin{equation}\label{twistedch}
 \sum_{i=1}^r  \frac1m_i (q_i+2 y \gamma_i^{a } )
\in {\mathbb Z}
\end{equation}
for each twisted sector $y$. Thus, from (\ref{twistedpro})
we see that supersymmetry imposes a further constraint on
${ \gamma_i}^{a }$, namely
\begin{equation}\label{susyconst}
 \sum_{i=1}^r  \frac1m_i \gamma_i^{a }
\in {\mathbb Z}
\end{equation}
(This is the usual $2\beta_0\cdot\Gamma
\in {\mathbb Z}
$ condition of
\cite{gepner}).

The full modular invariant closed partition function reads
\beqa
\lefteqn{\sum_{\vec\alpha,{\vec{\bar\alpha}}}
{\cal N}_{{\vec \alpha};{\vec{ \bar \alpha} }}
 \chi^{G, susy}_{\vec\alpha}
{\chi}^{G, susy~*}_{{\vec {\bar \alpha}}} = } \nonumber \\
& = \sum_{\vec\alpha,{\vec{\bar\alpha}}}
 {\cal N} _{ {\vec \alpha}; { \vec { \bar \alpha}}  }
\frac{1}M  \sum_{x,y}   e^{-2i \pi  x {\frac{ \Gamma  }{m}}
( {\vec q }+ \Gamma   y ) }
\frac{1}M  \sum_{\bar x, \bar y}   e^{-2i\pi{ {\bar x}  \frac{ \Gamma
}{m}({\vec  {\bar q }}+
 { \Gamma }{\bar  y}) }}
  \chi^{ susy}_{{\vec\alpha +2y \Gamma }} {\bar
\chi}^{susy}_{{\vec {\bar \alpha} +2 {\bar  y} { \Gamma } }}
\label{modpf}
\eeqa

The left-right symmetric way in which we managed to express  the closed
partition function permits to immediately write down the associated
projected Klein bottle
amplitude:
\begin{equation}  \label{kbproamp}
{\cal Z}_K^G (it)  =
\sum_{\vec\alpha} {{\cal K}_{{\vec \alpha}}
\chi^{G,susy}_{{\vec \alpha }}(2it)}=
\frac{1}M  \sum_{\vec\alpha}{\cal K}_{\vec\alpha}\sum_{x,y}   e^{-2i \pi  x
{\frac{ \Gamma
}{m}}
( {\vec q }+ \Gamma   y ) }
\chi^{ susy}_{{\vec \alpha +2y \Gamma }}
\end{equation}
where ${\cal K}_{\vec\alpha}={\cal N}_{\vec\alpha{\vec{ \alpha}}}$,
from where we can proceed as above in order to obtain the open string sector.

Notice that moddings with no twisted sectors at all are  possible.
This is indeed the case
when  (\ref{twistedpro}) has the unique solution $y=0$. The modding is
thus freely acting
 and leads, essentially, to a reduction in the number of states. This is
the case, for instance, mentioned in the $3 ^5 $ case above for the
modding $\Gamma=(-2,-1,0,1,2)$.

However, the presence of G-twisted sectors generically leads
to a  set of characters which is different from the ones present in the
non projected theory producing,
$e.g.$, different tadpole cancellation equations.
Therefore we expect both the closed and open string sectors to
be sensibly modified by the modding. We consider  a simple example  in the $1^6$ model below.

\subsection{Phase modding $1^6$}

 Inequivalent projections are given by
\beqa
{{ \Gamma ^1 }}& =& (1,1,1,0,0,0)\\
{{\Gamma ^2}} &= & (1,1,1,-1,-1,-1)
\eeqa
Since $\Gamma ^2= 0 ~ mod ~ 3$ here, (\ref{twistedpro}) reduces to
the requirements
\beqa\label{chproj}
q_1+q_2+q_3&= &0 \,  mod \, 3\\
q_1+q_2+q_3 -q_4-q_5-q_6&= &0 \,  mod \, 3
\eeqa
for $ \Gamma ^1$ and $  \Gamma ^2 $ respectively, for all twisted
sectors $y=0,1,2$.
Let us concentrate in the first modding in the diagonal case.

The projection $q_1+q_2+q_3 = 0$ implies that  the only allowed 
 supersymmetric characters (see tables B.8 $-$ B.10 in Appendix B) are
(no
permutations here) \footnote{Notice that the permuted character
$(0,0)^3(1,-1)^3$
and corresponding twists are
also allowed by (\ref{chproj}). However, when all $y$ twists are
considered they lead
to equivalent characters and must not be counted twice.}
\beqa
& & (0,0)^6 \\ \nonumber
& &(1,-1)^3 (0,0)^3 \\ \nonumber
& &(1,1)^3 (0,0)^3 \\ \nonumber
& & (0,0)^3 { \underline {(0,0)(1,-1)(1,1)}}  \nonumber
\label{modchar13}
\eeqa

When the spacetime part is included they lead to a vector, two massless
matter,
(instead of the original
20 massless states) and 6 massive characters respectively.

 More explicitly, let us denote by
\beqa
NS_1 &\equiv &|s; {  {(0,0,0)^3(1,1,0)^3}},Q_{int}=1>\\ \nonumber
NS_2 &\equiv &|s; {{(1,-1,0)^3(0,0,0)^3}},Q_{int}=-1>\\ \nonumber
(2,1)R_1 &\equiv &|(2,1); {{(0,1,1)^3(0,-1,-1)^3}},Q_{int}=0> \nonumber
\eeqa
the  massless scalars and fermions contained in the $(1,-1)^3 (0,0)^3$
supersymmetric character.
Similarly for $(1,1)^3 (0,0)^3 $ we have
\beqa
NS_3 &\equiv &|s; { (1,1,0)^3 {(0,0,0)^3}},Q_{int}=1>\\ \nonumber
NS_4 &\equiv &|s; {(0,0,0)^3(1,-1,0)^3},Q_{int}=-1>\\ \nonumber
(2,1)R_2 &\equiv &|(2,1); {{(0,-1,-1)^3(0,1,1)^3}},Q_{int}=0> \nonumber
\eeqa
and for the vector $(0,0)^3 (0,0)^3 $
\beqa
V  &\equiv &|(2,2); {  {(0,0,0)^3(0,0,0)^3}},Q_{int}=1>\\ \nonumber
(1,2)R_3 &\equiv &|(1,2); {{(0,1,1)^3(0,1,1)^3}},Q_{int}=1>\\ \nonumber
(1,2)R_4 &\equiv &|(1,2); {{(0,-1,-1)^3(0,-1,-1)^3}},Q_{int}=-1> \nonumber
\eeqa

\noindent{\bf Closed sector}

The massless states in the closed sector are obtained by coupling,
diagonally, the above states for
left and right sectors  and by keeping invariant combinations under $\Omega $
left-right exchange.

The first two characters lead to $ 8 ({\bf 2},{\bf 1}) $ (from the 16
fermions
$ R_i NS_j $ ), 13 scalars $ 13 ({\bf 1},{\bf 1}) $
($4,6,2,1$ from$ NS_i NS_i $, $ NS_i NS_j $,$ R_i R_i $ and
$ R_i R_j $, $i\neq j$, respectively) and one tensor multiplet (from
$R_1R_2$).
This is the content of $3 H + T$.
Interestingly enough the number of tensors plus hypers adds up to 4 instead of 20.
This is an indication that the original $K3$ of the unmodded $1 ^6$ theory
{\it became a torus} after dividing by the discrete symmetry.

 The third character produces an  $N=1$ supergravity
multiplet when V$-$V  coupling is considered.  However
couplings of the form $V-(2,1)R_{1(2)} $ are now allowed leading, in particular,
 to two states of the type $ 2(3,2)$, signaling the presence of extra
gravitini as expected in a torus compactification.
Actually, it can be checked that all massless states finally arrange into
one $N=2$ supergravity multiplet
\beq
 (3, 3) + (1, 3)  + (3,1)+ 2(2, 3)+2(3,2)+(1,1)
+ 4(2,2)+2(1,2)+ 2(2,1)
\eeq
plus four $N=2$ vector multiplets
\beq
 (2, 2) + 4(1, 1)  + 2(2,1)+ 2(1, 2)
\eeq
\noindent
{\bf Open sector}

The modded Klein bottle amplitude contains the supersymmetric characters
obtained from
(\ref{modchar13}) and can be written as
\beq
{\cal Z}^G_K(it) =
 \frac12\{\chi_{(0,0)}  ^3
\left ( \chi_{(0,0)} +
\chi_{(1,-1)} + \chi_{(1,1)}\right )^3 \}^{susy}(2it)
\eeq
which reads, in transverse channel,
\beqa
\tilde{\cal Z}^G_K(il) =
9  ~\frac12& \{ &
\chi_{(0,0)}^6 +
 \chi_{(1,-1)}^3
\chi_{(0,0)}^3 +
  \chi_{(1,1)}^3
\chi_{(0,0)}^3 + \nonumber\\
&+& \underline{ \{\chi_{(0,0)}
\chi_{(1,-1)}
\chi_{(1,1)} \}
\chi_{(0,0)} ^3}\}^{susy}(il)
\eeqa
The following partition functions in the direct and transverse channels of
the cylinder and M\"obius strip provide a solution with D-brane
 gauge group $SO(n_2) \times U(n_1)$
\beqa
{\cal Z}_C(it) &=&\{(n_1^2 + \frac12n_2^2 )\sum_j \chi_{(0,0)}^3 X_j +
(\frac12n_1^2 + n_1n_2) \sum_j\underline  {
\chi_{(0,0)}\chi_{(1,-1)}\chi_{(1,1)}}X_j
\}^{susy}(it)
\nonumber\\
\tilde{\cal Z}_C(il) &=&\frac12 \{ (2n_1+n_2)^2 [
\chi_{(0,0)}^6 +
 \chi_{(1,-1)}^3
\chi_{(0,0)}^3 + \chi_{(1,1)}^3
\chi_{(0,0)}^3] \nonumber\\
&& + (n_1-n_2)^2
\underline {\chi_{(0,0)}
\chi_{(1,-1)}
\chi_{(1,1)}}
\chi_{(0,0)}^3 \}^{susy}(il) \nonumber\\
{\cal Z}_M(it) &=&\sum_j(-1)^{N_{(1,1)}}\{ -\frac12n_2
 \hat\chi_{(0,0)}^3\hat X_j
- \frac12n_1
\underline{\{\hat\chi_{(0,0)}
\hat \chi_{(1,-1)}
\chi_{(1,1)}\}} \hat X_j \}^{susy}(it) \nonumber\\
\tilde{\cal Z}_M(is) &=&\frac12 \{-(2n_1+n_2)
 [\chi_{(0,0)}^6 +
 \chi_{(1,-1)}^3
\chi_{(0,0)}^3 +\chi_{(1,1)}^3
\chi_{(0,0)}]^3 \nonumber\\
&&- (n_1-n_2) e^{-\frac {i\pi}3}
\underline {\{\chi_{(0,0)}
\chi_{(1,-1)}
\chi_{(1,1)}\}
[\chi_{(0,0)}]^3\}}^{susy}(il) \nonumber
\eeqa
where $\sum_i X_i$ denotes the sum over all possible products of three
characters and underlining denotes the sum over permutations.

The tadpole cancellation condition leads in this case to $2n_1+n_2=8$. There are two
massless
hypermultiplets transforming  in adjoint representations of $SO(n_2)$ and  $U(n_1)$ respectively.

It is interesting to  notice that  the modding of the
diagonal ${\bf (1_A) ^6}$ theory we are considering leads to the same Klein
bottle
amplitude as the ${\bf (1_C)^3(1_A)^3}$ coupling.
Nevertheless, closed sectors are different.

\subsection{Cyclic Permutations }
Cyclic permutation symmetries in Gepner and coset heterotic models
were studied in
\cite{cyclic1,cyclic2,aaan}. Permutation boundary states were considered
recently in \cite{recknagelpb}.

Let us start, for simplicity, with the piece of the  internal sector
 built up from  $M$  ($M$ a prime number) identical conformal blocks.

Following \cite{aaan}  we  introduce a formal projection operator $P$
over identical states of the $N=2$ superconformal algebra
(not necessarily primary states), such that
$P$ acting on a tensor product of states produces a vanishing
result unless all states have equal charges and weights.
After dividing by this permutation symmetry  the following
``character" can be defined
\beqa
\chi_{invar}(\tau) &=& (P+{\frac{1-P}M} )\chi^M (\tau)=
\frac1M \chi^ M+{\frac{M-1}M} P\chi^M = \\
 &=& \frac1M  \chi ^M (\tau)+\frac{M-1}M
\chi(M\tau)
\eeqa
where we indicate with a superscript $M$ that the character contains the
product of $M$ identical blocks. Also,
$P\chi= {\rm Tr} Pe^{2i\pi \tau(L_0-c/{24})}$
formally indicates that the traces must be computed by
simultaneously considering the same state in all blocks, such that
$P\chi^M(\tau)=\chi(M\tau)$ is the character of just one block but
evaluated at $M\tau $.
Each of these states is counted once. The term $(1-P)\over {M}$
corresponds to the case when at least one state in a block is
different from the others.
Since  this state could belong to any of the $M$ blocks we must divide by $M$
in order to obtain just one full symmetric state.
We see that the invariant, untwisted part, will produce the same
 result as the
original partition function but where the states related by a permutation of
the $M$ blocks are counted just once.

 Consider the closed partition function.
Due to the presence of a {\it fixed point} contribution $\chi(M\tau)$ this
partition function is no longer invariant under
modular transformations and  $P$-twisted sectors
must be added.
By starting with the original modular invariant
partition function $Z^M(\tau)$
with  $M$ identical blocks we finally obtain
\beq
Z_{new}(\tau,\bar \tau)={Z^{M}(\tau,\bar\tau)\over {M}}+{{M-1}\over M}
Z(M\tau,M \bar \tau)+
{{M-1}\over {M}} \sum_{n=0}^{M-1} Z({{\tau+n}\over M},{{\bar\tau+n}\over M}
) \label{znewex}
\eeq

Modular invariance can be checked by noticing that
twisted sectors
\beq
{{M-1}\over {M}} \sum_{n=0}^{M-1} Z({{\tau+n}\over M},{{\bar \tau+n}\over M})
\label{ptwisted}
\eeq
and ${{M-1}\over M} Z(M\tau,M \bar \tau)$ are related among themselves to
the same
expressions evaluated in $-1/{\tau }$ by combinations of $SL(2,Z)$ actions on
the argument.
\footnote{More explicitly, by choosing
 \beq \g =
\bmat{cc}a& b\\ c& d \emat \quad \quad  =
 \bmat{cc} \frac{1+lm}M &m \\ l & M\emat
\label{sl2} \eeq
with $ l, m $ chosen such that $ \frac{1+l m}M \in {\mathbb Z} $,
 it is easy to see that
$ \g (\tau ', z')=
(\frac{a\tau' +b}{c\tau'+d},\frac{z'}{c\tau'+d})$ and then, for
$ \tau '=\frac {\tau M}{1-l\tau}$, we have
$ \g (\tau ')= \frac{\tau+m} M $, as required.}

Therefore, if fields of the original  theory are known, (\ref{znewex})
allows to compute the spectrum in the modded closed string theory.
Recall that, apart from the term $ {Z^{M}(\tau,\bar\tau)\over {M}}$ containing the full
  original partition function, in  the other terms M blocks  have been
replaced by just one.
In particular it can be  shown \cite{cyclic2}  that twisted sectors
can be interpreted as the partition function of a new $N=2$
superconformal field theory with central charge $\hat c=Mc$
($c$ is the central charge of each one of the identical theories)
and Virasoro generators given in terms of those of the original theory by
\beqa
{\hat L_m}& =&{L_{mM}\over{M}}+{c (M^{2}-1)\over{24 M}}\delta_{0,m}\\
{\hat G}_{r}^{\pm} &=& {1\over{\sqrt{M}}}G_{rM}^{\pm}\\
{\hat J_m}&=& J_{mM} .
\label{ncft}
\eeqa
Similar expressions are valid
for the right movers.

Thus,  weights and charges of the (twisted) primary states of the new
theory are obtained from the original ones as
\beq
h_{new}={{h+m}\over{M}}+{c (M^{2}-1)\over{24 M} }\label{hnew}
\eeq
\beq
Q_{new}=Q
\label{qnew}
\eeq
where $m$ is the level of the descendant field.

The sum over $n$ in (\ref{ptwisted}) imposes the constraint $h+m-\bar
h-\bar
m=0 \quad mod \quad M$.

In order to construct the partition function for the full D
 dimensional
string, the spacetime
sector and the other $r-M$ Gepner  blocks must be included, and the
 characters  must be supersymmetrized in the usual way
 \cite{cyclic1,cyclic2,aaan}.
Namely, the full, non modded partition function reads
\begin{equation}
{\cal Z}_T (\tau, {\bar \tau})=   \sum_{ {\vec\alpha}, {\vec
{\bar{\alpha}}}}
\chi^{susy}_{\vec \alpha}(\tau )
{\cal N}^{ {\vec\alpha}{\vec{\bar\alpha}}}
\chi^{susy~*}_{\vec{\bar \alpha}}({\bar
\tau })
\label{2bpfsusy}
\end{equation}
where ${\vec \alpha}$ is a  $d+r-M+M$ dimensional vector index
and   the full
character is
schematically   given by
\beq
\chi^{susy}_{\vec\alpha}(\tau )= \frac1m \sum_{n,p}
\left [\chi_{0}(\tau,z_{n,p})\right ]^d
\prod_{i=d+1}^{d+r-M}\chi_{{\alpha }_{i}}(\tau,z_{n,p})
 \prod_{j=d+r-M+1}^{d+r}\chi^M_{{\alpha}_j} (\tau,z_{n,p})
\label{orchar}
\eeq
 with $ z_{n,p}=\frac{ n }2 \tau + \frac{p}2$ (and similarly for the right
sector).
Again, the superscript $M$ in the
last character indicates a product of $M$ characters corresponding to
primary fields of $M$
identical blocks
(encoded in the vector $ {{\vec \alpha}_M}$ denoting the
last $M$ entries of $\vec\alpha$). 

The full modular invariant projected partition function is  thus
{ \samepage
\beqa
{\cal Z}_{new}^{susy}(\tau)& = & {\frac{1}M} \sum_{n,p}
{\cal Z}^{st}(\tau,z_{n,p})
{\cal Z}^{r-M}(\tau,z_{n,p})Z^M(\tau,z_{n,p})\\ \nonumber
& &  + {\frac{M-1}{M}}\sum_{n,p}{\cal Z}^{st}(\tau , z_{n,p}
){\cal Z}^{r-M}(\tau ,z_{n,p} ){\cal Z}(M\tau,M z_{n,p})\\
 \nonumber
& & +{\frac{M-1}{M}}\sum_{n,p}\sum_{m=0}^{M-1} {\cal
Z}^{st}(\tau,z_{n,p})
 {\cal Z}^{r-M}(\tau,z_{n,p}){\cal Z}(\frac{\tau+m}M,z_{n,p})
\label{znew}
\eeqa
}

The first term is nothing but the original partition function divided
by $M$.
The second one is the fixed point contribution where $M$ identical blocks
are replaced by just one evaluated at $M\tau$.  The sum of both
terms accounts for the permutation invariant contributions as discussed
above.
The last term is the twisted sector contribution which, as discussed
 in (\ref{ncft}), contains a {\it  new} conformal field theory where
 conformal weights and charges are given in  (\ref{qnew}).

Notice that fixed point and twisted contributions
are built up from $r-M+1$ internal blocks.
 The index vector ${\vec\alpha}$ must now be replaced by a``collapsed"
index $\vec\alpha^\prime$ and
the modular invariant coupling for such terms is
\begin{equation}
{\cal N}^{
{\vec\alpha^{\prime}}{\vec{\bar\alpha^\prime}}}=\prod_{i=1}^d{\cal
N}^{\vec\alpha_i\vec{\bar\alpha}_i}
\prod_{j=d+1}^{d+r-M}{\cal N }^{
{\alpha}_{j}{\bar\alpha}_{j}} {\cal N }^{ {\alpha}_M { \bar\alpha}_{M}}
\end{equation}

We present a sample computation below.

 {}From (\ref{znew}) we can now immediately obtain  the Klein bottle
partition function.
We must just keep the left-right invariant
piece and evaluate it at $2 Im\tau$ (see (\ref{kbdir})).
In particular  the $m$ dependence in the twisted sector drops out
and the factor $M$ in the denominator cancels out.
Therefore, the Klein bottle amplitude we are led to is
\beqa
{\cal Z}_{K}= \frac12 \sum_{\vec\alpha}{\cal K}_{\vec\alpha}
\chi_{\vec\alpha}^{susy}(2it)=\frac1{2M} \sum_{\vec\alpha}{\cal
K}_{\vec\alpha}
 \chi_{\vec\alpha}^{susy} &+&
  {\frac{M-1}{2M} }\sum_{\vec\alpha^\prime}{\cal K}_{\vec\alpha^\prime}
\chi^{fix}_{\vec\alpha^\prime}+
\nonumber\\
&+&
\frac{(M-1)}2 \sum_{\vec\alpha^\prime}{\cal K}_{\vec\alpha^\prime}
\chi^{twisted}_{\vec\alpha^\prime}
\label{kbperm}
\eeqa
where $\chi^{susy}_{\vec\alpha}$ is the original supersymmetric
  character
introduced in (\ref{orchar}),

\beq
\chi^{fix}_{\vec\alpha^\prime}=\frac1m
\sum_{n,p} \left [\chi_{0}(\tau,z_{n,p})\right ]^d
\prod_{i=d+1}^{d+r-M}\chi_{{\alpha }_{i}}(\tau,z_{n,p})
 \chi_{{ \alpha}_M} (2itM,Mz_{n,p})
\label{fixchar}
\eeq
and

\beq
\chi^{twisted}_{\vec\alpha^\prime}=\frac1m \sum_{n,p}
\left [\chi_{0}(2it,z_{n,p})\right ]^d \prod_{i=d+1}^{d+r-M}\chi_{{\alpha
}_{i}}(2it,z_{n,p})
 \chi_{{ \alpha}_M} ( \frac{2it}M,z_{n,p})
\label{twistedchar}
\eeq
($z_{n,p}$ is defined as above with $\tau \rightarrow 2it$).
 Interestingly enough, a similar result
was obtained in \cite{recknagelpb} in terms of boundary states.

Once the amplitude from the Klein bottle is obtained we can follow the
usual procedure to build up the open string sector.
It is worth noticing that when the characters are expressed in terms of
$l=1/2t$, by using the modular transformations given in (\ref{kbtrans})
in
order to
obtain the transverse channel amplitude, fixed and twisted characters
do exchange, namely
\begin{equation}
\chi^{fix}_{\vec\alpha^\prime}(2it)= S_{ \vec\alpha^\prime
{\vec{\beta}^\prime} }\chi^{twisted}_{\vec\beta^\prime}(il)
\end{equation}

Therefore, the Klein bottle amplitude in the transverse channel reads
\beqa
{\tilde {\cal Z}}_{K} =\frac1{2M} \sum_{\vec\alpha} S_{ \vec\alpha {\vec
\beta}
}{\cal K}^{\vec\alpha } \chi^{susy}_{\vec\beta}&+&
  \frac{(M-1)}2 \sum_{\vec\alpha^\prime}
S_{ \vec\alpha^\prime \vec\beta^\prime }{\cal K}^{\vec\beta^\prime}
\chi^{fix}_{\vec\alpha^\prime} \nonumber \\
& +& {\frac{M-1}{2M} }\sum_{\vec\alpha^\prime} S_{ \vec\alpha^\prime
\vec\beta^\prime }{\cal K}^{\vec\beta^\prime}
\chi^{twisted}_{\vec\alpha^\prime}
\label{kbpermtrans}
\eeqa
Recall that the factors in front are different from (\ref{kbperm}).

The first term will lead to the same tadpole structure as the original,
non
permuted, theory.
We do not expect new tadpole contributions
to be generated from the fixed point term. In fact,
if such contribution exists it is already contained in the first term.
The last piece, instead, will contain new states, involving  charges and
conformal weights given
in (\ref{hnew}) and (\ref{qnew}), and could produce new tadpole
cancellation conditions.
 Let us consider some examples.

\medskip

\noindent
{\bf$ \bullet ~~M=3$ permutations in diagonal $1^6$ model.}

Consider permutations of the first three blocks in  the diagonal $1^6$
theory.

In the original theory 20 massless hypermultiplets encoded in
the characters \break $ {\underline {(0,0,0)^3(1,1,0)^3} }$ (and positive charge
states
$ {\underline {(0,0,0)^3(1,-1,0)^3} }$)  are present.
(Recall that  underlining denotes
 all possible permutations of the underlined blocks).
When we divide out by permutations of the first three blocks, the
untwisted sector
contribution requires permuted states to be identified (counted
just once), thus we are left with
\beqa
&& (0,0,0)^3(1,1,0)^3\\
&& (1,1,0)^3 (0,0,0)^3\\
&& \{(1,1,0)^2  (0,0,0)\}{\underline {(0,0,0)^2  (1,1,0)}}  \\
&& \{(1,1,0)  (0,0,0)^2 \}{\underline {(0,0,0)  (1,1,0)^2}}
\eeqa
(and similarly those with opposite charge) for  $1+1+3+3=8$
hypermultiplets when coupled to the same states in the right moving sector.

We must also include twisted sector contributions.

The conformal weight of a state is the sum of the spacetime and
internal conformal weights. If we are interested in massless matter
 states we must look for all states with $\Delta _{int}=1/2 $
where the first three theories are now replaced by
 the new conformal
theory in (\ref{qnew}). We thus
see, from the expression for $\Delta _{new}$ in (\ref{qnew}), that
masslessness requires $m=0$.
For such $m=0$ we obtain that
\beqa
(l,q,s) & \to & ( \Delta, \Delta _{new}, Q _{new}=Q)\\ \nonumber
(1,1,0) &\to & ( 1/6, 1/6, -1/3)\\\nonumber
(1,-1,-2) &\to & ( 2/3, 1/3, -2/3)\\\nonumber
(1,-1,0) &\to & ( 1/6, 1/6, 1/3)\\\nonumber
(1,1,2) &\to & ( 2/3, 1/3, 2/3)\nonumber
\eeqa
which lead to the massless odd charge combinations
\beqa
\{(1,1,0)\}_{new} {\underline {(1,1,0)(1,1,0)(0,0,0)}} \\\nonumber
\{(1,-1,-2)\}_{new} {\underline {(1,1,0)(0,0,0)(0,0,0)}} \\\nonumber
\{(1,-1,0)\}_{new} {\underline {(1,-1,0)(1,-1,0)(0,0,0)}}\\ \nonumber
\{(1,1,2)\}_{new} {\underline {(1,-1,0)(0,0,0)(0,0,0)}} \nonumber
\label{ctwistedsates}
\eeqa
6 massless states with total charge $-1$ (and other six with charge 1).
Since there is an extra factor
$M-1=2$, we have a total of 24 massless twisted states
that lead to 12 hypermultiplets when coupled to identical states in
the
right sector.
Untwisted and twisted states sum up to a total of 20 corresponding
$K3$ moduli, as expected.

It is worth stressing that care must be taken when considering field
identifications in the {\it new} theory. Fields that were equivalent in
the
original theory ($ Z(\tau  )$) are not in the twisted new $Z(\tau /M  )$
theory and this could lead to miscountings if not treated properly.
Namely, equivalences (\ref{id1}), (\ref{id2}) read now
\beqa
\{(l,q,s)\}_{new}\equiv \{(k-l,q+M(k+2),s+2M)\}_{new}\equiv
\{(l,q+2M(k+2),s\}_{new}\nonumber\\
\equiv  \{(l,q+,s+4M)\}_{new}
\label{identif}
\eeqa
For $M=3$ these equivalences do not lead to extra states. However, the
situation is different for instance when  $M=5$ permutations are considered.
In fact, in such a case it is easy to see that there are 4 
untwisted hypers and that the combinations
\beqa
\{(0,0,2)\}_{new} (0,0,0) \nonumber\\
\{(1,1,2)\}_{new} (1,-1,0)
\label{ctwistedsatesm5}
\eeqa
are massless twisted contributions. When coupled  diagonally
to the right movers we would obtain 8 twisted hypermultiplets (recall the
$M-1=4$ factor in front) instead of 18 as expected.
However, when the new equivalences above are taken into account we find
that,
when coupling left and right sectors in a  diagonal invariant manner, we
actually have
\beqa
\{(1,3,4)\}_{new} (0,0,0) &--& \{(1,3,4)\}_{new} (0,0,0) \\
\{(1,1,2)\}_{new} (1,1,0)& --& \{(1,1,2)\}_{new} (1,1,0) \\
\{(1,3,4)\}_{new} (0,0,0) &-- & \{(1,13,10)\}_{new} [(1,1,0)\equiv (0,2,2)] \\
\{(1,1,2)\}_{new} (1,1,0) &--& \{(1,21,4)\}_{new}[ (0,0,0)\equiv (1,3,2) ]
\eeqa
leading to 16 hypermultiplets as required.

\noindent
{\bf Open sector}

Let us sketch the construction of the open sector.

The amplitude from the Klein bottle in the transverse sector
is generically given in (\ref{kbpermtrans}).

 In our $1^6$ example the first
factor is just $1/M$ times  the original one, as given in (\ref{klein16}).
The ``twisted" contribution (originated in the direct channel {\it fixed
} point) is proportional  to
\beq
{\frac{M-1}{2M} }\sum_{n,p}\frac{(-1)^{n+p}}{2m} \left (
\chi_{(0,0);p,-n}(il)
\right )^3\left (
\chi_{(0,0);p,-n}( il/M) \right )
\eeq

Both terms will contribute to tadpoles.
A simple solution, as we did in Section 6, is to propose
 a similar partition function for the amplitudes from the cylinder and
M\"obius strip. Namely,
${\tilde {\cal Z}}_{C}= n^2={\tilde {\cal Z}}_{K}$ and
${\tilde {\cal Z}}_{M}= -2n {\tilde {\cal Z}}_{K}(il+1/2)$.
Thus, factorization is ensured and tadpole cancellation requires $n=8$ as
before.
Therefore, from the direct channel amplitudes we find again an
$SO(8)$
gauge group. Now the untwisted sector contributes with $4$ (2 in $M=5$)
hypermultiplets in $\bf {28}$ while the twisted sector generates
the extra
$6$ (8) (recall identifications in (\ref{identif})), in order to complete
a total of 10 hypermultiplets.

Thus, as advanced, we recover the same massless spectrum where
part of it comes from the invariant piece of the partition function while
the rest originates in the twisted  sector contributions.
In fact, this is expected by anomaly cancellation in $D=6$.
Even if  $\bf {28}$ is gauge anomaly free, ten such hypers
are needed in order to ensure absence of gravitational anomalies.

A similar computation for the $1^9$ model in $D=4$, which originally has
84 (left) states in the $\bf 6$ of
$SO(4)$, leads, for instance, to 12 and 6 untwisted and twisted states,
respectively, in the $M=7$ permutations case, leading to an effective
reduction of the number of states.
Recall that (this is a general result) due to the term 
${c (M^{2}-1)\over{24 M} }$ in (\ref{hnew}), there will not be direct
contributions from the new twisted sector to vector characters. In
principle it could contribute, indirectly, through tadpole cancellation.

As a last example let us consider the $3^5$ quintic diagonal $D=4$ model
where
permutation of all 5 theories is considered. The twisted sector
corresponds now to just one theory with charges and weights given in
(\ref{hnew}), (\ref{qnew}). {}From (\ref{hnew}) we notice that
$h_{new}=\frac35 + \frac15(h +
m)$
where $ h$ are the conformal weights
given in Tables B.8$-$B.10. We thus see that there are no massless twisted
states allowed. Moreover, we see that  odd total charge condition can
not be fulfilled and therefore there is no twisted sector at all.
Thus, 5 cyclic permutation twists act freely in this model.
The original 101 (charge 1) massless states reduce in this case to $21$.

\section{Non supersymmetric models}

It is interesting  to extend previous results in order to
incorporate anti-D branes \cite{sugi,au,aaads,abg}. While supersymmetry
will be preserved in the closed sector it will be generically
broken in the open sector. Antibranes differ from branes in that
they carry opposite $RR$ charges. Namely, when antibranes are
incorporated we should have, besides the D-brane $RR$ charges $
D_{\vec\alpha}=D_{\vec\alpha\vec\alpha^\prime} n_{\vec\alpha^\prime} $
(see (\ref{factori})), antibrane charges terms
of the form $ {\bar D}_{\vec\alpha}=-D_{\vec\alpha\vec\alpha^\prime}
w_{\vec\alpha^\prime}$ where
 $ w_{\vec\alpha^\prime}$ is the number of antibranes of "type"
$\vec\alpha^\prime$ on
which open
strings can end. Notice that
geometrical (conformal theory) terms $D_{\vec\alpha}$ are the same for
both branes and antibranes since they only differ in the sign of the $RR$
charge.

A first consequence of such inclusion is that tadpole cancellation equations
 (\ref{tadpolecg}) are modified and  they now read
\begin{equation}
O_{\vec\alpha} +D_{\vec\alpha}- {\bar D}_{\vec\alpha}=0 .
\label{tadpolecganti}
\end{equation}
for characters containing $RR$ massless fields.

In terms of the partition function,
antibranes are thus treated as branes (we must just replace
$n_{\vec\alpha}\rightarrow w_{\vec\alpha}$ everywhere)  but taking into
account
 that they have opposite signs in $RR$ transverse channel,
namely, in characters containing
$\vartheta
{\frac{\nu }{2} \atopwithdelims[] 0}$ in the spacetime sector.
Therefore, it is easier to first look at the transverse channel and
then study changes in the spectrum by transforming to the open string
direct channel.

Consider, for instance, the transverse cylinder amplitude
originated in direct channel amplitudes containing bosonic states.
 It must be of the form
\begin{equation}
(D_{\vec\alpha}+(-1) ^p {\bar D}_{\vec\alpha}) ^2 \chi_{\vec\alpha} 
(-\frac 1{\tau}, \frac p{2\tau}) =
[D_{\vec\alpha\vec\beta}(n_{\vec\beta}+(-1) ^p
w_{\vec\beta})] ^2 \chi _{\vec\alpha} (-\frac 1\tau, \frac{p}{2\tau})
\end{equation}
where 
$p$ odd (even)
corresponds to $RR$ ($NS NS$) closed bosonic
states. When transformed to the direct channel, it reads
\begin{equation}
[C^{\vec\alpha}_{\vec\alpha^\prime\vec\alpha^{\prime\prime}}
(n^{\vec\alpha^\prime}+(-1)
^p w^{\vec\alpha^\prime})(n^{\vec\alpha^{\prime\prime}}+(-1) ^p
w^{\vec\alpha^{\prime\prime}})] \chi _{\vec\alpha}
(\tau, \frac{p}{2} )
\end{equation}
Here $\chi_{\vec\alpha}$ is the product of characters denoted
$\chi^{\prime}_{\vec\alpha}$ in equation (\ref{gso}).

Also, since under the
$P$ transformation $\vartheta {{\alpha } \atopwithdelims[] {\beta}}
\rightarrow \vartheta {{\alpha } \atopwithdelims[] {\beta}} $,
the direct channel  amplitude from the M\"obius strip is
\begin{equation}
[M_{\vec\alpha\vec\alpha^\prime}(n^{\vec\alpha^\prime}+
w^{\vec\alpha^\prime})] \chi _{\vec\alpha}
(\tau, \frac{p}{2} ) .
\end{equation}

Thus, we observe that strings stretching between anti
branes produce the same spectrum as in the brane-brane sector.
However, when a  string stretches between a brane and an antibrane
a factor  $(-1) ^p  \chi _{\vec\alpha} (\tau, \frac{p}{2} )$
does appear. Interestingly enough, when the
sum over $p$ is performed, $\it even $ charge states instead of
$\it odd$ ones,
as required by supersymmetry, are now selected.
Therefore, as expected, supersymmetry is broken, in the open sector,
by the presence of antibranes. Moreover, even charge will now allow,
in particular,  (real) scalar tachyons 
charged under both branes and antibranes gauge groups
($C^{\vec\alpha}_{\vec\alpha^\prime\vec\alpha^{\prime\prime}}
n^{\vec\alpha^\prime}w^{\vec\alpha^{\prime\prime}} $).

Similar reasoning leads us to fermionic amplitudes from the cylinder and
 M\"obius strip
\beqa
&
&C^{\vec\alpha}_{\vec\alpha^\prime\vec\alpha^{\prime\prime}}
(n^{\vec\alpha^\prime}
+w^{\vec\alpha^\prime})(n^{\vec\alpha^{\prime\prime}}+
w^{\vec\alpha^{\prime\prime}})\chi _{\vec\alpha} (\tau, \frac\tau
2+\frac p2)
\\
& &M^{\vec\alpha}_{\vec\alpha^\prime}(n^{\vec\alpha^\prime}-
w^{\vec\alpha^\prime}) \chi _{\vec\alpha} (\tau, \frac \tau 2+ 
\frac{p}{2})
\eeqa
In particular we observe that, due to the minus sign in the
$O-{\bar D}$ M\"obius strip sector antisymmetric
$\Yasymm$ representations under the brane group become
symmetric $\Ysymm $   and vice versa.

\noindent
${\bf (1_A)^4 (1_C)^2 }$ \textit{example }

 {}From the  discussion above and by using results from section 6
for  $D-D$ and $O-D$ sectors we obtain 
\beqa
U(n_1)\times SO(n_2) \times [\, U(w_1)\times SO(w_2)\,] \quad
\eeqa
$DD$ and $\bar D \bar D$ gauge groups and the massless spectrum is given by

\beqa
\begin{array}{ccc}
{\bf DD+ {DO}+OD} & {\rm Fermions_+} & 4(1,\Yasymm ) + 4(\rm Adj,1)+  6(\Yasymm,1 )+6 (n_1,n_2) \\
         & {\rm Fermions_-} & (1,\Yasymm ) + ({\rm Adj},1) \\
& {\rm Compl. scalars} &  4(1,\Yasymm ) + 4(\rm Adj,1)+  6(\Yasymm,1 )+6 (n_1,n_2)  \\

   {\bf {{\bar D}{\bar D} +{\bar D}O}+ O {\bar D }}& {\rm Fermions_+} & 4(1,\Ysymm ) + 4(\Ysymm,1)+  6(\Ysymm,1)+6 (w_1,w_2) \\
         & {\rm Fermions_-} & (\Yasymm,1) + (1,{\rm Adj}) \\
& {\rm Compl. scalars} & 4(1,\Yasymm) + 4(\rm Adj,1)+  6(\Yasymm,1)+6 (w_1,w_2) \\

\bf{ D {\bar D}+{\bar D} D} &  {\rm Fermions_+} & (n_1,1;w_1,1) + (1,n_2;1,w_2) \\
         & {\rm Fermions_-} & 6(1,n_2;w_1,1) + 6(n_1,1;1,w_2) \\

& {\rm tachyons} &  (n_1,1;w_1,1)+(1,n_2;1,w_2) \\
\end{array}
\eeqa

It is easy to check that  all, gauge and gravitational  anomalies
(see closed sector above) cancel
if the following constraint
\begin{equation}
w_2-n_2+2(w_1-n_1)+8=0
\end{equation}
is satisfied. This is the tadpole cancellation condition
(\ref{tadpolecganti}), as expected.

We see that, due to lack of supersymmetry, scalar tachyons
generically appear, signaling instability of the vacuum (recall
that the closed sector is still supersymmetric). However, tachyon free
non supersymmetric vacua are still possible by choosing, for
instance $w_1=n_2=0$ . In this case  $U(n_1)\times SO(2 (n_1-4)) $
where the first (second) factor comes from brane$-$brane
(antibrane$-$antibrane) sector. The rank of the gauge group can be
arbitrarily high. However this is just a signal (see  \cite{au}
for similar examples in the orbifold context) that such vacua
should be interpreted as excitations of a stable supersymmetric
vacuum obtained when branes and antibranes annihilate to leave
$n_1=4$ D-branes. In principle, models where not all antibranes
could annihilate could exist in this context.

In fact, the quintic model provides us with such possibilities. Just
to illustrate this issue let us consider, as in example
(\ref{35example}) the non-vanishing coefficients $n_0, n_1, n_2$.
Tadpole cancellation conditions (\ref{Tadpoles35}) can be
rewritten as \beq n_0+n_1+2n_2 = 32 \qquad , \qquad n_1-n_0 = 8
\eeq where the first equation tells us that a total of 32 D-branes
is needed and the second one plays the role of twisted tadpole
cancellation equation \cite{au}. Notice that, while a supersymmetric model
with $n_0-n_1 < 8$ is certainly not allowed, a consistent
non-supersymmetric model can be built up. Namely, let us choose
$n_0 = 0, n_1 = 4$ and introduce a set of $w_0$ antibranes of type
$0$. Tadpole equations (\ref{tadpolecganti}) now read \beq
-w_0+4+2n_2=32 \qquad ; \qquad w_0 = 4 \eeq which allow for a
consistent, non-supersymmetric model with $SO(16)\times SO(4)$
brane-brane group and $SO(4)$ antibrane-antibrane gauge group .

Moreover, since branes and antibranes are of different kind, no tachyons
will be present.

 \section{Summary and outlook}
 In this work we have addressed the
construction of Type II B orientifolds, in $D=8,6$ and $ D=4$
dimensions, where the internal sector is built up from Gepner
models by extending some preliminary work on the subject.
An important step in this construction is the identification,
following original Gepner ideas, of the $N=1$
{\it supersymmetric } character $\chi^{susy}_{\vec\alpha}(\tau)$
in (\ref{susych}) for each moving
sector. In particular, once such characters are obtained, it
proves rather easy to, formally, implement moddings by phase or
permutation symmetries as presented in sections 7 and 8.

Of course, a serious limitation for computing explicit cases is that
a big number of characters must be taken into account. This number
generically increases with the number of internal dimensions and
with the level $k$ of the internal theory. In fact,  even if RR
tadpole cancellation requires to look only at massless states in
the transverse channel, factorization must be checked for all
massless and   massive states. In some models this requires to
consider  thousands of characters, a hard task even for a fast
computer.
Thus, special attention must be dedicated to
computational techniques reducing the number of characters to
handle.
For instance,  as we explicitly showed in $D=8$ examples,
computations are strongly simplified by the use of
\textit{reduced} modular transformation matrices taking into
account only the \textit{odd} charged states, instead of the direct
product of block transformation matrices. The use of a full
conformal algebra character (given in terms of $(l,q)$ instead of
$(l,q,s)$ ), also seems to present some advantages in concrete
computations of factorization and tadpole cancellation, by reducing
the number of states to deal with.

Moreover, besides its potential
phenomenological interest, modding by discrete symmetries allows
to reduce these numbers. For instance, the hundred  massless
(left)  matter characters in the
$3^5$ model could be reduced to 4 by
simultaneously
modding by phase and permutation symmetries. In addition to eight
dimensional models which were discussed rather exhaustively to be
compared with other constructions, explicit models in $D=6,4$
were mainly presented as examples. They illustrate how the number
of generations can be modified, how gauge symmetries are enhanced,
how phase moddings  could induce a topology change, etc.

We hope that this may  help to offer guidelines to handle specific models
with a phenomenological  or theoretical interest. For instance, it
seems interesting to reconsider cases like the  $3^5$ quintic. In
our example (see (\ref{cil35})) we have chosen to achieve
factorization by selecting the  same characters in the transverse
cylinder amplitudes than  those appearing in the Klein bottle
contribution.
This allowed us to easily show how high rank solutions can be obtained.
However, with such choice we also restricted ourselves to
symplectic and/or orthogonal groups since corresponding M\"obius strip
amplitudes are needed in order to complete squares. A possible
extension is to look for a generalization where, besides the
already considered terms, other characters, not present in the Klein
bottle partition function are
included in the cylinder amplitude which would thus lead to unitary
groups with presumably chiral matter content \cite{prog}

Open string version of 3-generations like heterotic Gepner model 
\cite{gep}, which involves both
modding by phases and cyclic 
permutation symmetries, could be interesting to study along
the lines of our work.

 Also, it would be nice to find realizations of non supersymmetric models
with antibranes  where complete brane-antibrane
annihilation, leading to stable supersymmetric vacuum is not allowed, for
instance, by tadpole cancellation conditions, as it is the case in
some orbifold compactifications \cite{aiq}.

Extentions of Gepner models including non-diagonal modular invariant
couplings \cite{huis2} or  more general coset models, 
like Kazama Suzuki constructions \cite{ks} 
should be possible to address along the above lines.
 We hope that our work, based on a partition function approach, could
help in clarifying connections with more geometrical interpretations.

\centerline{\bf Acknowledgments} We are grateful to L.E. Iba\~nez,
A. Uranga and I. Allekotte for stimulating  discussions and
suggestions. G.A. work is partially supported by ANPCyT grant
03-03403 and Fundaci\'on Antorchas and C.N. work by CONICET
grant PIP 0873 and UBACyT X805. E.A. work is supported by Fundaci\'on
Antorchas.

\section{Appendix A}

{\bf Characters of N=2 superconformal minimal models and N=2 strings}

In this Appendix we collect several properties of the characters which are
useful to work out various assertions contained in the main body of the
article.

Recall the definition of $\chi_{(l,q)}$ in (\ref{carss})
\beq
\chi _{l,q}(\tau ,z)\equiv {\rm Tr}_{{\cal H}_{(l,q)}}(e^{2\pi i(L_{0}-%
\frac{c}{24})\tau }e^{2\pi iJ_{0}z})=\chi _{(l,q,s)}(\tau ,z)+\chi
_{(l,q,s+2)}(\tau ,z) .
\eeq
Explicit expressions for these characters of the N=2 superconformal
minimal
models have been computed in references \cite{dobrev,matsuo,kiritsis} and
extensively used in \cite{gepner,jnnns}. In
terms of
Riemann $\vartheta $-functions they read \cite{nos}
\begin{equation}
\chi _{l,q}(\tau ,z )=\frac{\vartheta
{0 \atopwithdelims[] 0}%
(\tau ,z)\vartheta
{-\frac{l+1}{m}+\frac{1}{2} \atopwithdelims[] \frac{1}{2}}
(m\tau ,0)\eta ^{3}(m\tau )e^{\pi i(\frac{l+1}{m}-\frac{1}{2})}}{\eta
^{3}(\tau )\vartheta
{\frac{l+q+1-m}{2m} \atopwithdelims[] 0}
(m\tau ,z)\vartheta
{\frac{-l+q-1+m}{2m} \atopwithdelims[] 0}
(m\tau ,z)}
\end{equation}
where $m=k+2$, $\eta $ is the Dedekind function and the definitions and
properties of the $\vartheta $-functions can be found in reference
\cite{polchi}.

Consider the combinations $\chi _{l,q}^{\pm }(\tau ,z)\equiv
\chi _{l,q,s}(\tau ,z)\pm \chi
_{l,q,s+2}(\tau ,z)$ with $s=0,-1$ for NS, R respectively.
The following relations may be easily seen from the definition
\begin{eqnarray}
\chi _{l,q}^{\pm }(\tau ,z+\frac{1}{2}) &=&e^{\pi iQ_{l,q}}\chi _{l,q}^{\mp
}(\tau ,z)  \nonumber \\
\chi _{l,q}^{\pm}(\tau ,z+A) &=&e^{2\pi iQ_{l,q}A}\chi^{\pm} _{l,q}(\tau
,z)\qquad A\in Z
\label{112}
\end{eqnarray}
Using that $mQ_{l,q}$ is an integer number,
these properties allow to
implement the GSO projection on the
product of characters of N=2 strings
 as
\beq
\sum\limits_{p=0}^{2m-1}\frac{(-1)^{p}}{2m}[\chi _{\nu }(\tau ,z+\frac{p}{2%
})]^{d}\prod\limits_{i=1}^{r}\chi _{l_{i},q_{i}}(\tau ,z+\frac{p}{2}) =
\frac{\delta _{Q_{\vec{\alpha }},Z}}{2} (1-
e^{\pi iQ_{\vec{\alpha }}})[\chi _{\nu }(\tau
,z)]^{d}\prod\limits_{i=1}^{r}\chi _{l_{i},q_{i}}(\tau ,z)\} .
\eeq
It is obvious from this expression that the combinations of
characters surviving the GSO
projection are those
verifying that
$\sum\limits_{i=1}^{r}Q_{l_{i},q_{i},s_{i}}+\sum\limits_{j=1}^{d}Q_{j
}$ is an odd integer number.

It is also convenient to express the twisted characters
$\chi _{l,q+n}(\tau ,z)$ in terms of
\break $\chi _{l,q}(\tau ,z)$. This can be done by shifting $z$ as
follows
\begin{equation}
\chi _{l,q+n}(\tau ,z)=e^{2\pi i(n^{2}\tau\frac{c}{24}+n\frac{c}{6}z)}\chi
_{l,q}(\tau ,z+\frac{n}{2}\tau ) .
\label{115}
\end{equation}
As mentioned in the text, twisting by
$n=2(k+2)$ one recovers the original character at $n=0$, except for
$l=\frac{k}{2}$ when $k$ is even. In this case, a round trip requires
$n=(k+2)$.

Finally, supersymmetry and GSO projection can both be implemented as
\beqa
&&\chi _{\vec{\alpha}}^{susy}(\tau ,z)= \\
&&\sum\limits_{n,p\;{\rm mod\;2m}}\;\frac{%
(-1)^{n+p}}{2m}e^{2\pi i(n^{2}\tau\frac{c}{24}+n\frac{c}{6}z)}\left [ \chi
_0(\tau
,z+\frac n2 \tau +\frac{p}{2})\right ]^{d}\prod\limits_{i=1}^{r}\chi
_{l_{i},q_{i}}(\tau ,z+\frac{n%
}{2}\tau +\frac{p}{2}) \label{jisusy} \nonumber
\eeqa
with $c=12$.

These supersymmetric characters can be split as
\begin{equation}
\chi _{\vec{\alpha}}^{susy}(\tau ,z)=\chi _{\vec{\alpha}}^{NS}(\tau ,z)-\chi
_{\vec{\alpha}}^{R}(\tau ,z). \label{chisusy}
\end{equation}
where
\beq
\chi _{\vec{\alpha}}^{NS} =\sum_{even~n=0}^{2m-2}\chi _{\vec{\alpha}%
^{(n)}}\qquad ,  \qquad
\chi _{\vec{\alpha}}^{R} =\sum_{odd~n=1}^{2m-1}\chi _{\vec{\alpha}%
^{(n)}}
\eeq
These two blocks contain states with identical charges and conformal
weights, therefore (\ref{chisusy}) implies $\chi
_{\vec{\alpha}}^{susy}(\tau
,z)\equiv 0$.

An alternative decomposition of the supersymmetric characters
is the following
\begin{equation}
\chi _{\vec{\alpha}}^{susy}(\tau ,z)=\frac 12  ( {(}\chi
_{\vec{\alpha}}^{+}(\tau ,z)-\chi
_{\vec{\alpha}}^{-}(\tau ,z) {)}
\end{equation}
where
\begin{eqnarray}
\chi _{\vec{\alpha}}^{+}(\tau ,z) &\equiv &
\chi _{\vec{\alpha}%
}^{NS^{+}}(\tau ,z)-\chi _{\vec{\alpha}}^{R^{+}}(\tau ,z)  \nonumber \\
\chi _{\vec{\alpha}}^{NS^{+}}(\tau ,z)
&=&\sum\limits_{n=0\;(even)}^{2m-2}\delta _{Q_{\vec{\alpha}},Z}\;[\chi
_{\nu +n}^{NS^{+}}(\tau ,z)]^{d}\prod_{i=1}^{r}\chi _{\alpha
_{i}+n}^{NS^{+}}(\tau ,z).  \nonumber \\
\chi _{\vec{\alpha}}^{R^{+}}(\tau ,z)
&=&\sum\limits_{n=1\;(odd)}^{2m-1}\delta _{Q_{\vec{\alpha}},Z}\;[\chi
_{\nu +n}^{NS^{+}}(\tau ,z)]^{d}\prod_{i=1}^{r}\chi _{\alpha
_{i}+n}^{NS^{+}}(\tau ,z).
\end{eqnarray}
and
\begin{eqnarray}
\chi _{\vec{\alpha}}^{-}(\tau ,z) &\equiv &\chi _{\vec{\alpha}%
}^{NS^{-}}(\tau ,z)-\chi _{\vec{\alpha}}^{R^{-}}(\tau ,z)  \nonumber \\
\chi _{\vec{\alpha}}^{NS^{-}}(\tau ,z)
&=&\sum\limits_{n=0\;(even)}^{2m-2}\delta _{Q_{\vec{\alpha}},Z}\;
e^{i\pi Q_{\vec\alpha ^{(n)}}}
[\chi
_{\nu +n}^{NS^{-}}(\tau ,z)]^{d}\prod_{i=1}^{r}\chi _{\alpha
_{i}+n}^{NS^{-}}(\tau ,z)  \nonumber \\
\chi _{\vec{\alpha}}^{R^{-}}(\tau ,z)
&=&\sum\limits_{n=1\;(odd)}^{2m-1}\delta _{Q_{\vec{\alpha}},Z}\;
[\chi
_{\nu +n}^{NS^{-}}(\tau ,z)]^{d}\prod_{i=1}^{r}\chi _{\alpha
_{i}+n}^{NS^{-}}(\tau ,z).
\end{eqnarray}
\noindent
{\bf Modular transformations of supersymmetric characters}

In order to study the modular transformations of
$\chi_{\vec\alpha}^{susy}$ it is
convenient to introduce the following notation
\begin{eqnarray}
\chi _{l,q;n,p}(\tau ,z) &\equiv &e^{2\pi i(n^{2}\frac{c}{24}+n\frac{c}{6}(z+%
\frac{p}{2}))}\chi _{l,q}(\tau ,z+\frac{n}{2}\tau +\frac{p}{2}) \\
\chi _{\nu;n,p}(\tau ,z) &=&e^{2\pi
i(n^{2}\frac{c_{st}}{24}+n\frac{c_{s-t}}{6}%
(z+\frac{p}{2}))}\chi _{\nu}(\tau ,z+\frac{n}{2}\tau +\frac{p}{2})
\end{eqnarray}
{}From (\ref{tm}) it follows that
\begin{eqnarray}
\chi _{l,q;n,p}(-\frac{1}{\tau },\frac{z}{\tau }) &=&e^{2\pi i\frac{cnp}{12}%
}e^{2\pi i\frac{z^{2}c}{6\tau }}\sum\limits_{{\ }l^{\prime },q^{\prime
}}S_{l,q;l^{\prime },q^{\prime }}\chi _{l^{\prime },q^{\prime };p,-n}(\tau
,z) \\
\chi _{l,q;n,p}(-\frac{1}{\tau },\frac{z}{\tau }) &=&e^{2\pi i\frac{cnp}{12}%
}e^{2\pi i\frac{z^{2}c}{6\tau }}\sum\limits_{{\ }l^{\prime },q^{\prime
}}S_{l,q;l^{\prime },q^{\prime }}^{-1}\chi _{l^{\prime },q^{\prime
};-p,n}(\tau ,-z) .
\end{eqnarray}

The {\sf S} modular transformation of $\chi^{susy}_{\vec\alpha}$ may be
obtained
 multiplying the $S_{l,q;l^{\prime },q^{\prime }}$
matrix elements of each individual theory. For the spacetime characters
$S$ is the unit matrix with an extra
factor $(-i\tau )^{d}$.

Notice that $S$ exchanges $n$ and $p$ and thus {\sf S}:
$\chi ^{NS^{+}}\longleftrightarrow $
$\chi ^{NS^{+}},\chi ^{NS^{-}}\longleftrightarrow $ $\chi ^{R^{+}}$ and
$\chi^{R^{-}}\longleftrightarrow $ $\chi ^{R^{-}}$. Moreover, modular
invariance implies that it is not possible to
achieve supersymmetry without GSO projection and vice versa.

The {\sf S} modular transformation on the product of characters is as
follows

\begin{eqnarray}
&&\chi _{\vec{\alpha}}^{NS/R}(-\frac{1}{\tau },\frac{z}{\tau })=(-i\tau
)^{-d}e^{2\pi i\frac{z^{2}c}{6\tau }}\sum_{\vec{\alpha}^{\prime }}
\left [\prod_{i=1}^{r}S_{(i)}\right]
_{\vec{\alpha}\vec{\alpha}^{\prime }}
\chi _{\vec{\alpha}^{\prime }}^{+/-}(\tau ,z)
\\
&&\chi _{\vec{\alpha}}^{+/-}(-\frac{1}{\tau },\frac{z}{\tau })=(-i\tau
)^{-d}e^{2\pi i\frac{z^{2}c}{6\tau }}\sum_{\vec{\alpha}^{\prime }}
\left [\prod_{i=1}^{r}S_{(i)}\right]
_{\vec{\alpha}\vec{\alpha}^{\prime }}
\chi _{\vec{\alpha}^{\prime }}^{NS/R}(\tau ,z).
\end{eqnarray}
 The sum over
$\vec{\alpha}^{\prime }$ runs over all vectors with components
$\alpha_{i}=(l_{i},q_{i})$ in the standard range. Recalling the identity
(\ref{identifi}) and noticing that
$\left [\prod_{i=1}^{r}S_{(i)}\right]
_{\vec{\alpha}\vec{\alpha}^{\prime (n) }}
= \left [\prod_{i=1}^{r}S_{(i)}\right]
_{\vec{\alpha}\vec{\alpha}^{\prime }} $,
for all $\vec{\alpha }$ such that $Q_{\vec{\alpha }}$ is an integer
number, a matrix ${ S}$ may be defined to act on the independent characters
$\chi^{susy}_{\vec\alpha}$ as follows
\beq
{S}_{\vec{\alpha}\vec{\delta}}= \frac{m}{2^{\epsilon _{\vec{\delta}}}}
 \left [\prod_{i=1}^{r}S_{(i)}\right]
_{\vec{\alpha}\vec{\delta} }
\eeq
where
$\epsilon _{\vec{\delta}} = 1 (0)$
if $k_{i}$ is even for all $i$ and
 $\vec{\delta}$ is short (otherwise) and $\epsilon _{\vec{\delta}} = 0$
if $ k_{i}$ is odd for all $i$.
The rank of ${S}$ is given by the number of independent
supersymmetric characters.

This ${S}$ matrix is not symmetric when there is a short vector.
Applying it twice one obtains
\begin{equation}
\chi _{\vec{\alpha}}^{NS/R}(\tau
,z)={S}_{\vec{\alpha}\vec{\alpha}^{\prime
}}^{2}\chi _{\vec{\alpha}^{\prime }}^{NS/R}(\tau ,-z),
\end{equation}
which, together with (\ref{chcon})
implies
${S}^{2}=C$,
$C$ being the charge conjugation matrix.

Regarding the {\sf T} transformation one can show that
\beq
\chi_{\vec\alpha}^{NS^\pm}(\tau + 1, z) = e^{2\pi i (\Delta_{\vec\alpha} -
\frac
c{24}-\frac {Q_{\vec\alpha}}2)}
\chi_{\vec\alpha}^{NS^\mp}(\tau, z) \quad ; \quad
\chi_{\vec\alpha}^{R^\pm}(\tau+1,z)=
e^{2\pi i(\Delta_{\vec\alpha}-\frac
{Q_{\vec\alpha}}2)}\chi_{\vec\alpha}^{R^\pm}(\tau, z) \nonumber \\
\nonumber
\eeq
and
\beq
\chi _{\vec{\alpha}}^{NS/R}(\tau +1,z)=
e^{2\pi i(\Delta _{\vec{\alpha}}-\frac{Q_{\vec{\alpha}%
}}{2})}
\chi _{\vec{\alpha}%
}^{NS/R}(\tau ,z) .
\end{equation}
One may think of the phase $e^{2\pi i(\Delta
_{\vec{\alpha}}-\frac{Q_{\vec{\alpha}%
}}{2})}$ as the diagonal element of a matrix
\begin{equation}
T_{\vec{\alpha}}\equiv
e^{2\pi i(\Delta _{\vec{\alpha}}-\frac{Q_{\vec{\alpha}%
}}{2})}
\delta _{\vec{\alpha}\vec{\alpha}^{\prime }} .
\end{equation}

Note that the transformation
{\sf T}$^{(2)}:\tau \rightarrow \tau +2$ can be realized by $T^2$ as
\begin{eqnarray}
\chi _{\vec{\alpha}}^{NS^{+}/NS^{-}}(\tau +2,z) &=&e^{4\pi i\Delta _{\vec{%
\alpha}}}\chi _{\vec{\alpha}}^{NS^{+}/NS^{-}}(\tau ,z)  \nonumber \\
\chi _{\vec{\alpha}}^{R^{+}/R^{-}}(\tau +2,z) &=&e^{4\pi i\Delta _{\vec{%
\alpha}}}\chi _{\vec{\alpha}}^{R^{+}/R^{-}}(\tau ,z)
\end{eqnarray}
The diagonal elements are the phases
$e^{4\pi i(\Delta_{\vec{\alpha}}-\frac {Q_{\vec\alpha}}2)}$,
which reduce to  $e^{4\pi i\Delta
_{\vec{\alpha}}}$ when acting on non vanishing characters ($i.e.$ those
with
integer $Q_{\vec\alpha}$).

\medskip
{\bf The {\sf P} transformation}

The characters in the direct and transverse channels of the M\"obius strip
are related by the transformation {\sf P}: $it+\frac{1}{2} \rightarrow \frac{%
i}{4t}+\frac{1}{2}$. This can be generated from the modular transformations
{\sf S} and {\sf T} as
\begin{equation}
{\rm {\sf P}}={\rm {\sf TST}}^{2}{\sf S}
\end{equation}
and it squares to the identity, similarly as the {\sf S} transformation,
namely
\begin{equation}
{\sf P}^{2}={\sf S}^{2}={\sf 1} .
\end{equation}

There exists a matrix $P$ which performs
this
transformation on the characters $\chi _{\vec{\alpha}}^{R/NS}$. In terms of
the ${ S}$ and $T$ matrices, $P$ reads
\begin{equation}
P=T{S}T^{2}{S}^{-1} \quad ,
\end{equation}
and it can be shown that
\begin{equation}
P=T{S}T^{2}{S}^{-1}=T{S}^{-1}T^{2}{S} \quad .
\end{equation}
This  matrix  relates characters with different arguments as
\begin{equation}
\chi _{\vec{\alpha}}^{NS/R}(\frac{\tau -1}{2\tau -1},-\frac{z}{2\tau -1}%
)=(1-2\tau )^{-d}e^{2\pi i\frac{z^{2}c}{3(2\tau -1)}} \sum_{{\vec\alpha}%
^\prime}P_{\vec{\alpha}\vec{\alpha}^{\prime }}\chi _{\vec{\alpha}^{\prime
}}^{NS/R}(\tau ,z)  \label{MatrizP1}
\end{equation}

The following action on the characters is easy to see
\begin{equation}
\sum_{\vec\alpha^\prime , \vec\alpha^{\prime\prime}}P_{\vec{\alpha}\vec{%
\alpha}^{\prime \prime }}P_{\vec{\alpha}^{\prime \prime }\vec{\alpha}%
^{\prime }}\chi _{\vec{\alpha}^{\prime }}^{NS/R}(\tau ,z)= \chi _{\vec{\alpha%
}}^{NS/R}(\tau ,-z)
\end{equation}
so that
\begin{equation}
\sum_{\vec\alpha^\prime}(P^{2})_{\vec{\alpha}\vec{\alpha}^{\prime }}\chi _{%
\vec{\alpha}^{\prime }}^{NS/R}(\tau ,z)=\chi _{\vec{\alpha}}^{NS/R}(\tau
,-z)
\end{equation}
and from (\ref{chcon}) one may show that
\begin{equation}
(P^{2})_{\vec{\alpha}^{\prime }\vec{\alpha}} = C_{\vec\alpha
\vec\alpha^\prime} .
\end{equation}

The character $\chi_{\vec{\alpha}}(it+\frac{1}{2})$ contains an
expansion
in powers of
 $q\equiv e^{-2\pi t}$ multiplied by a phase $e^{\pi
i(\Delta _{\vec{\beta}}^{GSO}-\frac{1}{2})}$. Recalling (\ref{pesossusy}),
this phase is equal to $\pm e^{\pi i(\Delta
_{\vec{\alpha}}-\frac{Q_{\vec{\alpha}%
}}{2})}$ which squares to $T_{\vec{\alpha}}$, thus we denote it
 $T_{\vec{\alpha}}^{(\frac{1}{2})}$. Extracting this phase the
character becomes real. It is convenient
to
work in the basis of real characters $\hat{\chi}_{\vec{\alpha}}$ defined
as
\begin{equation}
\hat{\chi}_{{\vec{\alpha}}^{(n)}}^{R/NS}(it+\frac{1}{2},0)\equiv e^{-\pi
i(\Delta _{\vec{\alpha}}-\frac{Q_{\vec{\alpha}}}{2})}\chi^{R/NS}
_{{\vec{\alpha}}%
^{(n)}}(t) .
\label{chihat}
\end{equation}
The $\hat{{\sf P}}$ transformation connecting direct and transverse
real M\"{o}bius amplitudes is now performed by the matrix
\begin{equation}
\hat{P}=T^{(-1/2)}{S}T^{2}{S}^{-1}T^{(1/2)}
\end{equation}
where
\begin{equation}
T_{\vec{\alpha}\vec{\alpha}}^{(1/2)}=e^{\pi i(\Delta _{\vec{\alpha}}-\frac{%
Q_{\vec{\alpha}}}{2})}
\end{equation}

Notice that $T^{(1/2)}_{\vec\alpha^{(n)}\vec\alpha^{(n)}} = \pm
T^{(1/2)}_{\vec\alpha\vec\alpha}$ and hence the real characters
$\hat\chi_{\vec\alpha}$
change under twisting as $\hat\chi_{\vec\alpha^{(n)}}(it+\frac 12) = \pm
\hat\chi_{\vec\alpha}(it+\frac 12)$. Consequently we may choose one $%
\vec\alpha$ and $T^{(1/2)}$ will be the matrix corresponding to that choice.

Therefore  the characters in the direct and
transverse channels are related as
\beq
\hat\chi^{NS/R} _{\vec{\alpha}} (it+\frac{1}{2})
=(2it)^{d}\hat P_{\vec{\alpha}%
\vec{\alpha}^{\prime }}\hat\chi _{\vec{\alpha}^{\prime
}}^{NS/R}(\frac{i}{4t}+%
\frac{1}{2}) .
\eeq

\section{Appendix B}

In this appendix we list the GSO projected combinations of states
contained in the characters of some Gepner models.

In the spacetime columns we write the Weyl weights of the spinor and
vector representations of the little group. Massive states will gather
in representations of the full group. For example, the models in $D=8$
have $SO(6)$ as little group and $SO(7)$ as full group, and the
weights given are those of $SO(6)$ (even for massive states).

\noindent
{\bf {Spectrum of states of $1^3$}}

The GSO projected combinations of states contained in the characters
of the $1^3$ Gepner model
are given in the following tables:

\[
\begin{tabular}{||l||l||l||l||l||l||l||l||}
\hline\hline
 ${\rm highest ~weight ~state}$ & {\rm spacetime} & $\Delta
_{st}$ & $Q_{st}$ & $\Delta _{int}$ & $Q_{int}$ & $\Delta $ & $Q$ \\
\hline\hline
  $(0,0,0)^{3}$ & $\underline{(\pm 1,0,0)}$ & $\frac{1}{2}$ & $%
\pm 1$ & $0$ & $0$ & $\frac{1}{2}$ & $\pm 1$ \\ \hline\hline
 $(0,1,1)^{3}$ & $
\begin{array}{c}
\underline{(-\frac{1}{2}, \frac{1}{2}, \frac{1}{2})} \\
(-\frac{1}{2},-\frac{1}{2},-\frac{1}{2})
\end{array}
$ & $\frac{3}{8}$ & $
\begin{array}{c}
\frac{1}{2} \\
-\frac{3}{2}
\end{array}
$ & $\frac{1}{8}$ & $\frac{1}{2}$ & $\frac{1}{2}$ & $
\begin{array}{c}
1 \\
-1
\end{array}
$ \\ \hline\hline
 $(1,-1,0)^{3}$ & $(0,0,0)$ & $0$ & $0$ & $\frac{1}{2}$ & $1$ &
$\frac{1}{2}
$ & $1$ \\ \hline\hline
$(1,1,0)^{3}$ & $(0,0,0)$ & $0$ & $0$ & $\frac{1}{2}$ & $-1$ &
$\frac{1}{2}
$ & $-1$ \\ \hline\hline
$(0,-1,-1)^{3}$ & $
\begin{array}{c}
( \frac{1}{2}, \frac{1}{2}, \frac{1}{2}) \\
\underline{(-\frac{1}{2},-\frac{1}{2}, \frac{1}{2})} 
\end{array}
$ & $\frac{3}{8}$ & $
\begin{array}{c}
\frac{3}{2} \\
-\frac{1}{2}
\end{array}
$ & $\frac{1}{8}$ & $-\frac{1}{2}$ & $\frac{1}{2}$ & $
\begin{array}{c}
1 \\
-1
\end{array}
$ \\ \hline\hline
\end{tabular}
\]

{\centerline {{\bf TABLE B.1}:
GSO projected states in $\chi^{susy}_{(0,0)^3}$}}

\medskip
\noindent

{\footnotesize{
\[
\begin{tabular}{||l||l||l||l||l||l||l||l||}
\hline\hline
${\rm highest~ weight~ state}$
& {\rm spacetime} & $\Delta
_{st}$ & $Q_{st}$ & $\Delta _{int}$ & $Q_{int}$ & $\Delta $ & $Q$ \\
\hline\hline
  $(0,0,0)(1,-1,2)(1,1,0)$ & $(0,0,0)$ & $0
$ & $0$ & $\frac{5}{6}$ & $-1$ & $\frac{5}{6}$ & $-1$ \\ \hline\hline
 $(0,0,0)(1,-1,0)(1,1,2)$ & $(0,0,0)$ & $0$ & $0$ & $\frac{5}{6}$ & $1$ &
$%
\frac{5}{6}$ & $1$ \\ \hline\hline
 $(0,0,0)(1,-1,0)(1,1,0)$ & $\underline{(\pm 1,0,0)}$ &
 $\frac{1}{2}$ & $\pm 1$ &
$\frac{1}{%
3}$ & $0$ & $\frac{5}{6}$ & $\pm 1$
\\ \hline\hline
 $(0,1,1)(1,0,-1)(0,-1,-1)$ & $
\begin{array}{c}
( \frac{1}{2}, \frac{1}{2}, \frac{1}{2}) \\
\underline{(-\frac{1}{2},-\frac{1}{2}, \frac{1}{2})} 
\end{array}
$ & $\frac{3}{8}$ & $
\begin{array}{c}
\frac{3}{2} \\
-\frac{1}{2}
\end{array}
$ & $\frac{11}{24}$ & $-\frac{1}{2}$ & $\frac{5}{6}$ & $
\begin{array}{c}
1 \\
-1
\end{array}
$ \\ \hline\hline
 $(0,1,1)(1,0,1)(0,-1,-1)$ & $
\begin{array}{c}
(-\frac{1}{2},-\frac{1}{2},-\frac{1}{2}) \\
\underline{(-\frac{1}{2}, \frac{1}{2}, \frac{1}{2})}
\end{array}
$ & $\frac{3}{8}$ & $
\begin{array}{c}
-\frac{3}{2} \\
\frac{1}{2}
\end{array}
$ & $\frac{11}{24}$ & $\frac{1}{2}$ & $\frac{5}{6}$ & $
\begin{array}{c}
-1 \\
1
\end{array}$ \\ \hline\hline
  $(1,-1,2)(1,1,0)(0,0,0)$ & $(0,0,0)$ & $0
$ & $0$ & $\frac{5}{6}$ & $-1$ & $\frac{5}{6}$ & $-1$ \\ \hline\hline
 $(1,-1,0)(1,1,2)(0,0,0)$ & $(0,0,0)$ & $0$ & $0$ & $\frac{5}{6}$ & $1$ &
$%
\frac{5}{6}$ & $1$ \\ \hline\hline
 $(1,-1,0)(1,1,0)(0,0,0)$ & $\underline{(\pm 1,0,0)}$ &
 $\frac{1}{2}$ & $\pm 1$ &
$\frac{1}{%
3}$ & $0$ & $\frac{5}{6}$ & $\pm 1$ \\ \hline\hline
 $(1,0,-1)(0,-1,-1)(0,1,1)$ & $
\begin{array}{c}
( \frac{1}{2}, \frac{1}{2}, \frac{1}{2}) \\
\underline{(-\frac{1}{2},-\frac{1}{2}, \frac{1}{2})} 
\end{array}
$ & $\frac{3}{8}$ & $
\begin{array}{c}
\frac{3}{2} \\
-\frac{1}{2}
\end{array}
$ & $\frac{11}{24}$ & $-\frac{1}{2}$ & $\frac{5}{6}$ & $
\begin{array}{c}
1 \\
-1
\end{array}
$ \\ \hline\hline
 $(1,0,1)(0,-1,-1)(0,1,1)$ & $
\begin{array}{c}
(-\frac{1}{2},-\frac{1}{2},-\frac{1}{2}) \\
\underline{(-\frac{1}{2}, \frac{1}{2}, \frac{1}{2})}
\end{array}
$ & $\frac{3}{8}$ & $
\begin{array}{c}
-\frac{3}{2} \\
\frac{1}{2}
\end{array}
$ & $\frac{11}{24}$ & $\frac{1}{2}$ & $\frac{5}{6}$ & $
\begin{array}{c}
-1 \\
1
\end{array}
$ \\ \hline\hline
 $(1,1,0)(0,0,0)(1,-1,-2)$ & $(0,0,0)$ & $0$ & $0$ & $\frac{5}{6}$ & $-1$
&
$\frac{5}{6}$ & $-1$ \\ \hline\hline
 $(1,1,2)(0,0,0)(1,-1,0)$ & $(0,0,0)$ & $0$ & $0$ & $\frac{5}{6}$ & $1$
&
$\frac{5}{6}$ & $1$ \\ \hline\hline
 $(1,1,0)(0,0,0)(1,-1,0)$ & $\underline{(\pm 1,0,0)}$ &
 $\frac{1}{2}$ & $\pm 1$ &
$\frac{1}{%
3}$ & $0$ & $\frac{5}{6}$ & $\pm 1$ \\ \hline\hline
 $(0,-1,-1)(0,1,1)(1,0,-1)$ & $
\begin{array}{c}
( \frac{1}{2}, \frac{1}{2}, \frac{1}{2}) \\
\underline{(-\frac{1}{2},-\frac{1}{2}, \frac{1}{2})} 
\end{array}
$ & $\frac{3}{8}$ & $
\begin{array}{c}
\frac{3}{2} \\
-\frac{1}{2}
\end{array}
$ & $\frac{11}{24}$ & $-\frac{1}{2}$ & $\frac{5}{6}$ & $
\begin{array}{c}
1 \\
-1
\end{array}
$ \\ \hline\hline
 $(0,-1,-1)(0,1,1)(1,0,1)$ & $
\begin{array}{c}
(-\frac{1}{2},-\frac{1}{2},-\frac{1}{2}) \\
\underline{(-\frac{1}{2}, \frac{1}{2}, \frac{1}{2})}
\end{array}
$ & $\frac{3}{8}$ & $
\begin{array}{c}
-\frac{3}{2} \\
\frac{1}{2}
\end{array}
$ & $\frac{11}{24}$ & $\frac{1}{2}$ & $\frac{5}{6}$ & $
\begin{array}{c}
-1 \\
1
\end{array}
$ \\ \hline\hline
\end{tabular}
\]}}

{\centerline {{\bf TABLE B.2}: GSO projected states in
$\chi^{susy}_{(0,0)(1,-1)(1,1)}$}}

\medskip
~

\medskip
\noindent
{\bf {Spectrum of states of $2^2$}}

All the representations of the $k=2$ minimal model are obtained
by twisting the pairs $(0,0,0)$ $;$ $(0,0,2)$ and $(1,-1,0)$ ;
$(1,-1,2)$, namely

\[
.
\begin{tabular}{||l||l||l||l||}
\hline\hline
$n$ & ${\rm Representation}$ & $\Delta $ & $Q$ \\ \hline\hline
$0$ & $(0,0,0)$ & $0$ & $0$ \\ \hline\hline
$1$ & $(0,1,1)$ & $\frac{1}{16}$ & $\frac{1}{4}$ \\ \hline\hline
$2$ & $(0,2,2)\sim (2,-2,0)$ & $\frac{1}{4}$ & $\frac{1}{2}$ \\ \hline\hline
$3$ & $(2,-1,1)$ & $\frac{9}{16}$ & $\frac{3}{4}$ \\ \hline\hline
$4$ & $(2,0,2)^{\ast }$ & $1$ & $\pm 1$ \\ \hline\hline
$5$ & $(2,1,-1)$ & $\frac{9}{16}$ & $-\frac{3}{4}$ \\ \hline\hline
$6$ & $(2,2,0)$ & $\frac{1}{4}$ & $-\frac{1}{2}$ \\ \hline\hline
$7$ & $(2,3,1)$ & $\frac{1}{16}$ & $-\frac{1}{4}$ \\ \hline\hline
\end{tabular}
\begin{tabular}{||l||l||l||l||}
\hline\hline
$n$ & ${\rm Representation}$ & $\Delta $ & $Q$ \\ \hline\hline
$0$ & $(0,0,2)\;\sim (2,\pm 4,\pm 4)$ & $\frac{3}{2}$ & $\pm 1$ \\
\hline\hline
$1$ & $(0,1,3)\sim (1,-2,-3)\;$ & $\frac{17}{16}$ & $-\frac{3}{4}$ \\
\hline\hline
$2$ & $(2,-2,-2)\;$ & $\frac{3}{4}$ & $-\frac{1}{2}$ \\ \hline\hline
$3$ & $(2,-1,-1)$ & $\frac{9}{16}$ & $-\frac{1}{4}$ \\ \hline\hline
$4$ & $(2,0,0)$ & $\frac{1}{2}$ & $0$ \\ \hline\hline
$5$ & $(2,1,1)$ & $\frac{9}{16}$ & $\frac{1}{4}$ \\ \hline\hline
$6$ & $(2,2,2)$ & $\frac{3}{4}$ & $\frac{1}{2}$ \\ \hline\hline
$7$ & $(2,3,3)$ & $\frac{17}{16}$ & $\frac{3}{4}$ \\ \hline\hline
\end{tabular}
\]

\[
.
\begin{tabular}{||l||l||l||l||}
\hline\hline
$n$ &${\rm Representation}$ & $\Delta $ & $Q$ \\ \hline\hline
$0$ & $(1,-1,0)$ & $\frac{1}{8}$ & $\frac{1}{4}$ \\ \hline\hline
$1$ & $(1,0,1)$ & $\frac{5}{16}$ & $\frac{1}{2}$ \\ \hline\hline
$2$ & $(1,1,2)$ & $\frac{5}{8}$ & $\frac{3}{4}$ \\ \hline\hline
$3$ & $(1,2,3)$ & $\frac{17}{16}$ & $1$ \\ \hline\hline
$4$ & $(1,-1,-2)$ & $\frac{5}{8}$ & $-\frac{3}{4}$ \\ \hline\hline
$5$ & $(1,0,-1)$ & $\frac{5}{16}$ & $-\frac{1}{2}$ \\ \hline\hline
$6$ & $(1,1,0)$ & $\frac{1}{8}$ & $-\frac{1}{2}$ \\ \hline\hline
$7$ & $(1,2,1)$ & $\frac{1}{16}$ & $0$ \\ \hline\hline
\end{tabular}
\begin{tabular}{||l||l||l||l||}
\hline\hline
$n$ & ${\rm Representation}$ & $\Delta $ & $Q$ \\ \hline\hline
$0$ & $(1,-1,-2)$ & $\frac{5}{8}$ & $-\frac{3}{4}$ \\ \hline\hline
$1$ & $(1,0,-1)$ & $\frac{5}{16}$ & $-\frac{1}{2}$ \\ \hline\hline
$2$ & $(1,1,0)$ & $\frac{1}{8}$ & $-\frac{1}{4}$ \\ \hline\hline
$3$ & $(1,2,1)$ & $\frac{1}{16}$ & $0$ \\ \hline\hline
$4$ & $(1,-1,0)$ & $\frac{1}{8}$ & $\frac{1}{4}$ \\ \hline\hline
$5$ & $(1,0,1)$ & $\frac{5}{16}$ & $\frac{1}{2}$ \\ \hline\hline
$6$ & $(1,1,2)$ & $\frac{5}{8}$ & $\frac{3}{4}$ \\ \hline\hline
$7$ & $(1,2,3)$ & $\frac{17}{16}$ & $1$ \\ \hline\hline
\end{tabular}
\]

{\centerline {{\bf TABLE B.3}: Representations of $k=2$ minimal model}}

\medskip

{}From the definition of $\chi_{\vec\alpha} ^{susy}$
the following identities among characters hold
\begin{eqnarray}
\chi _{(0,0)^{2}}^{susy} &\equiv &\chi _{(2,-2)^{2}}^{susy}\equiv \chi
_{(2,0)^{2}}^{susy}\equiv \chi _{(2,2)^{2}}^{susy}  \nonumber \\
\chi _{(0,0)(2,0)}^{susy} &\equiv &\chi _{(2,-2)(2,2)}^{susy}\equiv \chi
_{(2,0)(0,0)}^{susy}\equiv \chi _{(2,2)(2,-2)}^{susy}  \nonumber \\
\chi _{(1,-1)(1,1)}^{susy} &\equiv &\chi _{(1,1)(1,-1)}^{susy}
\end{eqnarray}

The GSO projected combinations of states in the $2^2$ Gepner model are
given in the following
tables.

\[
\begin{tabular}{||l||l||l||l||l||l||l||l||}
\hline\hline
  ${\rm Internal~Theory}$ & {\rm spacetime} & $\Delta _{st}$ & $%
Q_{st}$ & $\Delta _{int}$ & $Q_{int}$ & $\Delta $ & $Q$ \\ \hline\hline
 $(0,0,0)^{2}$ & $\underline{(\pm 1,0,0)}$ & $%
\frac{1}{2}$ & $\pm 1$ & $0$ & $0$ & $\frac{1}{2}$ & $\pm 1$ \\ \hline\hline
 $(0,1,1)^{2}$ & $
\begin{array}{c}
\underline{(-\frac{1}{2}, \frac{1}{2}, \frac{1}{2})} \\
(-\frac{1}{2},-\frac{1}{2},-\frac{1}{2})
\end{array}
$ & $\frac{3}{8}$ & $
\begin{array}{c}
\frac{1}{2} \\
-\frac{3}{2}
\end{array}
$ & $\frac{1}{8}$ & $\frac{1}{2}$ & $\frac{1}{2}$ & $
\begin{array}{c}
1 \\
-1
\end{array}
$ \\ \hline\hline
 $(0,2,2)^{2}$ & $(0,0,0)$ & $0$ & $0$ & $\frac{1}{2}$ & $1$ &
$\frac{1}{2}$
& $1$ \\
 $(2,2,0)^{2}$ & $(0,0,0)$ & $0$ & $0$ & $\frac{1}{2}$ & $-1$ &
$\frac{1}{2}
$ & $-1$ \\ \hline\hline
 $(2,3,1)^{2}$ & $
\begin{array}{c}
( \frac{1}{2}, \frac{1}{2}, \frac{1}{2}) \\
\underline{(-\frac{1}{2},-\frac{1}{2}, \frac{1}{2})} 
\end{array}
$ & $\frac{3}{8}$ & $
\begin{array}{c}
-\frac{1}{2} \\
\frac{3}{2}
\end{array}
$ & $\frac{1}{8}$ & $-\frac{1}{2}$ & $\frac{1}{2}$ & $
\begin{array}{c}
-1 \\
1
\end{array}
$ \\ \hline\hline
\end{tabular}
\]

{\centerline {{\bf TABLE B.4}: GSO projected states in $\chi^{susy}_{(0,0)^2}$}}

\medskip
\noindent

\[
\begin{tabular}{||l||l||l||l||l||l||l||l||}
\hline\hline
${\rm Internal~Theory}$ & {\rm spacetime} & $\Delta _{st}$ & $%
Q_{st}$ & $\Delta _{int}$ & $Q_{int}$ & $\Delta $ & $Q$ \\ \hline\hline
 $(0,0,0)(2,0,0)$ & $\underline{(\pm 1,0,0)}$ &
 $\frac{1}{2}$ & $\pm 1$ & $\frac{1}{2}$ & $0$ & $1$ & $\pm 1$ \\ \hline\hline
 $(0,0,0)(2,0,2)$ & $(0,0,0)$ & $0$ & $0$ & $1$ & $\pm 1$ & $1$ & $\pm 1$
\\ \hline\hline
 $(0,1,1)(2,1,1)$ & $
\begin{array}{c}
\underline{(-\frac{1}{2}, \frac{1}{2}, \frac{1}{2})} \\
(-\frac{1}{2},-\frac{1}{2},-\frac{1}{2})
\end{array}
$ & $\frac{3}{8}$ & $
\begin{array}{c}
\frac{1}{2} \\
-\frac{3}{2}
\end{array}
$ & $\frac{5}{8}$ & $\frac{1}{2}$ & $1$ & $
\begin{array}{c}
1 \\
-1
\end{array}
$ \\ \hline\hline
 $(0,1,1)(2,1,-1)$ & $
\begin{array}{c}
( \frac{1}{2}, \frac{1}{2}, \frac{1}{2}) \\
\underline{(-\frac{1}{2},-\frac{1}{2}, \frac{1}{2})} 
\end{array}
$ & $\frac{3}{8}$ & $
\begin{array}{c}
-\frac{1}{2} \\
\frac{3}{2}
\end{array}
$ & $\frac{5}{8}$ & $-\frac{1}{2}$ & $1$ & $
\begin{array}{c}
-1 \\
1
\end{array}
$ \\ \hline\hline
 $(0,2,2)(2,2,0)$ & $\underline{(\pm 1,0,0)}$ &
 $\frac{1}{2}$ & $\pm 1$ & $%
\frac{1}{2}$ & $0$ & $1$ & $\pm 1$ \\ \hline\hline
 $(0,2,2)(2,2,2)$ & $(0,0,0)$ & $0$ & $0$ & $1$ & $1$ & $1$ & $1$ \\
\hline\hline
 $(2,-2,-2)(2,2,0)$ & $(0,0,0)$ & $0$ & $0$ & $1$ & $-1$ & $1$ & $-1$ \\
\hline\hline
 $(2,-1,-1)(2,3,1)$ & $
\begin{array}{c}
( \frac{1}{2}, \frac{1}{2}, \frac{1}{2}) \\
\underline{(-\frac{1}{2},-\frac{1}{2}, \frac{1}{2})} 
\end{array}
$ & $\frac{3}{8}$ & $
\begin{array}{c}
-\frac{1}{2} \\
\frac{3}{2}
\end{array}
$ & $\frac{5}{8}$ & $-\frac{1}{2}$ & $1$ & $
\begin{array}{c}
-1 \\
1
\end{array}
$ \\ \hline\hline
 $(2,-1,1)(2,3,1)$ & $
\begin{array}{c}
\underline{(-\frac{1}{2}, \frac{1}{2}, \frac{1}{2})} \\
(-\frac{1}{2},-\frac{1}{2},-\frac{1}{2})
\end{array}
$ & $\frac{3}{8}$ & $
\begin{array}{c}
\frac{1}{2} \\
-\frac{3}{2}
\end{array}
$ & $\frac{5}{8}$ & $\frac{1}{2}$ & $1$ & $
\begin{array}{c}
1 \\
-1
\end{array}
$ \\ \hline\hline
\end{tabular}
\]

{\centerline {{\bf TABLE B.5}: GSO projected states in
$\chi^{susy}_{(1,-1)(1,1)}$}}

\medskip
\noindent

\[
\begin{tabular}{||l||l||l||l||l||l||l||l||}
\hline\hline
 ${\rm Internal~Theory}$ & {\rm spacetime} & $\Delta _{st}$ & $%
Q_{st}$ & $\Delta _{int}$ & $Q_{int}$ & $\Delta $ & $Q$ \\ \hline\hline
$(1,-1,0)(1,1,0)$ & $\underline{(\pm 1,0,0)}$ &
$\frac{1}{2}$ & $\pm 1$ & $\frac{1}{4}$ & $0$ & $\frac{3}{4}$ & $\pm 1$ \\
\hline\hline
$(1,-1,2)(1,1,0)$ & $(0,0,0)$ & $0$ & $0$ & $\frac{3}{4}$ & $1$ &
$\frac{3%
}{4}$ & $1$ \\ \hline\hline
 $(1,-1,0)(1,1,2)$ & $(0,0,0)$ & $0$ & $0$ & $\frac{3}{4}$ & $1$ &
$\frac{3%
}{4}$ & $1$ \\ \hline\hline
 $(1,0,1)(1,2,1)$ & $
\begin{array}{c}
\underline{(-\frac{1}{2}, \frac{1}{2}, \frac{1}{2})} \\
(-\frac{1}{2},-\frac{1}{2},-\frac{1}{2})
\end{array}
$ & $\frac{3}{8}$ & $
\begin{array}{c}
\frac{1}{2} \\
-\frac{3}{2}
\end{array}
$ & $\frac{3}{8}$ & $\frac{1}{2}$ & $\frac{3}{4}$ & $
\begin{array}{c}
1 \\
-1
\end{array}
$ \\ \hline\hline
 $(1,0,-1)(1,2,1)$ & $
\begin{array}{c}
( \frac{1}{2}, \frac{1}{2}, \frac{1}{2}) \\
\underline{(-\frac{1}{2},-\frac{1}{2}, \frac{1}{2})} 
\end{array}
$ & $\frac{3}{8}$ & $
\begin{array}{c}
-\frac{1}{2} \\
\frac{3}{2}
\end{array}
$ & $\frac{3}{8}$ & $-\frac{1}{2}$ & $\frac{3}{4}$ & $
\begin{array}{c}
-1 \\
1
\end{array}
$ \\ \hline\hline
\end{tabular}
\]

{\centerline {{\bf TABLE B.6}: GSO projected states in
$\chi^{susy}_{(0,0);(2,0)}$}}

\medskip
\noindent
{\bf Spectrum of states of $4~1$}

The states in the $k=4$ minimal model are listed in the following tables
\[
.
\begin{tabular}{||l||l||l||l||}
\hline\hline
$n$ & ${\rm Representation}$ & $\Delta $ & $Q$ \\ \hline\hline
$0$ & $(0,0,0)$ & $0$ & $0$ \\ \hline\hline
$1$ & $(0,1,1)$ & $\frac{1}{12}$ & $\frac{1}{3}$ \\ \hline\hline
$2$ & $(0,2,2)\sim (4,-4,0)$ & $\frac{1}{3}$ & $\frac{2}{3}$ \\ \hline\hline
$3$ & $(4,-3,1)$ & $\frac{3}{4}$ & $1$ \\ \hline\hline
$4$ & $(4,-2,2)$ & $\frac{4}{3}$ & $\frac{4}{3}$ \\ \hline\hline
$5$ & $(4,-1,-1)\;$ & $\frac{13}{12}$ & $-\frac{1}{3}$ \\ \hline\hline
$6$ & $(4,0,0)$ & $1$ & $0$ \\ \hline\hline
$7$ & $(4,1,1)$ & $\frac{13}{12}$ & $\frac{1}{3}$ \\ \hline\hline
$8$ & $(4,2,2)$ & $\frac{4}{3}$ & $\frac{2}{3}$ \\ \hline\hline
$9$ & $(4,3,-1)$ & $\frac{3}{4}$ & $-1$ \\ \hline\hline
$10$ & $(4,4,0)$ & $\frac{1}{3}$ & $-\frac{2}{3}$ \\ \hline\hline
$11$ & $(4,5,1)$ & $\frac{1}{12}$ & $-\frac{1}{3}$ \\ \hline\hline
\end{tabular}
\begin{tabular}{||l||l||l||l||}
\hline\hline
$n$ & ${\rm Representation}$ & $\Delta $ & $Q$ \\ \hline\hline
$0$ & $(0,0,2)\;\sim (4,\pm 6,\pm 4)$ & $\frac{3}{2}$ & $\pm 1$ \\
\hline\hline
$1$ & $(4,-5,-3)\;$ & $\frac{13}{12}$ & $-\frac{2}{3}$ \\ \hline\hline
$2$ & $(4,-4,-2)\;$ & $\frac{5}{6}$ & $-\frac{1}{3}$ \\ \hline\hline
$3$ & $(4,-3,-1)\;$ & $\frac{3}{4}$ & $0$ \\ \hline\hline
$4$ & $(4,-2,0)$ & $\frac{5}{6}$ & $\frac{1}{3}$ \\ \hline\hline
$5$ & $(4,-1,1)$ & $\frac{13}{12}$ & $\frac{2}{3}$ \\ \hline\hline
$6$ & $(4,0,\pm 2)$ & $\frac{3}{2}$ & $\pm 1$ \\ \hline\hline
$7$ & $(4,1,-1)$ & $\frac{13}{12}$ & $-\frac{2}{3}$ \\ \hline\hline
$8$ & $(4,2,0)$ & $\frac{5}{6}$ & $-\frac{1}{3}$ \\ \hline\hline
$9$ & $(4,3,1)$ & $\frac{3}{4}$ & $0$ \\ \hline\hline
$10$ & $(4,4,2)$ & $\frac{5}{6}$ & $\frac{1}{3}$ \\ \hline\hline
$11$ & $(4,5,3)$ & $\frac{13}{12}$ & $\frac{2}{3}$ \\ \hline\hline
\end{tabular}
\]

\[
.
\begin{tabular}{||l||l||l||l||}
\hline\hline
$n$ & ${\rm Representation}$ & $\Delta $ & $Q$ \\ \hline\hline
$0$ & $(2,0,0)$ & $\frac{1}{3}$ & $0$ \\ \hline\hline
$1$ & $(2,1,1)$ & $\frac{5}{12}$ & $\frac{1}{3}$ \\ \hline\hline
$2$ & $(2,2,2)$ & $\frac{2}{3}$ & $\frac{2}{3}$ \\ \hline\hline
$3$ & $(2,\pm 3,\pm 3)$ & $\frac{13}{12}$ & $\pm 1$ \\ \hline\hline
$4$ & $(2,-2,-2)$ & $\frac{2}{3}$ & $-\frac{2}{3}$ \\ \hline\hline
$5$ & $(2,-1,-1)\;$ & $\frac{5}{12}$ & $-\frac{1}{3}$ \\ \hline\hline
\end{tabular}
\begin{tabular}{||l||l||l||l||}
\hline\hline
$n$ & ${\rm Representation}$ & $\Delta $ & $Q$ \\ \hline\hline
$0$ & $(2,0,\pm 2)\;$ & $\frac{5}{6}$ & $\pm 1$ \\ \hline\hline
$1$ & $(2,1,-1)\;$ & $\frac{5}{12}$ & $-\frac{2}{3}$ \\ \hline\hline
$2$ & $(2,2,0)\;$ & $\frac{1}{6}$ & $-\frac{1}{3}$ \\ \hline\hline
$3$ & $(2,3,1)\;$ & $\frac{1}{12}$ & $0$ \\ \hline\hline
$4$ & $(2,4,2)$ & $\frac{1}{6}$ & $\frac{1}{3}$ \\ \hline\hline
$5$ & $(2,-1,1)$ & $\frac{5}{12}$ & $\frac{2}{3}$ \\ \hline\hline
\end{tabular}
\]

{\centerline {{\bf TABLE B.7}: Representations in $k=4$ minimal model}}

\medskip
\noindent
{\bf Spectrum of states of $1^6$}

\[
\begin{tabular}{||l||l||l||l||l||l||l||l||}
\hline\hline
 ${\rm Internal~Theory}$ & ${\rm spacetime}$ & $\Delta
_{st}$ & $Q_{st}$ & $\Delta _{int}$ & $Q_{int}$ & $\Delta $ & $Q$
\\ \hline\hline
  $(0,0,0)^{6}$ & $\underline{(\pm 1,0)}$ & $\frac{1}{2}$ & $\pm
1 $ & $0$ & $0$ & $\frac{1}{2}$ & $\pm 1$  \\ \hline\hline
 $(0,1,1)^{6}$ & $\underline{(-\frac{1}{2}, \frac{1}{2})}$ & $\frac{1}{4}$ & $0$ & $\frac{1}{4}$ & $1$ & $%
\frac{1}{2}$ & $1$  \\ \hline\hline
 $(0,-1,-1)^{6}$ & $\underline{(-\frac{1}{2}, \frac{1}{2})}$ & $\frac{1}{4}$ & $0$ & $\frac{1}{4}$ & $-1$ & $%
\frac{1}{2}$ & $-1$  \\ \hline\hline
\end{tabular}
\]

{\centerline {{\bf TABLE B.8}: GSO projected states in $\chi^{susy}_{(0,0)^6}$}}

\medskip

\[
\begin{tabular}{||l||l||l||l||l||l||l||l||}
\hline\hline
  ${\rm Internal~Theory}$ & ${\rm spacetime}$ & $\Delta
_{st}$ & $Q_{st}$ & $\Delta _{int}$ & $Q_{int}$ & $\Delta $ & $Q$
 \\ \hline\hline
 $(0,0,0)^{4}(1,-1,2)(1,1,0)$ & $%
(0,0) $ & $0$ & $0$ & $\frac{5}{6}$ & $-1$ & $\frac{5}{6}$ & $-1$  \\
\hline\hline
 $(0,0,0)^{4}(1,-1,0)(1,1,2)$ & $(0,0)$ & $0$ & $0$ & $\frac{5}{6}$ & $1$ &
$\frac{5}{6}$ & $1$  \\ \hline\hline
 $(0,0,0)^{4}(1,-1,0)(1,1,0)$ & $\underline{(\pm 1,0)}$ & $\frac{1}{2}$ & $\pm 1$ & $\frac{1%
}{3}$ & $0$ & $\frac{5}{6}$ & $\pm 1$  \\ \hline\hline
$(1,-1,0)^{4}(1,1,0)(0,0,0)$ & $(0,0)$ & $0$ & $0$ & $\frac{5}{6}$ & $1$ &
$\frac{5}{6}$ & $1$ \\ \hline\hline
$(1,1,0)^{4}(0,0,0)(1,-1,0)$ & $(0,0)$ & $0$ & $0$ & $\frac{5}{6}$ & $-1$
& $\frac{5}{6}$ & $-1$ \\ \hline\hline
$(0,1,1)^{4}(1,0,-1)(1,-1,-1)$ & $(\frac{1}{2}, \frac{1}{2})$ & $\frac{1}{4}$ & $1$ & $\frac{7}{%
12}$ & $0$ & $\frac{5}{6}$ & $1$  \\ \hline\hline
$(0,1,1)^{4}(1,0,1)(1,-1,-1)$ & $(-\frac{1}{2},-\frac{1}{2})$ & $\frac{1}{4}$ & $-1$ & $\frac{7}{%
12}$ & $0$ & $\frac{5}{6}$ & $-1$ \\ \hline\hline
 $(0,1,1)^{4}(1,0,1)(1,-1,-1)$ & $\underline{(\frac{1}{2},-\frac{1}{2})}$ & $\frac{1}{4}$ & $0$ & $\frac{7}{%
12}$ & $1$ & $\frac{5}{6}$ & $1$  \\ \hline\hline
 $(0,-1,-1)^{4}(0,1,1)(1,0,1)$ & $(\frac{1}{2}, \frac{1}{2})$ & $\frac{1}{4}$ & $1$ & $\frac{7}{%
12}$ & $0$ & $\frac{5}{6}$ & $1$  \\ \hline\hline
 $(0,-1,-1)^{4}(0,1,1)(1,0,1)$ & $(-\frac{1}{2},-\frac{1}{2})$ & $\frac{1}{4}$ & $-1$ & $\frac{7}{%
12}$ & $0$ & $\frac{5}{6}$ & $-1$  \\ \hline\hline
$(0,-1,-1)^{4}(0,1,1)(1,0,-1)$ & $\underline{(\frac{1}{2},-\frac{1}{2})}$ & $\frac{1}{4}$ & $0$ & $\frac{7}{%
12}$ & $-1$ & $\frac{5}{6}$ & $-1$  \\ \hline\hline
\end{tabular}
\]
{\centerline {{\bf TABLE B.9}: GSO projected states in
$\chi^{susy}_{(0,0)^4(1,-1)(1,1)}$}}

\medskip

\[
\begin{tabular}{||l||l||l||l||l||l||l||l||}
\hline\hline
${\rm Internal~Theory}$ & ${\rm spacetime}$ & $\Delta
_{st}$ & $Q_{st}$ & $\Delta _{int}$ & $Q_{int}$ & $\Delta $ & $Q$
\\ \hline\hline
  $(0,0,0)^{3}(1,-1,0)^{3}$ & $(0,0)$ &
$0$ & $0$ & $\frac{1}{2}$ & $1$ & $\frac{1}{2}$ & $1$ \\ \hline\hline
$(1,1,0)^{3}(0,0,0)^{3}$ & $(0,0)$ & $0$ & $0$ & $\frac{1}{2}$ & $-1$ & $%
\frac{1}{2}$ & $-1$ \\ \hline\hline
$(0,-1,-1)^{3}(0,1,1)^{3}$ & $
\begin{array}{c}
( \frac{1}{2}, \frac{1}{2}) \\
(-\frac{1}{2},-\frac{1}{2}) 
\end{array}
$ & $\frac{1}{4}$ & $
\begin{array}{c}
+1 \\
-1
\end{array}
$ & $\frac{1}{4}$ & $0$ & $\frac{1}{2}$ & $
\begin{array}{c}
+1 \\
-1
\end{array}
$ \\ \hline\hline
\end{tabular}
\]

{\centerline {{\bf TABLE B.10}: GSO projected states in
$\chi _{(0,0)^{3}(1,-1)^{3}}^{susy}$
}}

\medskip
\noindent
{\bf Spectrum of states of $3^5$}

The states in the $k=3$ model may be classified in 2 groups: those with
$l=0,3$ and those with $l=1,2$. They are

\[
\begin{tabular}{||l||l||l||l||l||l||}
\hline\hline
${\rm Representation}$ & $\Delta $ & $Q$ & ${\rm Representation}$ &
$\Delta $ & $Q$ \\ \hline\hline
$(0,0,0)$ & $0$ & $0$ & $(0,0,2)$ & $\frac{3}{2}$ & $\pm 1$ \\ \hline\hline
$(3,-3,0)$ & $\frac{3}{10}$ & $\frac{3}{5}$ & $(3,-3,-2)$ & $\frac{4}{5}$ & $%
-\frac{2}{5}$ \\ \hline\hline
$(3,-1,\pm 2)$ & $\frac{6}{5}$ & $-\frac{4}{5},\frac{6}{5}$ & $(3,-1,0)$ & $%
\frac{7}{10}$ & $\frac{1}{5}$ \\ \hline\hline
$(3,1,0)$ & $\frac{7}{10}$ & $-\frac{1}{5}$ & $(3,1,\pm 2)$ & $\frac{6}{5}$
& $\frac{4}{5},-\frac{6}{5}$ \\ \hline\hline
$(3,3,2)$ & $\frac{4}{5}$ & $\frac{2}{5}$ & $(3,3,0)$ & $\frac{3}{10}$ & $-%
\frac{3}{5}$ \\ \hline\hline
\end{tabular}
\]

\[
\begin{tabular}{||l||l||l||l||l||l||}
\hline\hline
${\rm Representation}$ & $\Delta $ & $Q$ & ${\rm Representation}$ &
$\Delta $ & $Q$ \\ \hline\hline
$(1,-1,0)$ & $\frac{1}{10}$ & $\frac{1}{5}$ & $(1,-1,-2)$ & $\frac{3}{5}$ & $%
-\frac{4}{5}$ \\ \hline\hline
$(1,1,2)$ & $\frac{3}{5}$ & $\frac{4}{5}$ & $(1,1,0)$ & $\frac{1}{10}$ & $-%
\frac{1}{5}$ \\ \hline\hline
$(2,-2,2)$ & $\frac{7}{10}$ & $\frac{7}{5}$ & $(2,-2,0)$ & $\frac{1}{5}$ & $%
\frac{2}{5}$ \\ \hline\hline
$(2,0,\pm 2)$ & $\frac{9}{10}$ & $\pm 1$ & $(2,0,0)$ & $\frac{2}{5}$ & $0$
\\ \hline\hline
$(2,2,0)$ & $\frac{1}{5}$ & $-\frac{2}{5}$ & $(2,2,2)$ & $\frac{7}{10}$ & $-%
\frac{7}{5}$ \\ \hline\hline
\end{tabular}
\]

{\centerline {{\bf TABLE B.11}: Representations in $k=3$ minimal model}}

\medskip
\noindent
where both groups contain also the states obtained by twisting.

It is useful to list the GSO projected combinations with conformal weight
 $\frac{1}{2}$ in the NS sector, namely
\[
\begin{tabular}{||l||l||l||}
\hline\hline
${\rm Internal~Theory}$ & ${\rm spacetime}$ & ${\rm Character}$ \\ \hline\hline
$(0,0,0)^{5}$ & $\pm 1$ & $\chi _{(0,0)^{5}}$ \\ \hline\hline
$(0,0,0)^{3}(3,-3,0)(2,-2,0)$ & $0$ & $\chi _{(0,0)^{3}(3,-3)(2,-2)}$ \\
\hline\hline
$(0,0,0)^{2}(3,-3,0)(1,-1,0)^{2}$ & $0$ & $\chi
_{(0,0)^{2}(3,-3)(1,-1)^{2}}$
\\ \hline\hline
$(1,1,0)^{5}$ & $0$ & $\chi _{(2,0)^{5}}$ \\ \hline\hline
$(0,0,0)(1,-1,0)^{3}(2,-2,0)$ & $0$ & $\chi _{(0,0)(1,-1)^{3}(2,-2)}$ \\
\hline\hline
$(0,0,0)^{2}(1,-1,0)(2,-2,0)^{2}$ & $0$ & $\chi
_{(0,0)^{2}(1,-1)(2,-2)^{2}}$
\\ \hline\hline
\end{tabular}
\]

{\centerline {{\bf TABLE B.12}: GSO projected states with conformal weight
$\frac 12$}}

\medskip
\noindent
and their charge conjugated ones.

\end{document}